\documentclass[prd,twocolumn,preprintnumbers,showpacs,nofootinbib,superscriptaddress,notitlepage,floatfix]{revtex4-1}
\usepackage{hyperref}
\usepackage{graphicx}
\usepackage{amsmath}
\usepackage{amssymb}
\usepackage{amsfonts}
\usepackage{xcolor}
\usepackage{multirow}
\usepackage{slashed}
\usepackage[caption=false]{subfig}

\newcommand{\MS}{\overline{\mathrm{MS}}}
\newcommand{\RI}{\mathrm{RI}^\prime\mathrm{/MOM}}
\newcommand{\ns}{{\text{ns}}}
\newcommand\TMD{\mathrm{TMD}}

\DeclareRobustCommand{\eq}[1]{Eq.~\eqref{eq:#1}}

\DeclareRobustCommand{\fig}[1]{Fig.~\ref{fig:#1}}

\DeclareRobustCommand{\eq}[1]{Eq.~(\ref{eq:#1})}

\newcommand{\df}{\mathrm{d}}

\newcommand{\eps}{\epsilon}

\newcommand{\cO}{\mathcal{O}}

\newcommand{\nn}{\nonumber}

\newcommand{\bt}{\vec b_T}

\newcommand\bets{\begin{table*}}
\newcommand\eets[1]{\label{tb:#1}\end{table*}}

\begin{document}

\title{Lattice QCD calculation of the Collins-Soper kernel from quasi TMDPDFs}

\preprint{FERMILAB-PUB-21-326-T,MIT-CTP/5316}

\author{Phiala Shanahan}
 \email{phiala@mit.edu}
 \affiliation{Center for Theoretical Physics, Massachusetts Institute of Technology, Cambridge, MA, USA 02139}%
 
\author{Michael Wagman}%
 \email{mwagman@fnal.gov}
  \affiliation{Fermi National Accelerator Laboratory, Batavia, IL 60510, USA}  
  
\author{Yong Zhao}%
 \email{yong.zhao@anl.gov}
 \affiliation{Physics Division, Argonne National Laboratory, Lemont, IL 60439, USA}
 \affiliation{Physics Department, Brookhaven National Laboratory, Bldg. 510A, Upton, NY 11973, USA}

\begin{abstract}
     This work presents a lattice quantum chromodynamics (QCD) calculation of the nonperturbative Collins-Soper kernel, which describes the rapidity evolution of quark transverse-momentum-dependent parton distribution functions. The kernel is extracted at transverse momentum scales in the range 400~MeV~$< q_T < 1.7$~GeV in a calculation with dynamical fermions and quark masses corresponding to a larger-than-physical pion mass, $m_\pi=538(1)$~MeV. It is found that different approaches to extract the Collins-Soper kernel from the same underlying lattice QCD matrix elements yield significantly different results and uncertainty estimates, revealing that power corrections, such as those associated with higher-twist effects, and perturbative matching between quasi and light-cone beam functions, cannot be neglected.
\end{abstract}

\maketitle

\section{Introduction}

Transverse-momentum-dependent parton distribution
functions (TMDPDFs) describe the intrinsic transverse momentum $q_T$ of the partonic constituents of hadrons~\cite{Collins:1981uk,Collins:1981va,Collins:1984kg}. These non-perturbative functions can be accessed directly in high-energy scattering processes such as Drell-Yan production and semi-inclusive deep-inelastic scattering with small transverse hadron momentum $q_T$~\cite{Scimemi:2019cmh,Bacchetta:2019sam}, and indirectly through other processes such as studies of hadrons in jets~\cite{Buffing:2018ggv,Gutierrez-Reyes:2019vbx}. Significant efforts are underway to improve constraints on TMDPDFs both from current and planned experiments~\cite{Gautheron:2010wva,Dudek:2012vr,Aschenauer:2015eha,Accardi:2012qut,AbdulKhalek:2021gbh} and through theory calculations in the framework of lattice quantum chromodynamics (QCD)~\cite{Musch:2010ka,Musch:2011er,Engelhardt:2015xja,Yoon:2016dyh,Yoon:2017qzo,Shanahan:2019zcq,Shanahan:2020zxr,Zhang:2020dbb,Schlemmer:2021aij,Li:2021wvl} using approaches such as large-momentum effective theory (LaMET)~\cite{Ji:2013dva,Ji:2014gla,Ji:2020ect} or the Lorentz-invariant method based on ratios of TMDPDFs~\cite{Musch:2010ka}.

TMDPDFs $f_i^\text{TMD}(x,b_T,\mu,\zeta)$, defined for a parton of flavor $i$ in a given hadron state, are functions of the longitudinal momentum fraction $x$ of the parton, the Fourier conjugate $b_T$ of $q_T$, the virtuality scale $\mu$, and the rapidity scale $\zeta$ which is related to the hadron momentum. While the $\mu$-evolution of TMDPDFs is perturbative for perturbative scales $\mu$ and $\zeta$, the $\zeta$-evolution is governed by the Collins-Soper evolution kernel (or anomalous dimension) $\gamma_\zeta^i(\mu,b_T)$, which is nonperturbative for scales $b_T \sim q_T^{-1}\sim\Lambda^{-1}_\mathrm{QCD}$, even if both $\mu$ and $\zeta$ are perturbative. Constraints on the kernel $\gamma_\zeta^i(\mu,b_T)$ in the nonperturbative region are necessary in order to relate TMDPDFs determined from experiment or lattice QCD at different scales. 

Recently, it was shown in Refs.~\cite{Ebert:2018gzl,Ebert:2019okf,Ebert:2019tvc} that the Collins-Soper kernel can be calculated from ratios of quasi TMDPDFs at different hadron momenta, quantities which are both calculable in lattice QCD and which can be related to TMDPDFs~\cite{Ji:2014hxa,Ji:2018hvs,Ebert:2019okf,Ji:2019sxk,Ji:2019ewn}. This provides a pathway to first-principles QCD calculations of the kernel in the nonperturbative region, which will provide valuable complementary information to constraints from global analyses of experimental data. 
This prospect has motivated a series of proof-of-principle lattice QCD investigations of the Collins-Soper kernel both directly~\cite{Shanahan:2020zxr,Shanahan:2019zcq,Zhang:2020dbb,Schlemmer:2021aij,Li:2021wvl} through the approach of Refs.~\cite{Ebert:2018gzl,Ebert:2019okf,Ebert:2019tvc} and via related prescriptions~\cite{Vladimirov:2020ofp}. 

In this work, a direct calculation of the Collins-Soper kernel is presented, based on a lattice QCD study with dynamical fermions and quark masses corresponding to a larger-than-physical pion mass $m_\pi=538(1)$~MeV, and a single value of the lattice spacing and volume. The kernel is extracted at transverse momentum scales in the range 400~MeV~$< q_T < 1.7$~GeV and compared with phenomenological parametrizations and existing lattice QCD calculations. This analysis includes several advances over previous lattice QCD studies of the Collins-Soper kernel via the same approach. In particular, matching of quasi TMDPDFs and TMDPDFs is performed to one-loop order, the mixing of different TMDPDFs under renormalization is fully accounted for, and the analysis includes improved treatments of power corrections and systematic effects arising from the finite lattice volume and various statistical limitations of the calculation. It is found that different approaches to extract the Collins-Soper kernel from the same underlying lattice QCD matrix elements yield significantly different results and uncertainty estimates, revealing that power corrections, such as those associated with higher-twist effects, and perturbative matching between quasi and light-cone beam functions, cannot be neglected.

The method by which the Collins-Soper kernel can be computed following Refs.~\cite{Ebert:2018gzl,Ebert:2019okf} is detailed in Section~\ref{sec:formalism}. The lattice QCD calculation is reported in Section~\ref{sec:lattice}, while a summary and outlook is provided in Section~\ref{sec:outlook}.

\section{Quasi TMDPDFs and the Collins-Soper kernel}
\label{sec:formalism}

The quark Collins-Soper kernel $\gamma^q_\zeta(\mu, b_T)$ can be computed in lattice QCD from a ratio of nonsinglet quasi TMDPDFs $\tilde f_{\ns}^\TMD$ at different hadron momenta (taken in the $z$-direction) $P^z_i \gg \Lambda_\text{QCD}$~\cite{Ebert:2018gzl,Ebert:2019okf,Ji:2019ewn}:
\begin{align}\label{eq:gamma_zeta1}
	\gamma_\zeta^q(\mu, b_T) & 
= \frac{1}{\ln(P^z_1/P^z_2)}\nn\\
&\quad\times \ln \frac{C^\TMD_\ns (\mu,x P_2^z)\, \tilde f_{\ns}^\TMD(x, \bt, \mu, P_1^z)}	{C^\TMD_\ns (\mu,x P_1^z)\, \tilde f_{\ns}^\TMD(x, \bt, \mu, P_2^z)}\nn\\
&\quad +{\cal O}\Big({1\over (xP^zb_T)^2}\,,{\Lambda_{\rm QCD}^2\over (xP^z)^2}\Big)\,.
\end{align}
The perturbative matching coefficient $C^\TMD_\ns$ relates the quasi TMDPDFs, which are defined in terms of Euclidean-space matrix elements as detailed below, to the corresponding light-cone TMDPDFs through a factorization theorem based on an expansion in powers of the nucleon momentum~\cite{Ebert:2018gzl,Ebert:2019okf,Ji:2019sxk,Ji:2019ewn}. 
Additional nonperturbative factors related to the soft sector~\cite{Ji:2018hvs,Ebert:2019okf} cancel in the ratio; recently exploratory lattice QCD studies of these factors have been performed~\cite{Zhang:2020dbb,Li:2021wvl} following the approach proposed in Refs.~\cite{Ji:2019sxk,Ji:2019ewn}.
The flavor nonsinglet unpolarized quark quasi TMDPDF is defined as $\tilde f^{\mathrm{TMD}}_\ns=\tilde f^{\mathrm{TMD}}_u-\tilde f^{\mathrm{TMD}}_d$, where
\begin{align}\label{eq:quasiTMD}
&\tilde{f}_{i}^{\mathrm{TMD}}\big(x, \vec{b}_{T}, \mu, P^{z}\big)\equiv  \lim_{\substack{a\to 0 \\ \eta\to \infty}}
\int \frac{\mathrm{d} b^{z}}{2 \pi} e^{-\mathrm{i}b^{z}\left(x P^{z}\right)} \mathcal{Z}^{\MS}_{\gamma^4\Gamma}(\mu,b^z\!,a)\nonumber\\
&\qquad\qquad\times{P^z\over E_{\vec{P}}}\tilde{B}^{\Gamma}_{i}\big(b^{z},\vec{b}_{T},a,\eta,P^{z}\big) \tilde{\Delta}_{S}\left(b_{T},a,\eta\right).
\end{align}
Here $a$ denotes the lattice spacing, and $E_{\vec{P}}=\sqrt{\vec{P}^2+m_h^2}$ where $\vec{P}=P^z\vec{e}_z$ is the hadron three-momentum and $m_h$ is the hadron mass. The factor $\mathcal{Z}^{\MS}_{\gamma^4\Gamma}(\mu,b^z,a)$, where $\Gamma$ is a Dirac matrix label, renormalizes the quasi TMDPDF and matches it to the $\MS$ scheme at scale $\mu$~\cite{Constantinou:2019vyb,Ebert:2019tvc,Shanahan:2019zcq}, and  the quasi soft factor  $\tilde{\Delta}_S$~\cite{Ji:2014hxa,Ji:2018hvs,Ebert:2018gzl,Ebert:2019okf} and quasi beam function $\tilde{B}^\Gamma_i$ are both calculable in lattice QCD. Summation over $\Gamma$ is implied, accounting for operator mixing between quasi TMDPDFs with different Dirac structures resulting from the breaking of rotational and chiral symmetries in lattice QCD calculations~\cite{Constantinou:2019vyb,Shanahan:2019zcq,Green:2020xco,Ji:2021uvr}. Mixing with gluon operators is neglected in Eq.~\eqref{eq:quasiTMD}, but cancels in the nonsinglet combination of quasi TMDPDFs. It should be noted that the choice of the Dirac structure $\gamma^4$ in Eq.~\eqref{eq:quasiTMD} is not unique; the quasi TMDPDF with Dirac structure $\gamma^3$ can also be boosted onto $\gamma^+$ and thus be matched to the spin-independent TMDPDF in the infinite-momentum limit (in that case, the factor of $P^z/E_{\vec{P}}$ in Eq.~\eqref{eq:quasiTMD} is replaced by 1). While the notation is specialized to $\gamma^4$ for clarity throughout this exposition, numerical results are presented for both choices of Dirac structure in Sec.~\ref{sec:lattice}.

\begin{figure}[!t]
    \centering
    \includegraphics[width=0.7\columnwidth]{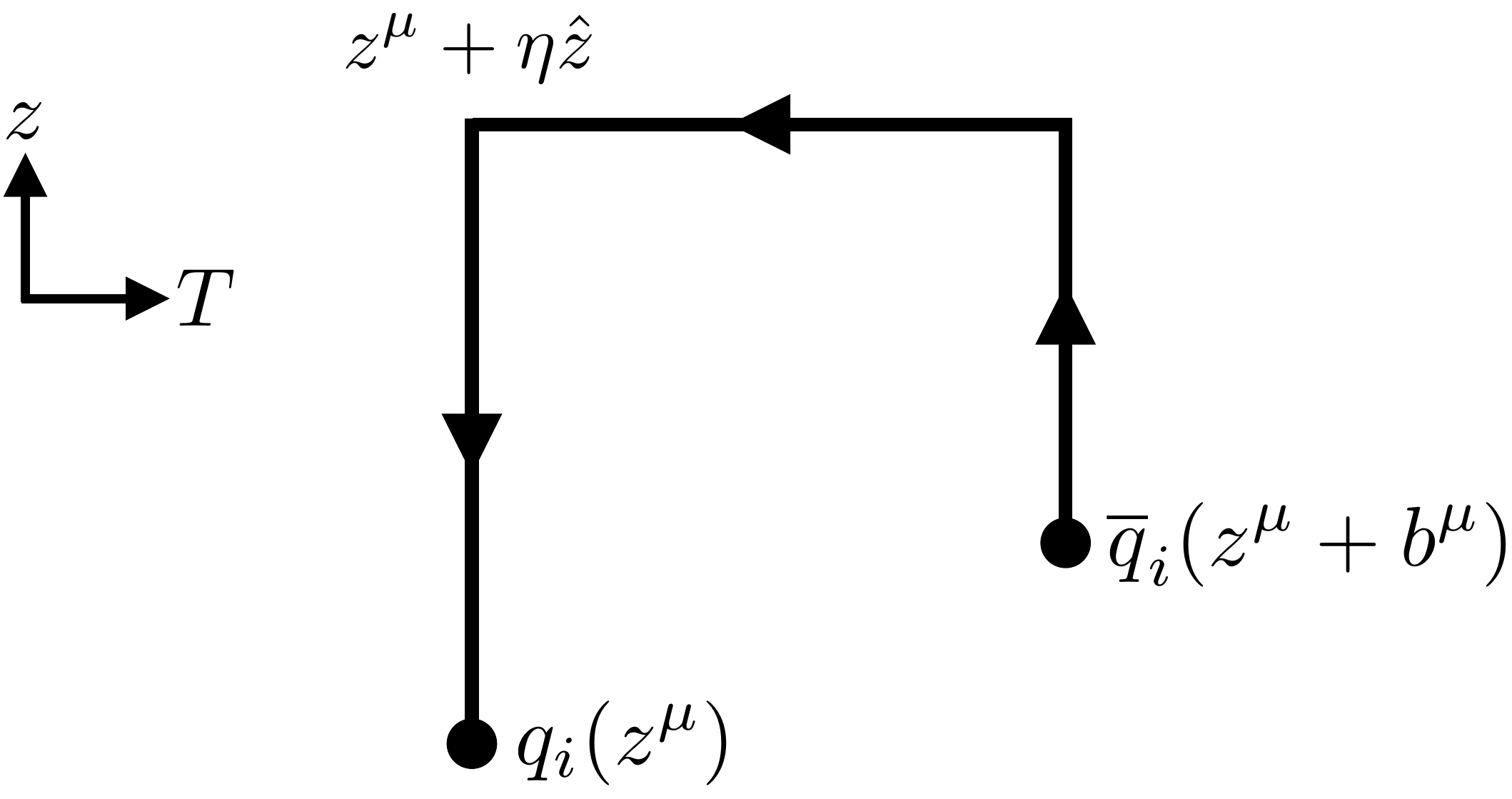}
    \caption{Diagrammatic representation of the Wilson line included in the operators $\mathcal{O}^i_\Gamma(b^\mu,z^\mu,\eta)$, see Eq.~\eqref{eq:op}.}
    \label{fig:staple}
\end{figure}

The quasi beam function $\tilde{B}^\Gamma_i$ is defined as the matrix element of a nonlocal quark bilinear operator with a staple-shaped Wilson line in a boosted hadron state:
\begin{align} \label{eq:qbeam}
\tilde B^\Gamma_{i}(b^z, \bt,a,\eta,P^z)
=&
\Bigl\langle h(P^z) \big| \mathcal{O}_\Gamma^{i}(b^\mu,0,\eta) \big| h(P^z) \Bigr\rangle\,,
\end{align}
where $h(P^z)$ denotes the state of hadron $h$ with four-momentum $P^{\mu}=(0,0,P^z,E_{\vec{P}})$. The operator $\mathcal{O}_\Gamma^{i}(b^\mu,0,\eta)$, depicted in \fig{staple}, is defined as
\begin{align}\nonumber
\mathcal{O}^{i}_\Gamma(b^\mu,z^\mu,\eta)\equiv &\ \bar q_i(z^\mu + b^\mu) \frac{\Gamma}{2} W_{\hat z}(z^\mu + b^\mu ;\eta-b^z) \\\nonumber
&\ \times 
W^\dagger_{T}(z^\mu + \eta \hat{z}; b_T) W^\dagger_{\hat z}(z^\mu;\eta) q_i(z^\mu)\\\label{eq:op}
\equiv&\ \bar q_i(z^\mu + b^\mu) \frac{\Gamma}{2}\widetilde{W}(\eta;b^\mu;z^\mu)q_i(z^\mu),
\end{align}
where $b^\mu = (\bt,b^z,0)$, and $W_{\hat \alpha}(x^\mu;\eta)$ denotes a Wilson line beginning at $x^\mu$ with length $\eta$ in the direction of ${\hat \alpha}$. The subscript $T$ denotes that the associated Wilson line is in a direction transverse to $\hat{z}^\mu=(0,0,1,0)$. 

In practice, it is useful to define a dimensionless `bare' nonsinglet quasi beam function:
\begin{align}
    \nonumber
B^\text{bare}_\Gamma(b^z, \bt,a,\eta,P^z)  \equiv & \frac{1}{2 E_{\vec{P}}}\left(\tilde B^\Gamma_{u}(b^z, \bt,a,\eta,P^z)\right. \\
\label{eq:qbeambare}& \left. \hspace{6mm}- \tilde B^\Gamma_{d}(b^z, \bt,a,\eta,P^z)\right),
\end{align}
as well as a modified $\MS$-renormalized quasi beam function $B^{\MS}_{\Gamma}$~\cite{Shanahan:2020zxr}:
\begin{align}\nonumber
   B^{\MS}_{\gamma^4}(\mu,b^z, \bt, a, \eta, P^z)& \equiv  Z_{\cO_{\gamma^4\Gamma}}^{\MS}(\mu,b^z, b_T^R,a,\eta) \\\label{eq:BMSbar}
    &\hspace{-25mm} \times \tilde{R}(b_T,b_T^R,a,\eta) B^{\text{bare}}_{\Gamma}(b^z, \bt, a, \eta, P^z).
\end{align}
Compared with the standard $\MS$-renormalized quasi beam function, this definition includes the additional factor $\tilde{R}$, described further below. The renormalization factor $Z_{\cO_{\gamma^4\Gamma}}^{\MS}$ is defined as the product of a regularization-independent momentum subtraction scheme ($\RI$) factor $Z^{\RI}_{\cO_{\gamma^4\Gamma}}$ and a perturbative matching factor to the $\overline{\text{MS}}$ scheme,  $\mathcal{R}^{\MS}_{\cO_{\gamma^4\Gamma}}$, which has been calculated at next-to-leading order in continuum perturbation theory with dimensional regularization ($D=4-2\eps$)~\cite{Constantinou:2019vyb,Ebert:2019tvc}:
\begin{align}\nonumber
    Z_{\cO_{\gamma^4\Gamma}}^{\MS}(\mu,b^z\!,b_T,a,\eta) = & \mathcal{R}^{\MS}_{\cO_{\gamma^4\Gamma}}(\mu,p_R,b^z,\vec{b}_T,\eta)\\\label{eq:ZOMS}
    &\times Z^{\RI}_{\cO_{\gamma^4\Gamma}}(p_R,b^z\!,\vec{b}_T,a,\eta).
\end{align}
In this expression, the dependence on the $\RI$ scale $p_R$ and on the direction of $\vec{b}_T$ cancels between $Z^{\RI}_{\cO_{\gamma^4\Gamma}}$ and $\mathcal{R}^{\MS}_{\cO_{\gamma^4\Gamma}}$ at all orders in perturbation theory (up to discretization artifacts).
The quasi beam function renormalization factor is related to the TMDPDF renormalization factor $\mathcal{Z}^{\MS}_{\gamma^4\Gamma}$ in Eq.~\eqref{eq:quasiTMD} as
\begin{align}\label{eq:separatedZ}
\mathcal{Z}_{\gamma^4\Gamma}^{\MS}(\mu,b^z\!,a) = Z_{\cO_{\gamma^4\Gamma}}^{\MS}(\mu,b^z\!,b_T,a,\eta) Z_{S}^{\MS}(\mu,b_T,a,\eta),
\end{align}
where $Z_{S}^{\MS}$ renormalizes the quasi soft factor $\tilde{\Delta}_S$. The $\eta$ and $b_T$-dependence on the right-hand side of \eq{separatedZ} describes linear power divergences proportional to $\eta/a$ and $b_T/a$ which cancel between the two terms.

To ensure that the matching factor $\mathcal{R}^{\MS}_{\cO_{\gamma^4\Gamma}}$ is in the perturbative region, both $Z^{\RI}_{\cO_{\gamma^4\Gamma}}$ and $\mathcal{R}^{\MS}_{\cO_{\gamma^4\Gamma}}$ should be computed at a scale $b_T^R\ll \Lambda_{\rm QCD}^{-1}$. In \eq{BMSbar}, this scale is taken to be distinct from $\vec{b}_T$ which is associated with the staple geometry of the operator defining the bare quasi beam function. As a result, $Z_{\cO_{\gamma^4\Gamma}}^{\MS}$ cannot completely cancel the ultraviolet (UV) divergence in the bare quasi beam function, and remnant linear divergences $\sim |b_T-b_T^R|/a$ appear in \eq{BMSbar}. The factor $\tilde{R}$ is included to cancel such divergences.
One possible choice of $\tilde{R}$ is~\cite{Shanahan:2019zcq}
\begin{align}\label{eq:R}
    &\tilde{R}(b_T,b_T^R,a,\eta) = \frac{{Z}_{\cO_{\gamma^4\gamma^4}}^{ \RI}(p_R=\tilde{p}_R,b^z=0,\vec{b}_T,a,\eta)}{{Z}_{\cO_{\gamma^4\gamma^4}}^{ \RI}(p_R=\tilde{p}_R,b^z=0,\vec{b}_T^R,a,\eta)}\,,
 \end{align}
defined for a fixed choice of $\tilde{p}_R$, and of the directions of $\vec{b}_T$ and $\vec{b}^R_T$. An alternative choice of $\tilde{R}$ is defined and used in Ref.~\cite{Ebert:2019tvc}.

In terms of the modified $\MS$-renormalized quasi beam functions, the Collins-Soper kernel may be computed as
\begin{align}\label{eq:finalCSexpression}
&\gamma^q_\zeta(\mu, b_T)  
= \frac{1}{\ln(P^z_1/P^z_2)}\ln\Biggr[ \frac{C^\TMD_\ns (\mu,x P_2^z)}{C^\TMD_\ns (\mu,x P_1^z)}\nn\\
&\! \times\! \frac{\int\! \df b^z e^{-ib^z\! xP_1^z} P_1^z \lim_{\substack{a\to 0 \\ \eta\to \infty}}B^{\MS}_{\gamma^4}(\mu,b^z, \bt, a, \eta, P_1^z)}
    {\int\! \df b^z e^{-ib^z\! xP_2^z}\!  P_2^z \lim_{\substack{a\to 0 \\ \eta\to \infty}}B^{\MS}_{\gamma^4}(\mu,b^z, \bt, a, \eta,  P_2^z)}\Biggr]\nn\\
&\qquad\qquad +{\cal O}\Big({1\over (xP^zb_T)^2}\,,{\Lambda_{\rm QCD}^2\over (xP^z)^2}\Big)\,.
\end{align}
Since $\tilde{\Delta}_{S}$ and its renormalization factor $Z_{S}^{\MS}$, as well as the factor $\tilde R$ included in the definition of $B^{\MS}_{\gamma^4}$, are independent of $b^z$, this expression is equivalent to that in Eq.~\eqref{eq:gamma_zeta1}. The specific choice of $\tilde R$ (including the choices of $\tilde{p}_R$ and $\bt$ and $\vec{b}^R_T$ orientations) does not affect the value of $\gamma^q_\zeta$.

\section{Numerical study}
\label{sec:lattice}

The Collins-Soper kernel is computed via Eq.~\eqref{eq:finalCSexpression} in a lattice QCD calculation with four dynamical quark flavors, using an ensemble of gauge field configurations generated by the MILC collaboration with the one-loop Symanzik improved gauge action and the highly improved staggered quark action. A single ensemble is studied, with a lattice volume of $L^3 \times T = 48^3 \times 64$, a lattice spacing corresponding to $a = 0.12$ fm, and sea quark masses tuned to approximately match the physical quark masses in nature; see Ref.~\cite{Bazavov:2012xda} for further details of the ensemble generation. Calculations are performed in a partially-quenched mixed-action setup, with the tree-level $\mathcal{O}(a)$-improved Wilson clover fermion action used for the valence quarks, with $\kappa=0.1241$ tuned such that the pion mass is $m_\pi=538(1)$~MeV. The gauge fields used in the calculation have been subjected to Wilson flow to flow-time $\mathfrak{t} = 1.0$~\cite{Luscher:2010iy}, to enhance the signal-to-noise ratio in the numerical results{\footnote{Note that the flowed gauge fields were also used for constructing $\slashed{D}$.}}. The following sections detail the computation of each element of $\gamma^q_\zeta$.

\subsection{Bare quasi beam functions}
\label{subsec:barebeam}

Quasi beam functions in a pion external state are computed from ratios of three-point and two-point correlation functions:
\begin{align}\nonumber
    &\mathcal{R}_\Gamma(t,\tau,b^\mu,a,\eta,P^z) \\\nonumber
    &\qquad= \frac{C_{\text{3pt}}^{\Gamma,u}(t,\tau,b^\mu,a,\eta,P^z\vec{e}_z)-C_{\text{3pt}}^{\Gamma,d}(t,\tau,b^\mu,a,\eta,P^z\vec{e}_z)}{C_\text{2pt}(t,P^z\vec{e}_z)}\\\label{eq:Rcal}
    &\qquad\xrightarrow{t\gg \tau \gg 0}B^\text{bare}_\Gamma(b^z, \bt,a,\eta,P^z) + \ldots,
\end{align}
where
\begin{align}\nonumber
   &C_{\text{3pt}}^{\Gamma,i}(t,\tau,b^\mu,a,\eta,\vec{P}=P^z \vec{e}_z) \\\nonumber 
   &\ = \sum_{\vec{x},\vec{z}}e^{i\vec{P}\cdot\vec{x}}\langle 0 | \pi_{\vec{P},S} (\vec{x},t) \mathcal{O}^i_\Gamma(b^\mu,(\vec{z},\tau),\eta) \pi_{\vec{P},S}^\dagger (0)|0\rangle\\
    &\xrightarrow{t\gg \tau \gg 0}\frac{Z_{\vec{P}}}{4 a E^2_{\vec{P}}}e^{-E_{\vec{P}}t}\tilde B^\Gamma_{i}(b^z, \bt,a,\eta,P^z)+\ldots
\end{align}
and
\begin{align}\nonumber\label{eq:twopt}
   C_\text{2pt}(t,\vec{P}) &= \sum_{\vec{x}}e^{i \vec{P}\cdot\vec{x}}\langle 0| \pi_{\vec{P},S} (\vec{x},t) \pi_{\vec{P},S}^\dagger (0)|0\rangle\\
    & \overset{t\gg 0}{\longrightarrow} \frac{Z_{\vec{P}}}{2 a E_{\vec{P}}}e^{-E_{\vec{P}}t}+\ldots.
\end{align}
In the construction of the correlation functions, momentum-smeared interpolating operators $\pi_{\vec{P},S}(\vec{x},t) = \overline{u}_{S(\vec{P}/2)}(\vec{x},t)\gamma_5d_{S(\vec{P}/2)}(\vec{x},t)$ are built from quasi local smeared quark fields $q_{S(\vec{P})}(\vec{x},t)$ constructed with 50 steps of iterative Gaussian momentum-smearing~\cite{Bali:2016lva} with smearing radius $\varepsilon = 0.2$. $Z_{\vec{P}}$ denotes the combination of overlap factors for the source and sink interpolating operators.

\begin{figure}[t!]
    \centering
     \includegraphics[width=\linewidth]{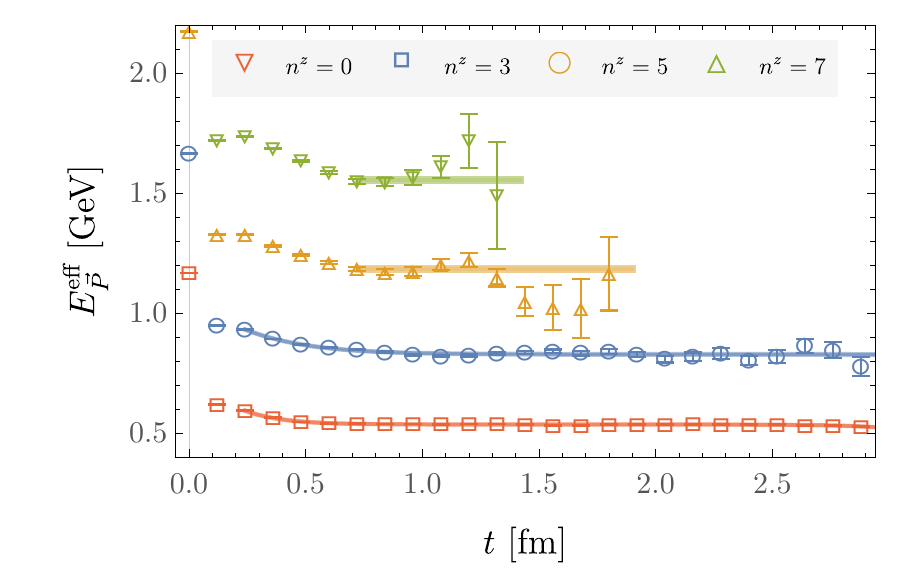}
    \caption{\label{fig:EMP}  Effective energy function defined in Eq.~\eqref{eq:Eeff} for pion states boosted in the $\hat{z}$-direction with $P^z = n^z 2\pi /L$. Shaded bands display the result of single-exponential fits to the two-point correlation functions with the largest two momenta, and two-exponential fits to those with the smallest two momenta, obtained as described in the text. }
\end{figure}

Two and three-point functions are constructed for three choices of pion boost $\vec{P}=P^z\vec{e}_z$, with $P^z=n^z 2\pi/L$ for $n_z\in\{3,5,7\}$. An effective energy function that asymptotes to $E_{\vec{P}}$ can be defined as
\begin{align}\label{eq:Eeff}\nonumber
  E^{\text{eff}}_{\vec{P}}(t) &= \frac{1}{a} \text{arccosh}{\left({\frac{C_\text{2pt}(t+a,\vec{P})+C_\text{2pt}(t-a,\vec{P})}{2C_\text{2pt}(t,\vec{P})}}\right)}\\
    & \overset{t\gg 0}{\longrightarrow} E_{\vec{P}} + \ldots,
\end{align}
where the ellipses denote exponentially-suppressed corrections from excited states.
This function is shown for the ensemble investigated here in Fig.~\ref{fig:EMP}, including the results of fits to the two-point correlation functions. The fits are performed as described in Appendix A of Ref.~\cite{Shanahan:2020zxr}, with the number of states in each fit chosen by maximizing an information criterion, and different choices of fit range sampled and combined in a weighted average. The relative deviations in the extracted energies from the continuum dispersion relation $E_{\vec{P}}=\sqrt{m_\pi^2+|\vec{P}|^2}$ are at most 5\% for all momenta studied, increasing with increasing $|\vec{P}|$, but consistent with the expected size of lattice artifacts.

\begin{ruledtabular}
\begin{table}[t]
	\begin{tabular}{ccccc}
		$n_z$ & $P^z$ [GeV] & $\eta/a$ & $n_\text{src} $ & $n_\text{cfg}$\\\hline
		3 & 0.65 & \{12,14\} & 4 & 96 \\
		3 & 0.65 & 23 & 16 & 100 \\
		5 & 1.1 &\{12,14\} & 4 & 449 \\
		7 & 1.5 &\{12,14\} & 16 & 596
	\end{tabular}
	\caption{\label{tab:measurements}
	Quasi beam functions are computed for $n_\text{src}$ source locations on each of $n_\text{cfg}$ configurations for each pion boost $P^z$. Matrix elements of operators with staple widths $\vec{b}_T$ in the positive $\hat{x}$ direction with $0\le|b_T|\le \eta$ and asymmetries $-\eta \le b^z\le 
	\eta$ are computed.
	}
\end{table}
\end{ruledtabular}

The ratio $\mathcal{R}_\Gamma$ defined in Eq.~\eqref{eq:Rcal} is constructed for all Dirac structures $\Gamma$ and a range of operators with different staple widths and asymmetries, detailed in Table~\ref{tab:measurements}. As indicated, the number of measurements is varied for the different boost momenta, to partially compensate for the differences in statistical noise. All ratios are computed for sink times $t\in\{7,9,11,13\}$ and with all operator insertion times $\tau$ such that $0<\tau<t$. The fitting procedure used to determine $B^\text{bare}_\Gamma$ at each set of parameters is precisely the same procedure as detailed in Appendix A of Ref.~\cite{Shanahan:2020zxr}. Several examples of the fits in $t$ and $\tau$ used to extract the quasi beam functions are shown in Appendix~\ref{app:ttaufits}.
An example of the resulting bare quasi beam functions, for particular choices of $b_T$ and $\eta$, is shown in Fig.~\ref{fig:bare_beam_eg}. Additional examples are provided in Appendix~\ref{app:barebeams}.

\begin{figure*}[t]
	\centering
	\includegraphics[width=0.49\linewidth]{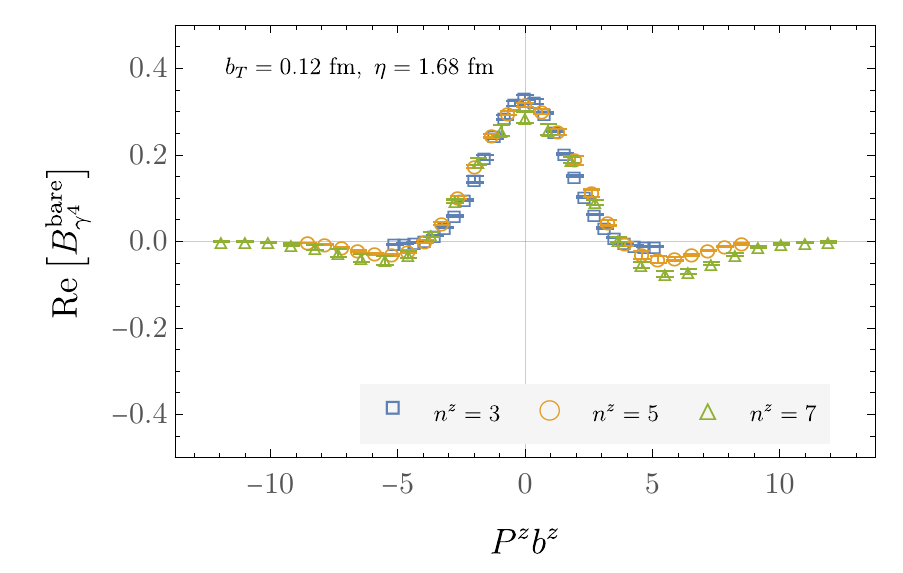}
	\includegraphics[width=0.49\linewidth]{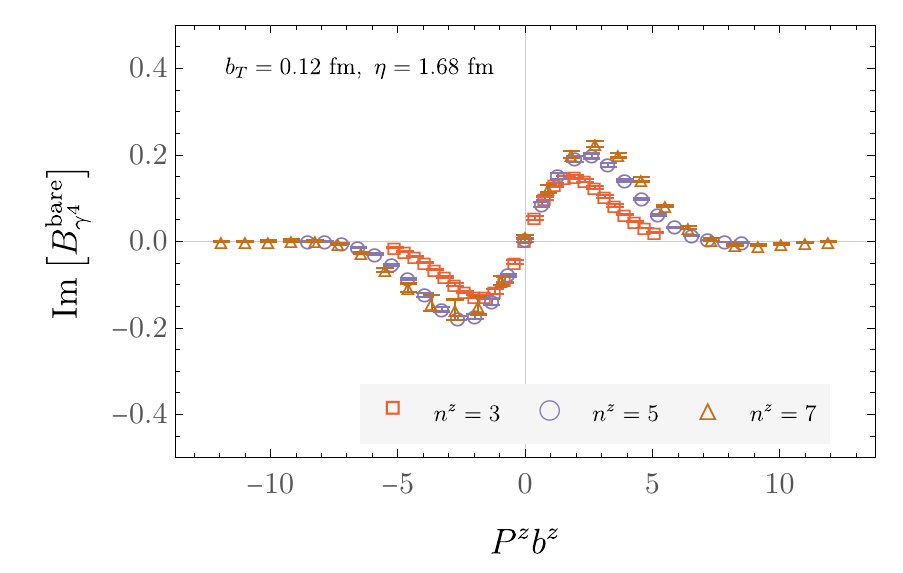} 
	\caption{ An example of a bare quasi beam function, computed as described in the text, for $b_T/a=1$ and $\eta/a=14$. Further examples are included in Appendix~\ref{app:barebeams}.
		\label{fig:bare_beam_eg}}
\end{figure*}

\subsection{Non-perturbative renormalization}
\label{subsec:renorm}

Computing the modified $\MS$-renormalized quasi beam function $B^{\MS}_{\Gamma}$ by Eq.~\eqref{eq:BMSbar} requires, in addition to the bare quasi beam functions, the $\RI$ renormalization factor $Z^{\RI}_{\cO_{\gamma^4\Gamma}}$. This factor enters the calculation of the renormalized quasi beam function both through the renormalization itself (Eq.~\eqref{eq:ZOMS}) and in the computation of the factor $\tilde{R}$ (Eq.~\eqref{eq:R}). The calculation of the nonperturbative renormalization undertaken here follows closely the presentation of Ref.~\cite{Shanahan:2019zcq}.

The matrix $Z^\text{$\RI$}_{\mathcal{O}_{\Gamma\Gamma'}}$ is defined by the renormalization condition 
\begin{equation}\label{eq:RIMOM}
   Z_q^{-1}(p_R) Z^\text{$\RI$}_{\mathcal{O}_{\Gamma\Gamma'}}(p_R)\Lambda^{\mathcal{O}_{\Gamma'}}_{\alpha\beta}(p) \big|_{p^\mu=p^\mu_R}=\Lambda_{\alpha\beta}^{\mathcal{O}_{\Gamma};\text{tree}}(p)\,,
\end{equation}
where dependence on the lattice spacing is left implicit and $\Lambda^{\mathcal{O}_{\Gamma};(\text{tree})}$ denotes the bare (tree-level) amputated Green's function of the operator $\mathcal{O}_\Gamma$ defined in Eq.~\eqref{eq:op} in an off-shell quark state in the Landau gauge:
\begin{equation}
   \Lambda^{\mathcal{O}_{\Gamma}} (p) =  S^{-1} (p) G^{\mathcal{O}_{\Gamma}}(p) S^{-1} (p) \,.
\end{equation}
Here $S(p)$ is the quark propagator projected to momentum $p$: 
\begin{align}
S_{\alpha\beta}(p,x)&=\sum_y e^{-ip\cdot y}\langle  q_\alpha(x)\bar q_\beta(y)\rangle, \\
S_{\alpha\beta}(p) &= \frac{1}{V}\sum_x e^{ip\cdot x} S_{\alpha\beta}(p,x),
\end{align}
and $G^{\mathcal{O}_{\Gamma}}$ denotes the non-amputated quark-quark Green's function with one insertion of $\cO_\Gamma$, which implicitly depends on the geometry of the staple-shaped Wilson line defining the operator:
\begin{align} \nonumber
   G^{\mathcal{O}_{\Gamma}}_{\alpha\beta} (p) &=\  \frac{1}{V}\sum_{x,y,z} {\rm e}^{ {\mathrm i} 
	p \cdot (x-y) } \langle q_\alpha (x) \cO_\Gamma (z+b,z) \bar{q}_\beta (y) \rangle, \\\label{Gdef}
	&=\ \frac{1}{V}\sum_{z} 
\langle\! \gamma_5 S^\dagger(p,\!b+z) \gamma_5 \widetilde{W}(\eta;b+z,\!z)\frac{\Gamma}{2} S(p,\!z) \!\rangle_{\alpha\beta},
\end{align}
where $V=L^3\times T$ is the lattice volume.
The quark wavefunction renormalization $Z_q$ is defined as
\begin{align}\nonumber
&Z_q(p_R) S(p)\big|_{p^2=p_R^2} = S^\text{tree}(p) \\\label{eq:Zq}
\implies &Z_q(p_R) =\ \frac{1}{12}\text{Tr}\left[S^{-1}(p)S^\text{tree}(p)\right]\bigg|_{p^2=p_R^2}\\\nonumber
& \hphantom{Z_q(p_R)}=\ \frac{ {\rm Tr} \left[ {\rm i} \sum_\lambda \gamma_\lambda 
	\sin (a p_\lambda)  S^{-1} (p) \right] }
{12 \sum_\lambda \sin^2 (a p_\lambda) } \bigg|_{p^2=p_R^2}.
\end{align}

From Eq.~\eqref{eq:RIMOM}, the matrix of $\RI$ renormalization factors may be computed as
\begin{equation}\label{eq:ZO}
   \left(Z^\text{$\RI$}_{\mathcal{O}_{\Gamma\Gamma'}}(p_R)\right)^{-1}  = \frac{\mathcal{V}^{\mathcal{O}_{\Gamma\Gamma'}}(p)}{6e^{i p_R\cdot b}Z_q(p_R) }\bigg|_{p^\mu=p^\mu_R},
\end{equation}
where $b^\mu$ denotes the separation of the endpoints of the nonlocal operator $\mathcal{O}_\Gamma$, the projected vertex function is defined as
\begin{equation}
   \mathcal{V}^{\mathcal{O}_{\Gamma\Gamma'}}(p)\equiv \text{Tr} \left[  \Lambda^{\mathcal{O}_{\Gamma}}(p) \Gamma' \right],
\end{equation}
and the replacement
\begin{align}\label{eq:RIMOM2}
\text{Tr}\left[ \Lambda^{\mathcal{O}_{\Gamma};\text{tree}}(p_R) \Gamma' \right]= 6e^{i p_R\cdot b}\delta^{\Gamma\Gamma'}
\end{align}
has been made.
It should be noted that since the operator $\mathcal{O}^{q}_{\Gamma}$ is nonlocal and frame dependent, different directions in $p_R^\mu$ constitute different renormalization schemes, related by finite renormalization factors. As such, $Z^\text{$\RI$}_{\mathcal{O}_{\Gamma\Gamma'}}$ depends on $p_R^\mu$ itself rather than only on its magnitude, which acts as the renormalization scale.

\begin{figure}
    \centering
    \includegraphics[width=\columnwidth]{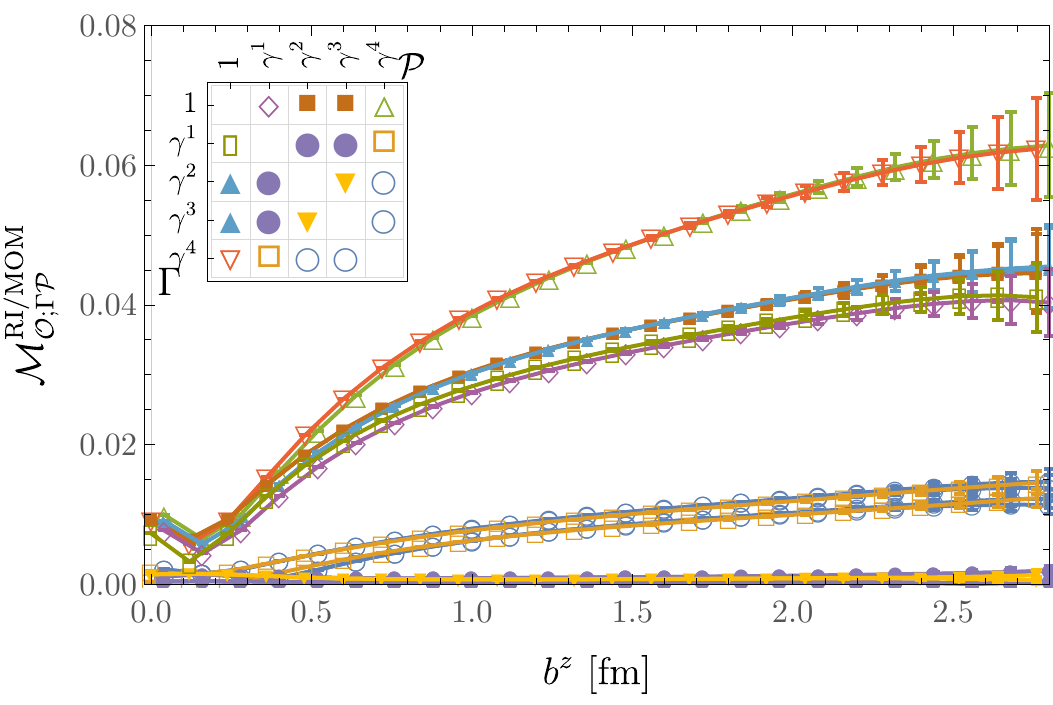}
    \caption{Submatrix of the $\RI$ mixing matrix $\mathcal{M}^\text{$\RI$}_{\mathcal{O}_{\Gamma\mathcal{P}}}$, defined in Eq.~\eqref{eq:mixeq}, computed for quark bilinear operators with straight Wilson lines ($b_T=0$) with various extents $b^z$, for momentum $n^\nu=(12,12,12,12)$ in lattice units. Points representing Dirac structures in the upper triangle of the mixing matrix are slightly offset on the horizontal axis for clarity.}
    \label{fig:straightvsbz}
\end{figure}

The complete $16 \times 16$ matrix of $\RI$ renormalization factors $Z^{\RI}_{\cO_{\Gamma\Gamma'}}$ is computed for the same set of operator geometries defined in Table~\ref{tab:measurements}, on $n_\text{cfg}=50$ gauge field configurations. For all operator geometries with $\eta/a\in\{12,14\}$, the renormalization factors are computed for a range of four-momenta tabulated in Tab.~\ref{tab:moms}, to enable an investigation of residual dependence on the renormalization scale (which would be canceled by an all-orders matching to the $\MS$ scheme) and discretization artifacts. For operator geometries with $\eta/a=23$, only the four-momentum with $n^\mu=(12,12,12,12)$ is used. An example of the resulting $\RI$ renormalization matrices is shown in Fig.~\ref{fig:straightvsbz}, which displays a subset of the off-diagonal renormalization factors normalized relative to the diagonal components:
\begin{align}\label{eq:mixeq}
   \mathcal{M}^\text{$\RI$}_{\mathcal{O}_{\Gamma\mathcal{P}}}
   \equiv& \frac{\text{Abs}[Z^\text{$\RI$}_{\mathcal{O}_{\Gamma\mathcal{P}}}]}{\frac{1}{16}\sum_i \text{Abs}[Z^\text{$\RI$}_{\mathcal{O}_{\Gamma_i\Gamma_i}}]}\,,
\end{align}
computed for quark bilinear operators with straight Wilson lines and a particular choice of momentum.

\begin{table}[t]
    \centering
    \begin{ruledtabular}
    \begin{tabular}{c|ccc}
       $n^\mu$  & $\sqrt{p^2}$ [GeV]  & $p^z$ [GeV] & $p^{[4]}/(p^2)^2$ \\ \hline
    (6,6,6,6) & 2.5 & 1.3 & 0.26 \\
    (6,6,6,9) & 2.7 & 1.3 & 0.26 \\
    (6,6,6,12) & 3.0 & 1.3 & 0.30 \\ \hline
    (8,8,8,8) & 3.3 & 1.7 & 0.26 \\
    (8,8,8,12) & 3.6 & 1.7 & 0.26 \\
    (8,8,8,16) & 4.0 & 1.7 & 0.30 \\ \hline
    (12,12,12,12) & 4.9 & 2.6 & 0.26 \\
    (12,12,12,18) & 5.4 & 2.6 & 0.25
    \end{tabular}
    \end{ruledtabular}
    \caption{Four-momenta used in the calculation of nonperturbative renormalization factors as described in the text, where $p^\mu$ is the four-momentum in physical units corresponding to $n^\mu$ in lattice units. The H(4) invariant $p^{[4]}$ is defined as $p^{[4]} = \sum_{\mu=1}^4 p_\mu^4$.}
    \label{tab:moms}
\end{table}

\begin{figure}[!t]
    \centering
    \includegraphics[width=0.46\textwidth]{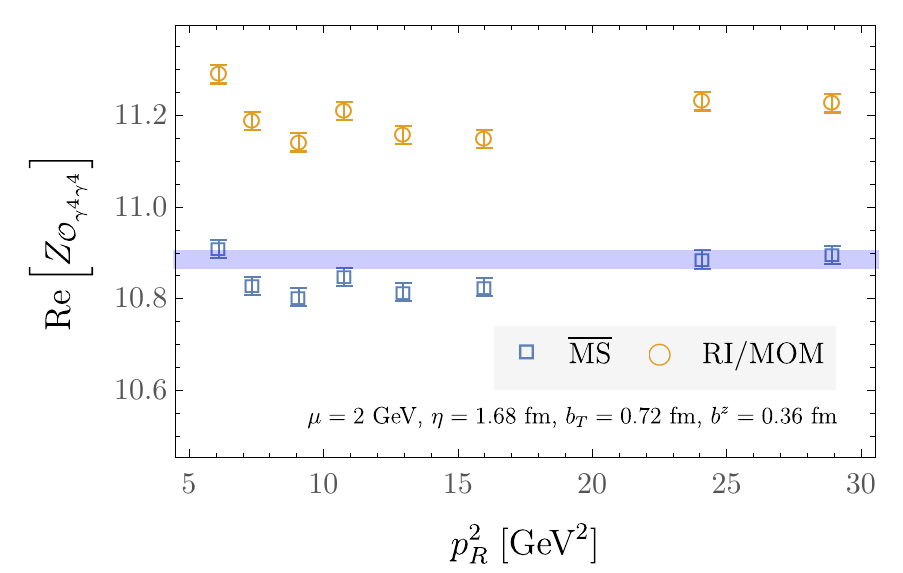}
    \caption{Example of numerical results for $Z^{\text{$\RI$}}_{\mathcal{O}_{\gamma^4\gamma^4}}$ (orange circles) and $Z^{\MS}_{\mathcal{O}_{\gamma^4\gamma^4}}$ (blue squares), computed using the different values of $p_R$ given in Table~\ref{tab:moms}, for operator geometry $\eta/a = 14$, $b^z/a = 3$, $b_T/a = 3$, at renormalization scale $\mu = 2$ GeV. The blue shaded band shows the value used in further analysis, obtained as described in the text.
    \label{fig:matching} }
\end{figure}

\begin{figure}[t]
    \subfloat[]{
        \centering
        \includegraphics[width=0.46\textwidth]{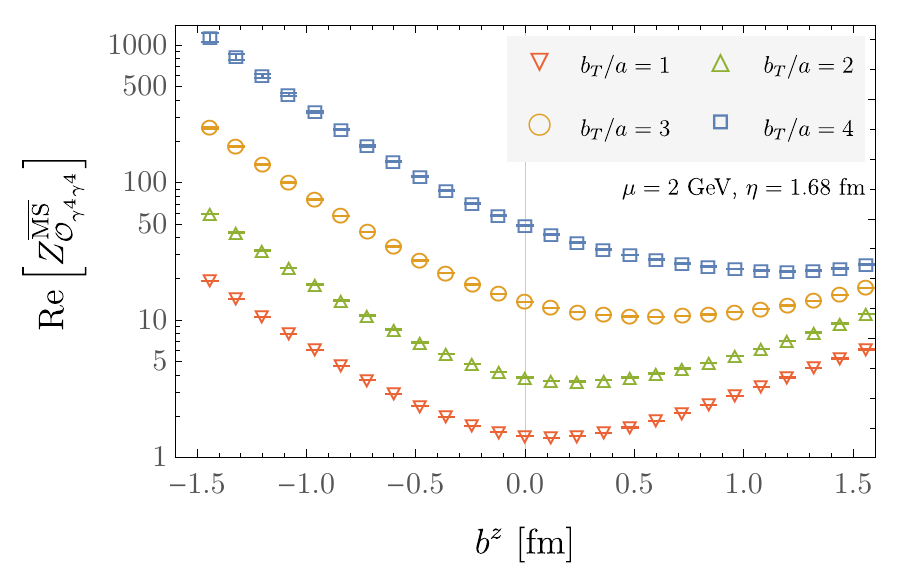}
        \label{fig:ZRvsBz}
        }\quad
    \subfloat[]{
        \centering
        \includegraphics[width=0.46\textwidth]{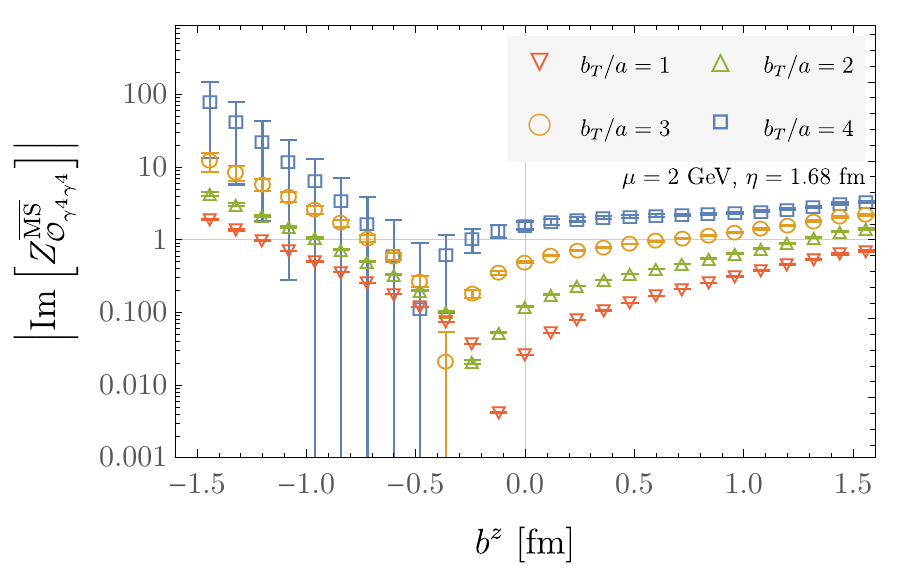}
        \label{fig:ZIvsBz} 
        }
        \caption{\label{fig:ZMSg4g4}
           Diagonal $\overline{\text{MS}}$ renormalization factors $Z^{\overline{\text{MS}}}_{\mathcal{O}_{\gamma^4\gamma^4}}$, computed for operator geometry $\eta/a = 14$ and various values of $b_T$ and $b^z$, at renormalization scale $\mu = 2$ GeV.}
\end{figure}

The $\MS$ renormalization factors are computed via Eq.~\eqref{eq:ZOMS} from the $\RI$ renormalization matrices and the perturbative matching factor $\mathcal{R}^{\MS}_{\cO_{\gamma^4\Gamma}}$, calculated at next-to-leading order in continuum perturbation theory with dimensional regularization ($D=4-2\eps$)~\cite{Constantinou:2019vyb,Ebert:2019tvc}. In this work, the scale $\tilde{p}_R$ is set to 4.9~GeV. Examples of the resulting $\MS$ renormalization factors are presented in Figs.~\ref{fig:matching} and \ref{fig:ZMSg4g4}. This renormalization is independent of $p_R$ up to discretization artifacts, two-loop perturbative  matching  corrections which are neglected here, and nonperturbative effects that vanish at asymptotically large $p^2_R$. While in principle one might fit to results generated at different $p_R$ values to constrain discretization effects, this is not feasible with the data generated here, and the covariance matrices for the nonlocal operators cannot be reliably estimated. The renormalization constants computed with $n^\mu=(12,12,12,12)$ are thus taken as best-estimates and used in the further analysis of the Collins-Soper kernel. For those operator structures where the $\RI$ renormalization factors have been computed at other choices of $p_R$, this yields indistinguishable results for the Collins-Soper kernel compared with results obtained using the weighted averaging procedure over momenta which is employed in Ref.~\cite{Shanahan:2019zcq}. The components of $\mathcal{M}^\text{$\RI$}_{\mathcal{O}_{\Gamma\mathcal{P}}}$ are of similar magnitude for both $\Gamma\in\{\gamma^3,\gamma^4\}$, which is consistent with the conclusion of Ref.~\cite{Ji:2021uvr} but is in contrast with the results of Refs.~\cite{Constantinou:2019vyb,Green:2020xco} which predict no mixing effects for $\Gamma=\gamma^3$ at ${\cal O}(a^0)$. The quasi beam functions with both Dirac structures are thus treated on equal footing in this work.

\subsection{Renormalized quasi beam functions}
\label{subsec:renbeam}

Modified $\MS$-renormalized quasi beam functions are computed via Eq.~\eqref{eq:BMSbar} from the bare quasi beam functions and $\MS$ renormalization factors calculated as described in Secs.~\ref{subsec:barebeam} and ~\ref{subsec:renorm}; the uncertainties of the two components are combined in quadrature. 
While for clarity all equations in this section are expressed for renormalized quasi beam functions defined with the Dirac structure $\gamma^4$, both $B^{\MS}_{\gamma^4}$ and $B^{\MS}_{\gamma^3}$ are computed and analyzed independently.

\begin{figure}[!t]
	\subfloat[Example of the $b^z$-dependence of the asymmetry in the renormalized quasi beam functions for a fixed choice of operator geometry with $\eta/a=14$, $b_T/a=1$.
	\label{fig:asymmetry_fit}]{
	\centering
	\includegraphics[width=0.46\textwidth]{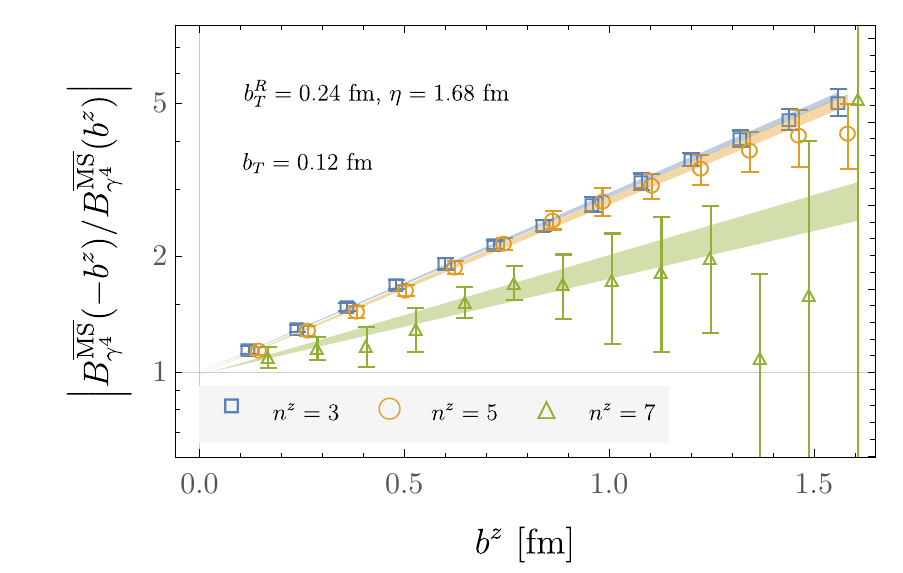}
	}\quad
	\subfloat[Example of the $b_T$-dependence of the asymmetry parameter $\Delta= -V(b_T)+V(b_T^R)$ for operators with $\eta/a=14$, fit to renormalized quasi beam functions $B^{\MS}_{\gamma^4}$ as described in the text. The approximate linear dependence of $\Delta$ on $b_T$ can be explained by a linear term in $V(b_T)$; the approximate independence on $n^z$ is expected, as $V(b_T)$ should be independent of the external state.
		\label{fig:asymmetry}]{
		\centering
	\includegraphics[width=0.46\textwidth]{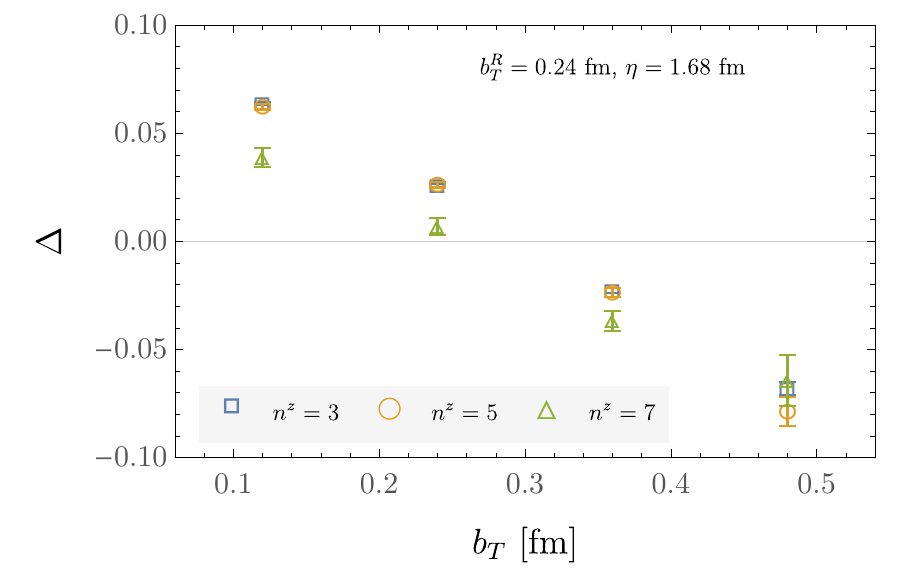}
	}
	\caption{Illustration of the asymmetry in the renormalized quasi beam functions computed in this work; additional examples are provided in Appendix~\ref{app:barebeams}.\label{fig:asymfig}}
\end{figure}

\begin{figure}[!t]
	\centering
	\subfloat[]{
   \includegraphics[width=0.46\textwidth]{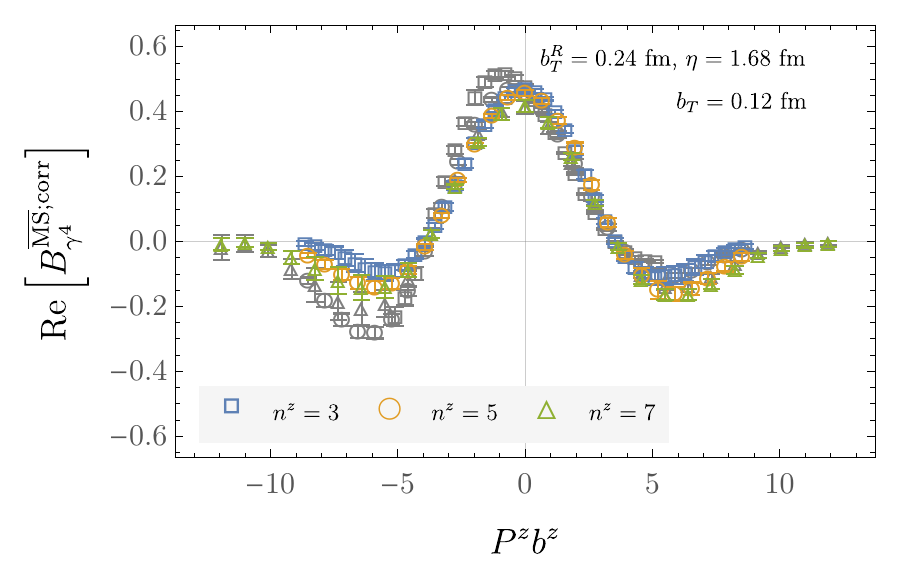}}\\
   \subfloat[]{
   \includegraphics[width=0.46\textwidth]{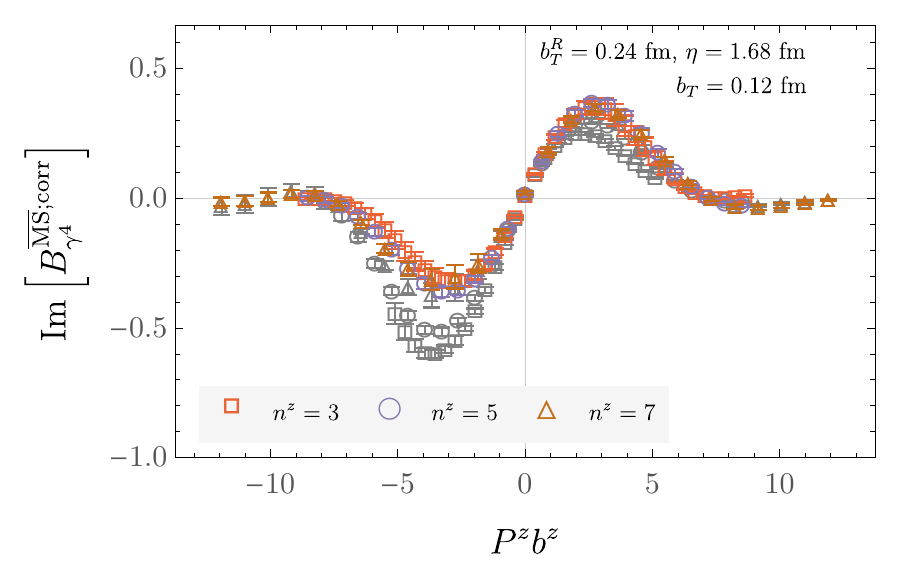}}
   \caption{ Comparison of an example of the modified $\MS$-renormalized quasi beam functions before (gray) and after (color) asymmetry correction via Eq.~\eqref{eq:asymcorr}, for beams functions computed with operator geometry $\eta/a=14$, $b_T/a=1$.
	\label{fig:asymmetry_comp}}
\end{figure}

The real and imaginary parts of the renormalized quasi beam functions should be symmetric and antisymmetric functions of $b^z$ respectively in the $\eta\rightarrow \infty$ limit. The numerical results obtained in this work, however, show significant departures from these expectations at finite $\eta$ as was also observed in the quenched calculation of Ref.~\cite{Shanahan:2019zcq}. The $b^z$-dependence of the asymmetry in the absolute value of the renormalized quasi beam functions is illustrated in Fig.~\ref{fig:asymmetry_fit}.
This asymmetry can be understood as an incomplete cancellation of the Wilson-line self-energy correction proportional to $e^{-V(b_T) \,b^z}$ between the $\MS$ renormalization factor and the bare quasi beam function, where $V(b_T)$ is the static quark-antiquark potential at distance $b_T$. Such an effect yields a $b^z$-dependent asymmetry proportional to $e^{-\Delta b^z}$, depending on an asymmetry parameter $\Delta= -V(b_T)+V(b_T^R)$.
This is in fact a good model of the asymmetry observed in the numerical calculations of this work; fits of $|B^{\MS}_{\gamma^4}(\mu,-b^z, \bt, a, \eta, P^z)|/|B^{\MS}_{\gamma^4}(\mu,b^z, \bt, a, \eta, P^z)|$ (and the analogous expression constructed from $B^{\MS}_{\gamma^3}$) to this functional form, fit separately for fixed values of $b_T$, $\eta$, and $P^z$, achieve acceptable goodness-of-fit with an average $\chi^2 / \text{d.o.f.} = 0.56$.
These fits, and the resulting values of the asymmetry parameter $\Delta$, are illustrated in Fig.~\ref{fig:asymfig} and in Appendix~\ref{app:barebeams}. Asymmetry-corrected modified $\MS$-renormalized quasi beam functions are thus defined as 
\begin{align}\nonumber
B^{\MS;\text{corr}}_{\gamma^4}(\mu,b^z,& \bt, a, \eta, P^z) =\ e^{\Delta(\bt,a,\eta,P^z)|b^z|}\\\nonumber
\times \left(\text{Re}\vphantom{B^{\MS}_{\gamma^4}}\right.& \!\!\left[B^{\MS}_{\gamma^4}(\mu,|b^z|, \bt, a, \eta, P^z)\right]\\\label{eq:asymcorr}
+&\left. \text{sign}(b^z)\,\text{Im}\!\! \left[B^{\MS}_{\gamma^4}(\mu,|b^z|, \bt, a, \eta, P^z)\right]\right).
\end{align}
Here, the uncertainties in the factor $e^{\Delta |b^z|}$ and the quasi beam functions are added in quadrature and, after asymmetry correction, the more precise results for beam functions with $b^z>0$ (which involve shorter Wilson lines) are mirrored to $b^z<0$. An example of the modified $\MS$-renormalized quasi beam functions before and after asymmetry correction is given in Fig.~\ref{fig:asymmetry_comp}.

\begin{figure}[!t]
    \subfloat[]{
        \centering
        \includegraphics[width=0.46\textwidth]{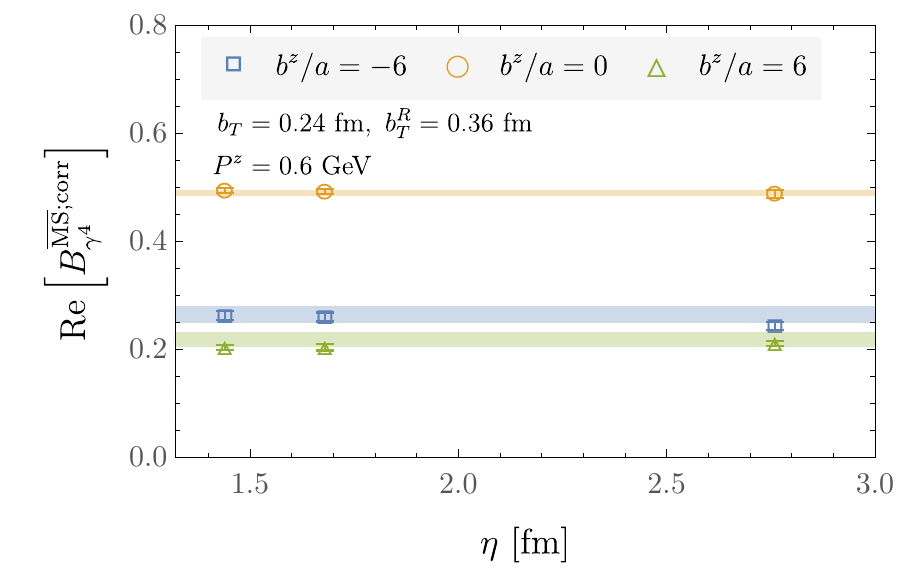}
        }\quad
    \subfloat[]{
        \centering
        \includegraphics[width=0.46\textwidth]{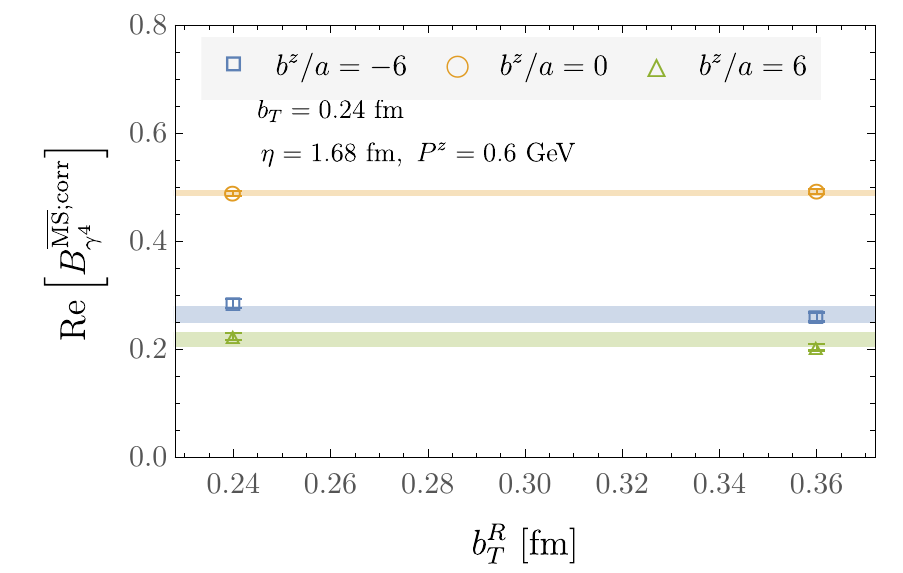}
        }\quad
        \caption{\label{fig:renorm_vs_bTR}  Example of the asymmetry-corrected modified $\MS$-renormalized quasi beam function $B^{\MS;\text{corr}}_{\gamma^4}$ defined in Eq.~\eqref{eq:BMSbar}. The horizontal shaded bands show the results of weighted averages of this quantity over choices of $b_T^R$ and $\eta$ (as a function of $b^z$, $b_T$, and $P^z$), as described in the text.}
\end{figure}

It is expected that the renormalized beam functions should be independent of the matching scale $b_T^R$. This expectation is satisfied within uncertainties for the numerical investigations of this work as illustrated in Fig.~\ref{fig:renorm_vs_bTR}; for use in the calculation of the Collins-Soper kernel, a weighted average~\cite{Aoki:2019cca} over possible choices of $b_T^R$ in the window $a \ll b_T^R \ll \Lambda_{\text{QCD}}^{-1}$ is thus implemented, performed precisely as detailed in Appendix C of Ref.~\cite{Shanahan:2020zxr}. Similarly, the asymmetry-corrected renormalized quasi beam functions do not depend on $\eta$ within uncertainties, and the formal extrapolation to $\eta \rightarrow \infty$ is implemented with an analogous weighted average. The resulting averaged values of the asymmetry-corrected modified $\MS$-renormalized quasi beam  function are denoted by $\overline{B}^{\MS}_{\gamma^4}(\mu,b^z,b_T,P^z)$, and have no dependence on $\eta$ or $b^R_T$; the dependence on $a$ is also dropped in this notation, as a single lattice spacing is used in this numerical study. Figures showing $\overline{B}^{\MS}_{\gamma^4}$ and $\overline{B}^{\MS}_{\gamma^3}$ are provided in Appendix~\ref{app:barebeams}.
The contributions to these quantities from mixing between operators with different Dirac structures is generally found to be less than $10\%$; see Fig.~\ref{fig:mixing_ratio} for an illustration. Additional examples are provided in Appendix~\ref{app:barebeams}. 

\begin{figure}[t]
 \subfloat[]{
        \centering
	\includegraphics[width=0.46\textwidth]{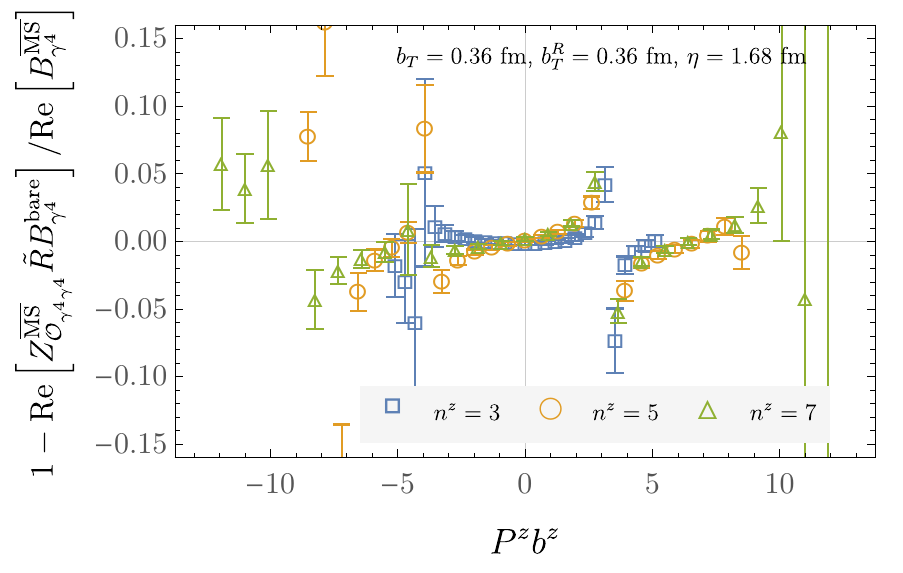}
	}\quad
	\subfloat[]{
        \centering
	\includegraphics[width=0.46\textwidth]{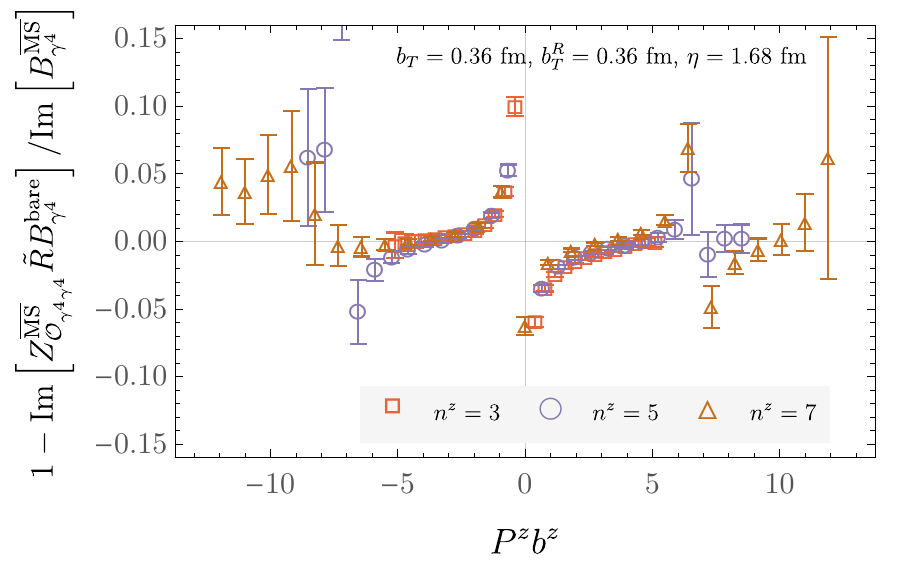}
	}
   \caption{Percentage contribution to the renormalized quasi beam functions from mixing of operators with different Dirac structures. Note that the ratios shown are outside of the plot range near the nodes of the beam functions; in this example the maximum mixing that is resolved from zero at greater than $2\sigma$ is $0.32(5)$, and occurs in the real component of the beam function for $n_z = 5$, $b_z/a = 5$.
      \label{fig:mixing_ratio}}
\end{figure}

\subsection{Collins-Soper kernel}
\label{sec:extr}

To determine the Collins-Soper kernel via Eq.~\eqref{eq:finalCSexpression} from the averaged asymmetry-corrected modified $\MS$-renormalized quasi beam functions $\overline{B}_{\gamma^4}^{\MS}$ and $\overline{B}_{\gamma^3}^{\MS}$, defined in the previous section, requires a Fourier transform of the quasi beam functions in $b^z$. As is clear from Fig.~\ref{fig:symm_beam}, however, the $b^z$-range of the data is not sufficient for the tails of the quasi beam functions at large $|b^z|$ to decay to plateaus consistent with zero, particularly at the largest $b_T$ and smallest $P^z$ values used in this numerical study. As such, it is to be expected that a discrete Fourier transform (DFT) will have significant systematic uncertainties from the truncation of the data in $P^z b^z$; details of a DFT analysis of the quasi beam functions are presented in Appendix~\ref{app:DFT}. Instead, Fourier transforms are taken after fitting the quasi beam functions to functional forms that allow extrapolation in $b^z$. This approach trades the systematic uncertainties of a DFT for model-dependence in the fit form used to extrapolate the quasi beam functions. While this model-dependence is difficult to quantify rigorously, this approach allows the physical expectation that the quasi beam functions should decay smoothly at large $|b^z|$ to be incorporated (a DFT effectively models the beam functions as zero outside the $b^z$-range in which lattice QCD results are computed).

In particular, the quasi beam functions are modeled as a sum of polynomials in $b^z$ times Gaussian functions, which provide an appropriate basis, since it is expected that the quasi beam functions should be analytic functions of $b^z$ (the $b^z\to0$ limit at fixed $b_T$ does not introduce additional divergences), and yield high-quality fits with few free parameters. Specifically, for each choice of $b_T$ and $P^z$, the real and imaginary parts of the quasi beam function $\overline{B}^{\MS}_{\gamma^4}$ (and independently $\overline{B}^{\MS}_{\gamma^3}$) are fit with even and odd functions of $b^z$ respectively, defined as
\begin{align}\label{eq:refit}
f_{\text{Re}}(\sigma,\{r_n\};b^z) &= \text{exp}[-(b^z)^2/(2 \sigma^2)] \sum_{n=0}^{n_\text{max}} r_n (b^z)^{2n}\\\label{eq:imfit}
f_{\text{Im}}(\sigma,\{r_n\};b^z) &= \text{exp}[-(b^z)^2/(2 \sigma^2)] \sum_{n=0}^{n_\text{max}} i_n (b^z)^{2n+1}.
\end{align}
The value of $n_{\text{max}}$ is chosen to minimize the Akaike information criterion (AIC)~\cite{1100705} and corresponds to $n_{\text{max}} \in \{2,3,4\}$ for all cases.
The fits with these optimal values of $n_{\text{max}}$ are of high quality in all cases, with an average $\chi^2 / \text{d.o.f.} = 0.41$.
The resulting models of the quasi beam functions are denoted $\hat{B}^{\MS}_{\gamma^4}$ (and, correspondingly $\hat{B}^{\MS}_{\gamma^3}$), and are illustrated in Fig.~\ref{fig:beam-fits} (with further examples provided in Appendix~\ref{app:barebeams}).

\begin{figure}[!t]
	\centering
	\subfloat[]{
   \includegraphics[width=0.46\textwidth]{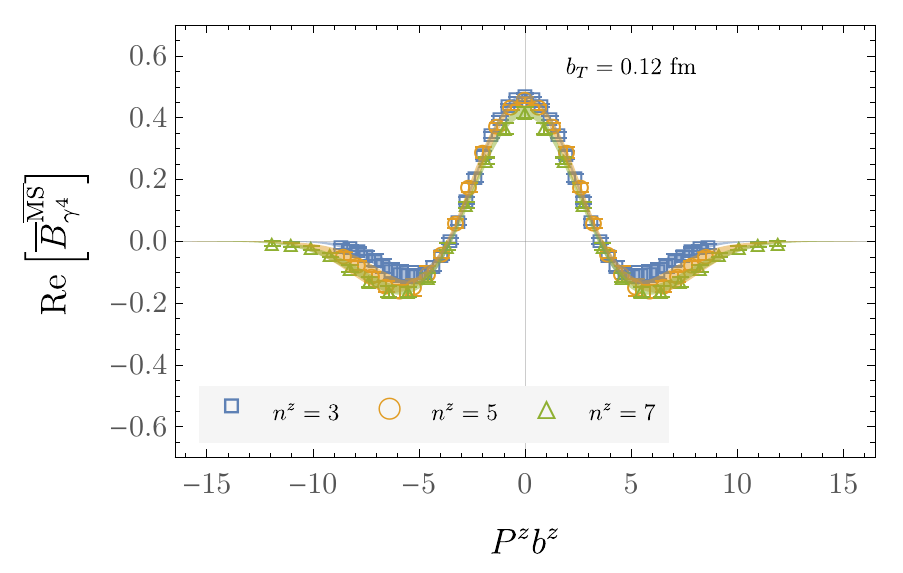}}\\
   \subfloat[]{
   \includegraphics[width=0.46\textwidth]{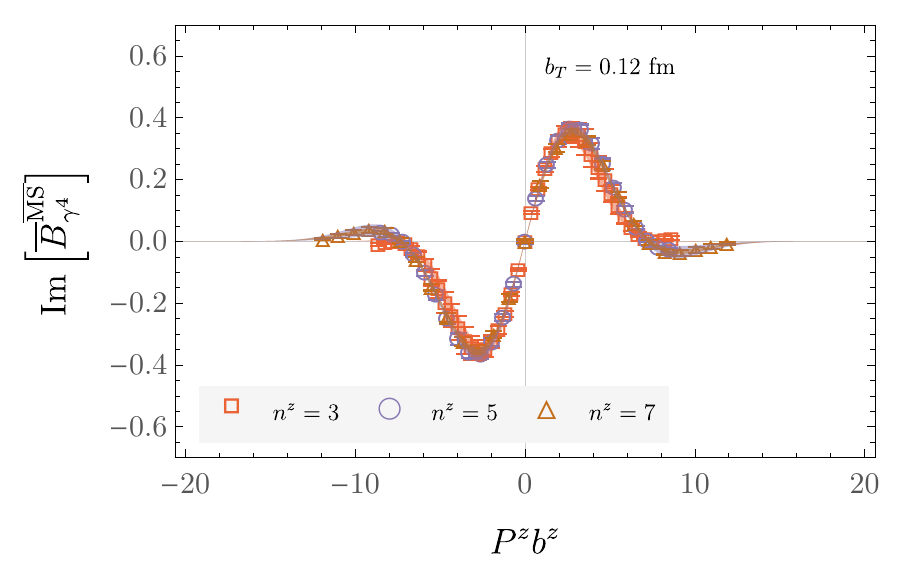}}
   \caption{ \label{fig:beam-fits}Example of fits by Eqs.~\eqref{eq:refit} and \eqref{eq:imfit} (shaded bands) to the real and imaginary parts of the quasi beam functions, for $b_T/a=1$. }
\end{figure}

Finally, in terms of the models $\hat{B}^{\MS}_{\gamma^4}$, the relation defining the Collins-Soper kernel in Eq.~\eqref{eq:finalCSexpression} is realized as
\begin{align}
\hat{\gamma}^q_\zeta(\mu, b_T; P_1^z,P_2^z,x)&
\equiv \frac{1}{\ln(P^z_1/P^z_2)}\ln\Biggr[ \frac{C^\TMD_\ns (\mu,x P_2^z)}{C^\TMD_\ns (\mu,x P_1^z)}\nn\\\label{eq:gammahat}
&\! \times\! \frac{\int\! \df b^z e^{-ib^z\! xP_1^z}  P_1^z \hat{B}^{\MS}_{\gamma^4}(\mu,b^z, b_T, P_1^z)}
    {\int\! \df b^z e^{-ib^z\! xP_2^z}\!  P_2^z \hat{B}^{\MS}_{\gamma^4}(\mu,b^z,b_T, P_2^z)}\Biggr],
\end{align}
which coincides with $\gamma^q_\zeta(\mu, b_T)$ up to power corrections such as higher-twist corrections in the factorization formula for the quasi TMDPDF, and discretization artifacts, which introduce the dependence on $P_1^z$, $P_2^z$, and $x$. One approach to determine $\gamma^q_\zeta(\mu, b_T)$ from $\hat{\gamma}^q_\zeta(\mu, b_T; P_1^z,P_2^z,x)$ is to model, fit, and subtract, these various artifacts. However, the most straightforward models of these effects do not provide good fits to the numerical data of this study, as detailed in Appendix~\ref{app:powerandtwist}. Instead, since the contamination in $\hat{\gamma}^q_\zeta$ will be different at each choice of $P_1^z$, $P_2^z$, and $x$, and the effects can be expected to be larger\footnote{The matching coefficient includes large logarithms of $xP^z_i$ at small $x$, while the quasi beam functions at $x\to0$ and $x\to1$ are sensitive to the long-range correlations in $b^z$ and are thus affected by the truncation of the data in $P^z b^z$. In addition, the power corrections are expected to be enhanced at small $x$.} at large and small values of $x$ and at small values of $P^z$, the variation of $\hat{\gamma}^q_\zeta$ over these choices is used to define an estimate of the systematic uncertainty.

\begin{figure}[!tp]
    \centering
    \includegraphics[width=0.46\textwidth]{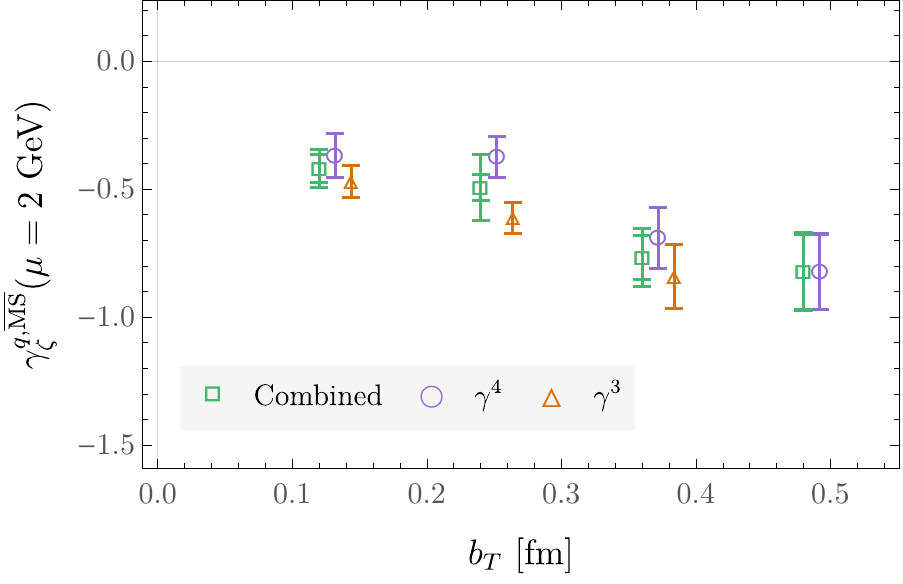}
    \caption{Collins-Soper kernel as a function of $b_T$, determined from $\hat{B}^{\MS}_{\gamma^4}$ (purple circles) $\hat{B}^{\MS}_{\gamma^3}$ (red triangles), and the final combined results of this work (green squares), computed as described in the text. For the latter points, the inner (outer) error bars show the first (quadrature-combined) uncertainties given in Table.~\ref{tab:CS}. No result computed from $\hat{B}^{\MS}_{\gamma^3}$ is shown at the largest $b_T$ value because in this case $\hat{\gamma}^q_\zeta$ cannot be fit to a constant with the method described in the text, as shown in Fig.~\ref{fig:CS_x_dependence_all}. }
    \label{fig:CSkernelg4g3}
\end{figure}

Precisely, a best value for the Collins-Soper kernel is determined from $\hat{\gamma}^q_\zeta$ via a multi-step procedure. First, the largest window of $x$ is determined for which the data for all choices of the pair $\{P_1^z,P_2^z\}$ are consistent with a common constant value. In practice this region is defined as the largest window in which a constant fit to the data at a set of discrete $x$ points has a $\chi^2/\text{d.o.f.}\le 1$. The central value and uncertainty are defined as the median and the 68\% confidence interval of the union of the bootstrap data in that $x$ window, including all $\{P_1^z,P_2^z\}$ pairs. The result of this procedure is robust to changes in the discretization of $x$, for sufficiently fine discretization scales (100 points spanning $0<x<1$ uniformly are used in the analysis presented). This procedure is performed separately for $\hat{\gamma}^q_\zeta$ determined from beam functions calculated with Dirac structures $\gamma^4$ and $\gamma^3$; examples of the resulting values are shown with $\hat{\gamma}^q_\zeta$ in Fig.~\ref{fig:CSxfits}, with the remainder presented in Appendix~\ref{app:barebeams}. The central values of the independent calculations with Dirac structures $\gamma^4$ and $\gamma^3$ are averaged, and the average uncertainty is added in quadrature with half the difference between the central values obtained using each Dirac structure, to yield the final results of this work which are shown as a function of $b_T$ in Fig.~\ref{fig:CSkernelg4g3}, and are tabulated in Table.~\ref{tab:CS}. 

\begin{ruledtabular}
\begin{table}[]
    \centering
    \begin{tabular}{ccccc}
        $b_T$ [fm] & 0.12 & 0.24 & 0.36 & 0.48 \\\hline
        $\gamma^{q,\MS}_\zeta$  & -0.419(53)(50)  & -0.49(5)(12)  & -0.76(9)(8) & -0.82(15) 
    \end{tabular}
    \caption{Collins-Soper kernel with $\mu = 2\text{ GeV}$ as a function of $b_T$. The first uncertainty is the average of that determined from calculations using $\hat{B}^{\MS}_{\gamma^4}$ and $\hat{B}^{\MS}_{\gamma^3}$ as described in the text, while the second is a systematic uncertainty computed as half the difference of the central values of the results obtained using quasi beam functions defined with the two Dirac structures.}
    \label{tab:CS}
\end{table}
\end{ruledtabular}

\begin{figure}[!tp]
	\centering
	\subfloat[$\hat{\gamma}^q_\zeta$ computed from quasi beam functions $\hat{B}^{\MS}_{\gamma^4}$.]{
   \includegraphics[width=0.45\textwidth]{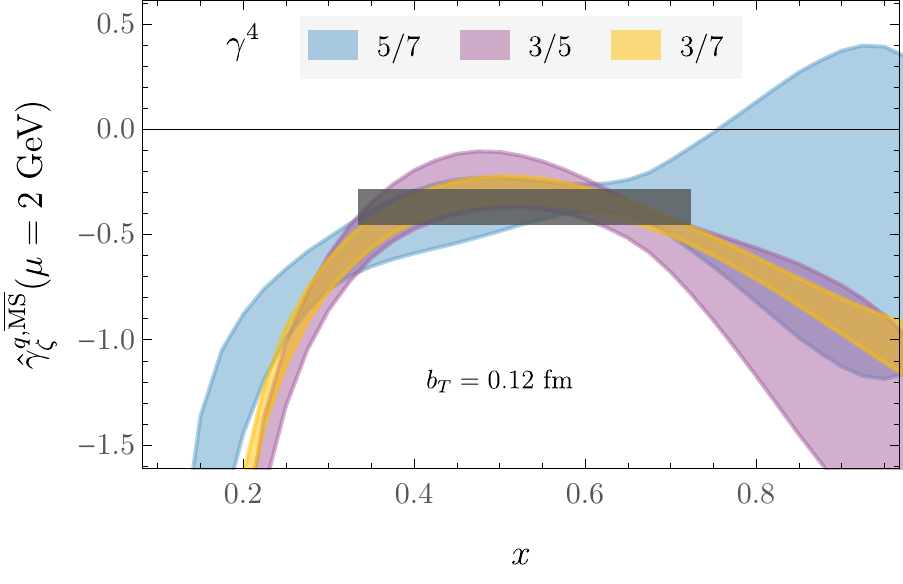}}\\
   \subfloat[$\hat{\gamma}^q_\zeta$ computed from quasi beam functions $\hat{B}^{\MS}_{\gamma^3}$.]{
   \includegraphics[width=0.45\textwidth]{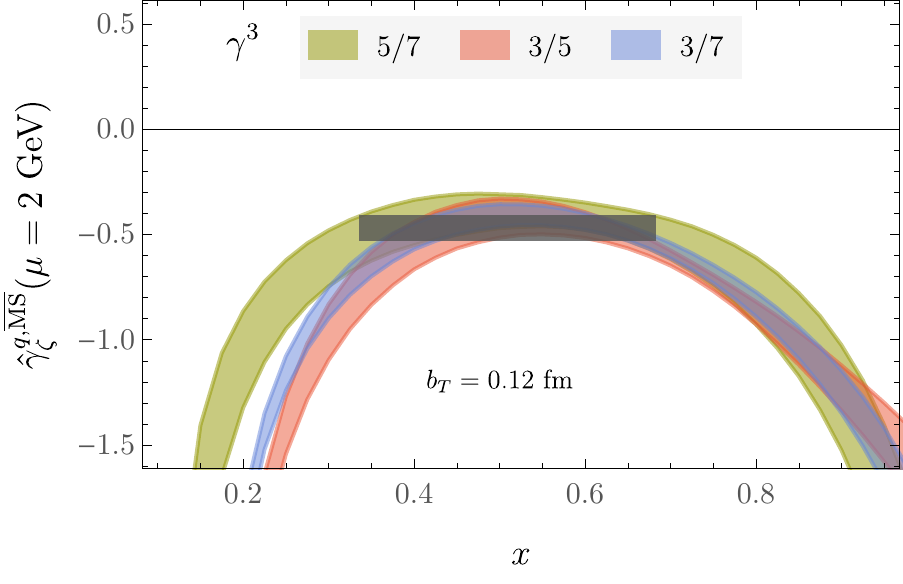}}
   \caption{\label{fig:CSxfits}$\hat{\gamma}^q_\zeta$, computed as defined in Eq.~\eqref{eq:gammahat} for all momentum pairs $\{P_1^z,P_2^z\}$, denoted by $P_1^z/P_2^z$ in the legend. The horizontal shaded band shows the fit window in $x$, as well as the total uncertainty of the best result, determined as described in the text.}
\end{figure}

\begin{figure*}[t]
	\centering
	\includegraphics[width=0.6\linewidth]{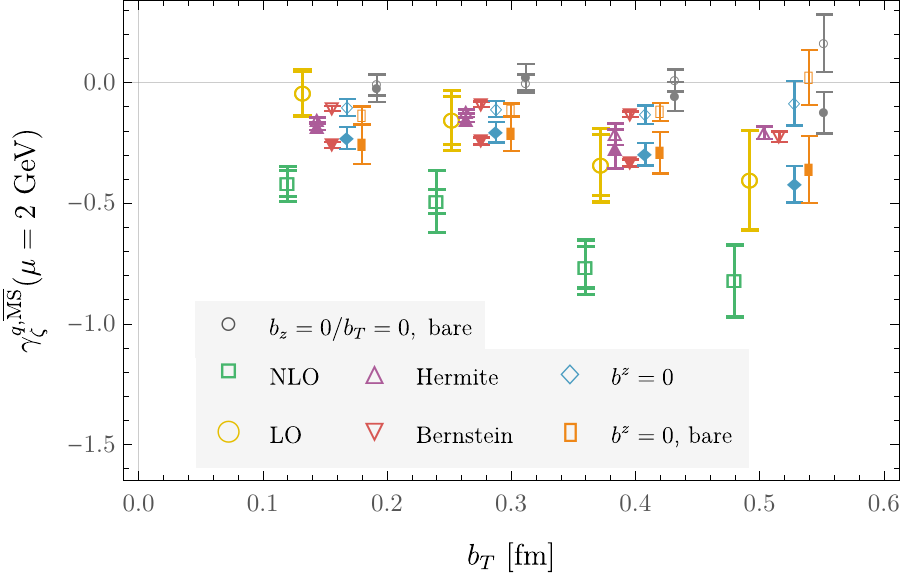}
	\caption{ $b_T$-dependence of the Collins-Soper kernel computed from the same quasi beam functions via the different approaches defined in Sec.~\ref{sec:extr}. All points other than the primary results of this work (``NLO") are offset on the horizontal axis for clarity. For the ``NLO" and ``LO" approaches, results computed based on quasi beam functions with Dirac structures $\gamma^4$ and $\gamma^3$ are combined as described in the text; the outer error bars include half the difference between the results with $\gamma^4$ and $\gamma^3$ combined in quadrature with the average uncertainty, shown by the inner error bars. For the other approaches the empty (filled) points show results obtained with Dirac structure $\gamma^4$ ($\gamma^3$). ``Hermite/Bernstein" points with Dirac structure $\gamma_3$ are not shown at $b_T/a=4$ because the corresponding fits of the $P^zb^z$-dependence of the relevant quasi beam functions were of poor quality, as described in the text. 
		\label{fig:CS_methods}}
\end{figure*}

In addition to the approach followed here, there are a number of alternative methods of extracting the Collins-Soper kernel that have been proposed or employed in other studies, for example: 
\begin{itemize}
\item ``LO": The perturbative matching coefficient $C^\TMD_\ns$ computed to leading-order (LO), instead of NLO, can be used in an analysis otherwise mirroring that presented here;
\item ``Hermite/Bernstein": As proposed in Ref.~\cite{Shanahan:2020zxr}, the $P^zb^z$-dependence of the quasi beam functions can be fit to models based on Hermite and Bernstein polynomial bases constructed to yield $x$-independent Collins-Soper kernels via Eq.~\eqref{eq:gammahat}, taking the LO value of the perturbative matching coefficient $C^\TMD_\ns$;
\item ``$b^z=0$": An approximation of the Collins-Soper kernel can be computed with LO matching using only the quasi beam functions evaluated at $b^z=0$ (this approach does not require a Fourier transform in $b^z$):
\begin{equation}
\hspace{9mm}[{\gamma}^q_\zeta(\mu, b_T)]^{b^z=0}
\equiv \frac{1}{\ln(P^z_1/P^z_2)}\ln\Biggr[
\frac{\overline{B}^{\MS}_{\gamma^4}(\mu,0, b_T,P_1^z)}
    { \overline{B}^{\MS}_{\gamma^4}(\mu,0,b_T, P_2^z)}\Biggr];
\end{equation}

\item ``$b^z=0$, bare": As proposed in Ref.~\cite{Zhang:2020dbb}, the same approach described for ``$b^z=0$" can be followed, using bare quasi beam functions ${B}^{\text{bare}}_{\gamma^4}$ rather than renormalized quasi beam functions (i.e., neglecting operator mixing between different Dirac structures);
\item ``$b^z=0/b_T=0$, bare": As proposed in Ref.~\cite{Li:2021wvl}, a variation of the `$b^z=0$" approach can be used, approximating the Collins Soper kernel as
\begin{align}
\hspace{4mm}[{\gamma}^q_\zeta&(\mu, b_T)]^{b^z=0/b_T=0}
\equiv \frac{1}{\ln(P^z_1/P^z_2)}\nn\\
&\times \ln\Biggr[
\frac{{B}^{\text{bare}}_{\gamma^4}(0, b_T,a,\eta, P_1^z){B}^{\text{bare}}_{\gamma^4}(0,0,a,\eta, P_2^z)}
    { {B}^{\text{bare}}_{\gamma^4}(0,b_T,a,\eta, P_2^z){B}^{\text{bare}}_{\gamma^4}(0,0,a,\eta, P_1^z)}\Biggr].
\end{align}
\end{itemize}
Each of these methods can be followed using the quasi beam functions computed in this work; a comparison of the results is provided in Fig.~\ref{fig:CS_methods}. For the ``LO" approach, the same procedure is followed to combine the results obtained using quasi beam functions with Dirac structures $\gamma^3$ and $\gamma^4$ as for the ``NLO" method which yields the main results of this work. 
For the other approaches the results obtained with the two Dirac structures are not always consistent at one standard deviation, and are shown separately; for $b_T /a=4$ no results for the ``Hermite/Bernstein" approach are shown with Dirac structure $\gamma^3$ as the model fits were of poor quality with $\chi^2/\text{d.o.f.}>2$.
In the case of the ``$b^z=0/b_T=0$, bare" approach, bare quasi beam functions with $\eta/a=14$, which is the largest extent studied in this work for all $P^z$, are used in the analysis.

\begin{figure*}[!t]
	\centering
	\subfloat[\label{fig:CS_comp_pheno}Comparison with the SV19~\cite{Scimemi:2019cmh} and Pavia19~\cite{Bacchetta:2019sam} phenomenological parameterizations and the next-to-next-to-next-to-leading order (N$^3$LO) perturbative result~\cite{Li:2016ctv,Henn:2019swt}.]{
	\includegraphics[width=0.45\textwidth]{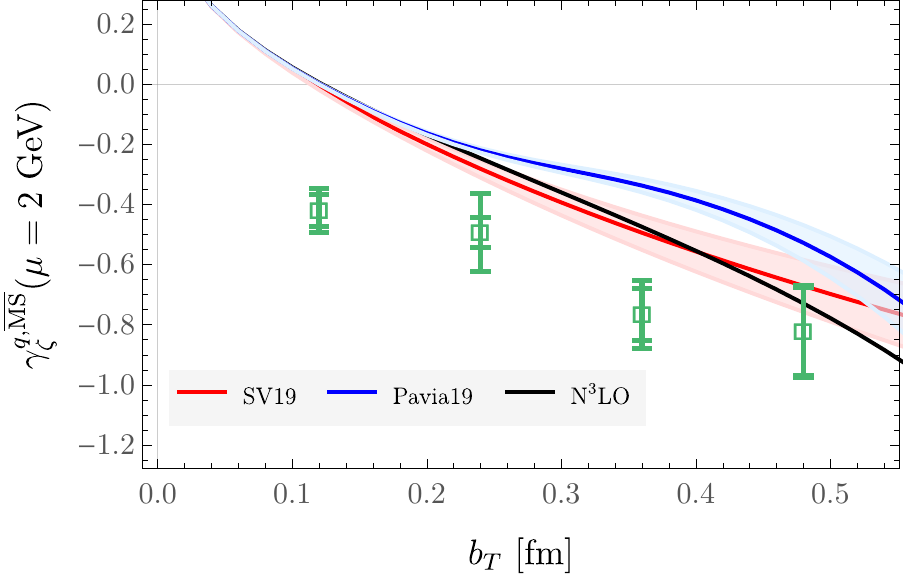}}
	\subfloat[	\label{fig:CS_comp_latt} Comparison with quenched results of Ref.~\cite{Shanahan:2020zxr} (SWZ), as well as results from the LPC~\cite{Zhang:2020dbb}, Regensburg/NMSU~\cite{Schlemmer:2021aij}, and ETMC/PKU~\cite{Li:2021wvl} collaborations. Different sets of points with the same color show different sets of results from the same collaboration.]{
	\includegraphics[width=0.45\textwidth]{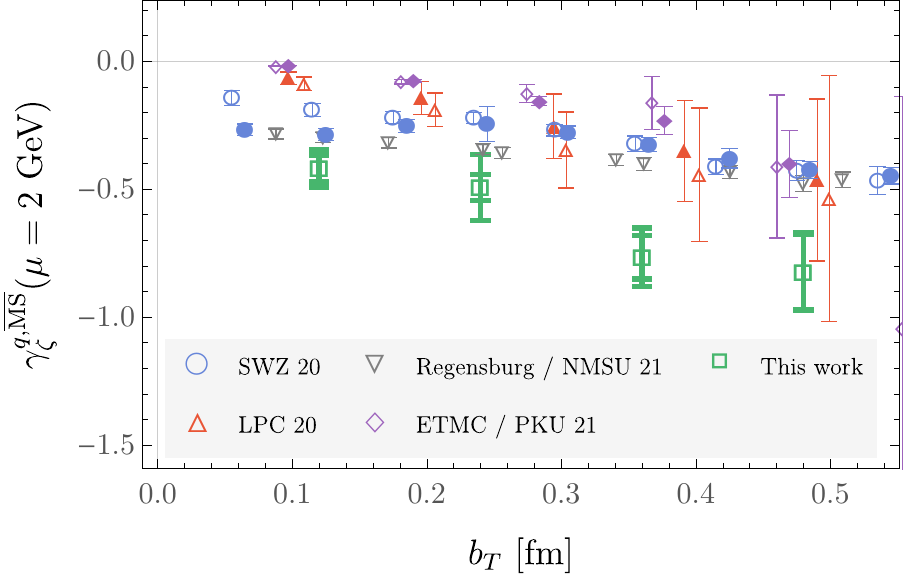}}
	\caption{ $b_T$-dependence of the Collins-Soper kernel as determined in this work (green squares in both panels) compared with (a) phenomenological results, and (b) the results of other lattice QCD calculations of this quantity.
		\label{fig:CS_comp}}
\end{figure*}

Clearly, although the same quasi beam functions are used, the Collins-Soper kernel determined via each of these approaches is very different, and many of the results are inconsistent with the best results of this study at several standard deviations, with uncertainties as much as an order of magnitude smaller. This is to be expected if the effects of higher-order matching, renormalization and mixing, and power corrections, are significant, as each of the approaches listed above treats one or more of these systematic effects differently than in the primary analysis presented here.

\section{Outlook}
\label{sec:outlook}

This work presents a determination of the Collins-Soper kernel from a dynamical lattice QCD calculation following the approach of Refs.~\cite{Ebert:2018gzl,Ebert:2019okf}.
Several systematic uncertainties remain to be addressed; in particular, the quark masses used correspond to an unphysically-large pion mass of $m_\pi=538(1)$~MeV, and the results are obtained using a single ensemble of gauge field configurations such that effects from the discretization and finite lattice volume cannot be fully quantified. A fully model-independent calculation will require these systematics to be addressed, lattice QCD calculations to be performed over a larger range of $P^zb^z$ to eliminate the need to extrapolate the quasi beam functions to large $|b^z|$ and enable the DFT approach to be used, and larger values of $P^z$ to be included to reduce the contributions from power corrections and higher-twist effects which dominate the uncertainties of this calculation. With these caveats in mind, the results of this work may be compared with phenomenological extractions of the Collins-Soper kernel, as shown in Fig.~\ref{fig:CS_comp_pheno}. The lattice QCD and phenomenological determinations are broadly consistent at large $b_T$, with clear deviations at the smallest $b_T$ values studied; discretization effects are expected to be largest at small $b_T$ and might be relevant for understanding this effect. It is clear that, while challenging to achieve computationally, future fully-controlled calculations by this approach have the potential to differentiate different models of the Collins-Soper kernel and will provide important input for the analysis of low-energy SIDIS data and the determinations of the TMDPDFs.

In considering the prospects for such future controlled determinations of the Collins-Soper kernel from lattice QCD, it is informative to contrast the results of this study with those of other lattice QCD investigations; a comparison of existing calculations~\cite{Shanahan:2020zxr,Zhang:2020dbb,Schlemmer:2021aij,Li:2021wvl} is provided in Fig.~\ref{fig:CS_comp_latt}. All dynamical calculations use quark masses resulting in similar values of the pion mass to that of the calculation presented here (ranging from the lightest ensemble with $m_\pi=350$~MeV in Ref.~\cite{Li:2021wvl} to $m_\pi=547$ MeV in Ref.~\cite{Zhang:2020dbb}), while the quenched calculation of Ref.~\cite{Shanahan:2020zxr}, in which the kernel should not depend on the valence quark masses since it is independent of the external state, is performed at $m_\pi=1.207$ GeV. 
Each calculation uses a slightly different approach to constrain the Collins-Soper kernel from quasi beam functions. In particular, the ``Hermite/Bernstein" approach is followed in Ref.~\cite{Shanahan:2020zxr} (``SWZ"), the calculation of Ref.~\cite{Zhang:2020dbb} (``LPC") uses the ``$b^z=0$, bare" approach, that of Ref.~\cite{Schlemmer:2021aij} (``Regensburg/NMSU") uses an approach similar to the ``$b^z=0$, bare" approach but with NLO matching, and Ref.~\cite{Li:2021wvl} (``ETMC/PKU) applies the ``$b^z=0/b_T=0$, bare" approach. While the various calculations exhibit similar dependence on $b_T$, there are some significant discrepancies between the numerical results, and a wide range of uncertainty estimates. Given the analysis of Sec.~\ref{sec:extr}, this is to be expected; even when the same quasi beam function data is used, following the various ``$b^z=0$" approaches and the approach presented here result in significant systematic differences, and significantly different uncertainty estimates. Since Refs.~\cite{Zhang:2020dbb,Schlemmer:2021aij,Li:2021wvl} all use somewhat larger maximum $P^z$ values than that of the present study, the effects of power corrections and higher-twist contamination can be expected to be smaller than those found in Sec.~\ref{sec:extr}, but these effects, together with the difference between NLO and LO matching illustrated in Appendix~\ref{app:NLOmatching}, could nevertheless be responsible for the discrepancies visible in Fig.~\ref{fig:CS_comp_latt}. Clearly, a fully-controlled determination of the Collins-Soper kernel from lattice QCD will require NLO or even higher-order matching or resummation and a treatment of power corrections that is more robust than that achieved in any of the studies performed to date. The analysis presented here constitutes an important step towards that goal, revealing clearly that these important sources of systematic uncertainty cannot be neglected.

\begin{acknowledgements}
The authors thank Will Detmold, Iain Stewart and Alexey Vladimirov for useful discussions, Xu Feng for providing numerical data for inclusion in Fig.~\ref{fig:CS_comp}, and Steven Gottlieb and the MILC collaboration for the use of the gauge configurations used in this project, which were generated at Indiana University on Big Red 2+ and Big Red 3. This research was supported in part by Lilly Endowment, Inc., through its support for the Indiana University Pervasive Technology Institute that provides the Big Red supercomputers.
This work is supported in part by the U.S. Department of Energy, Office of Science, Office of Nuclear Physics, under grant Contract Numbers DE-SC0011090, DE-AC02-06CH11357, DE-SC0012704, and within the framework of the TMD Topical Collaboration.
PES is additionally supported by the U.S. DOE Early Career Award DE-SC0021006, by a NEC research award, by the Carl G and Shirley Sontheimer Research Fund, and by the U.S. National Science Foundation under Cooperative Agreement PHY-2019786 (The NSF AI Institute for Artificial Intelligence and Fundamental Interactions, http://iaifi.org/).
This research used resources of the National Energy Research Scientific Computing Center (NERSC), a U.S. Department of Energy Office of Science User Facility operated under Contract No. DE-AC02-05CH11231, the Extreme Science and Engineering Discovery Environment (XSEDE) Bridges-2 at the Pittsburgh Supercomputing Center (PSC) through allocation TG-PHY200036, which is supported by National Science Foundation grant number ACI-1548562, and facilities of the USQCD Collaboration, which are funded by the Office of Science of the U.S. Department of Energy. This manuscript has been authored by Fermi Research Alliance, LLC under Contract No. DE-AC02-07CH11359 with the U.S. Department of Energy, Office of Science, Office of High Energy Physics.
The Chroma~\cite{Edwards:2004sx}, QLua \cite{qlua},  QUDA \cite{Clark:2009wm,Babich:2011np,Clark:2016rdz}, QDP-JIT \cite{6877336}, and QPhiX~\cite{10.1007/978-3-319-46079-6_30} software libraries were used in this work.
Data analysis used NumPy~\cite{harris2020array} and Julia~\cite{Julia-2017,mogensen2018optim}, and figures were produced using Mathematica \cite{Mathematica}.

\end{acknowledgements}

\appendix

\section{Fits to ratios of three and two point functions}
\label{app:ttaufits}

\begin{figure*}[tp]
	\centering
	\includegraphics[width=0.46\linewidth]{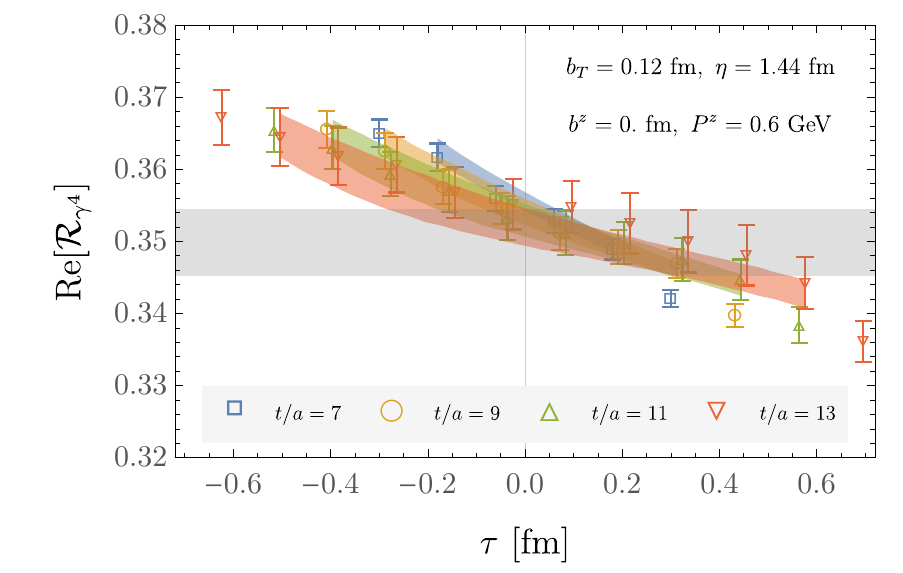}
	\includegraphics[width=0.46\linewidth]{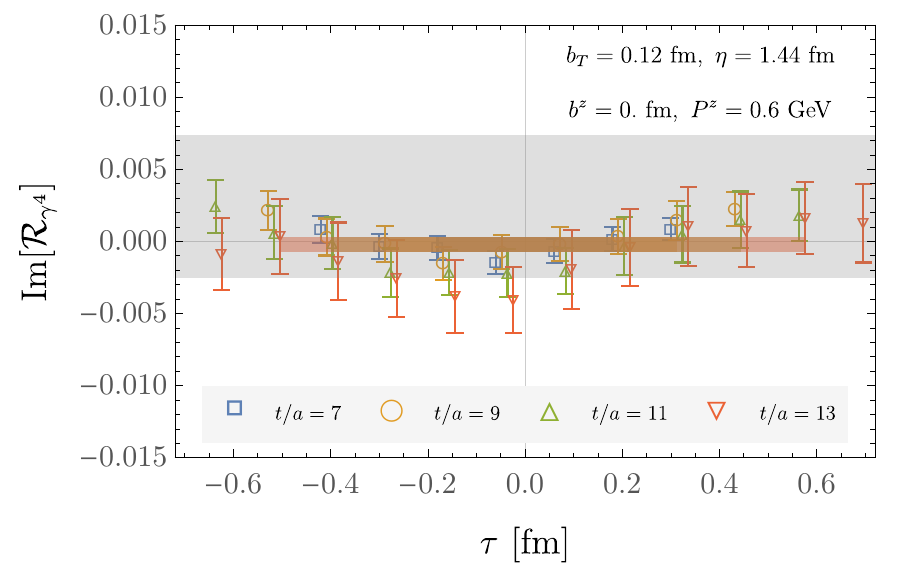} 
	\includegraphics[width=0.46\linewidth]{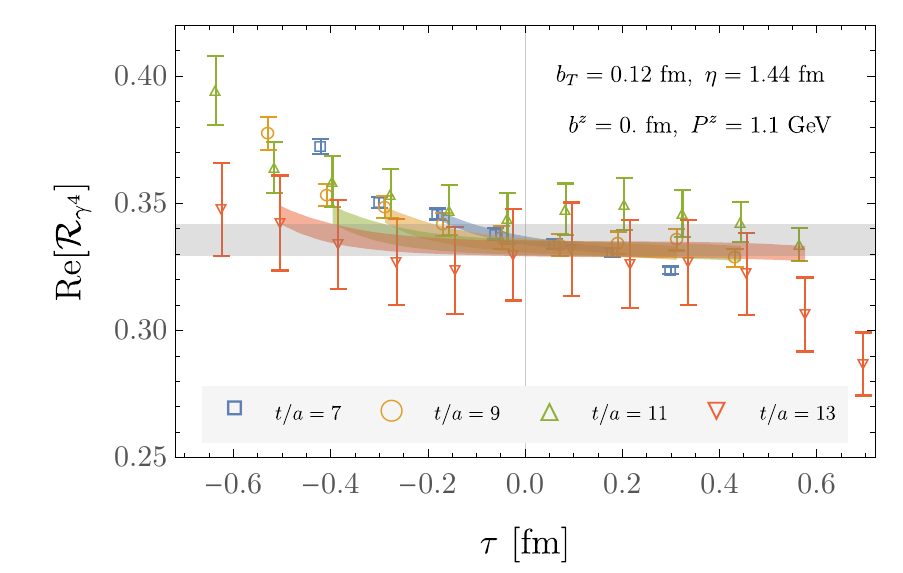}
	\includegraphics[width=0.46\linewidth]{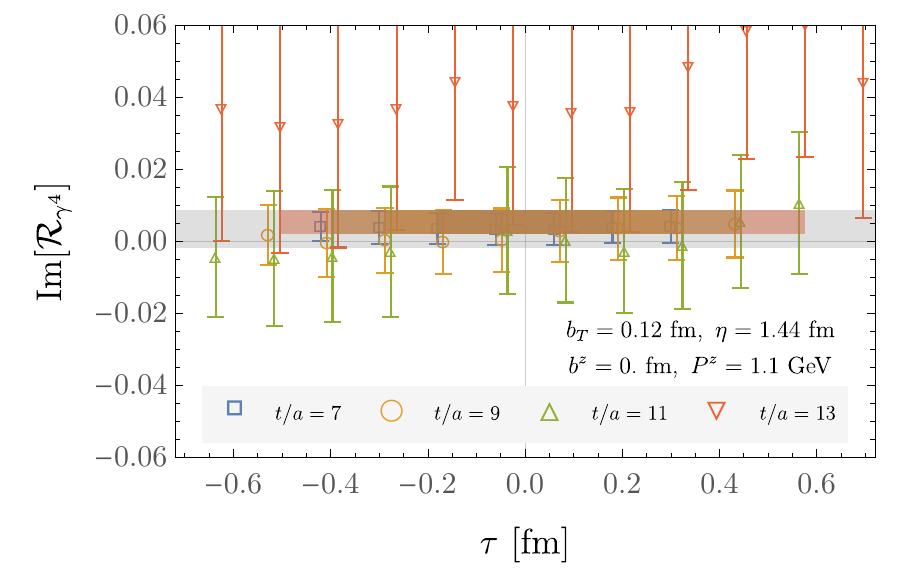} 
	\includegraphics[width=0.46\linewidth]{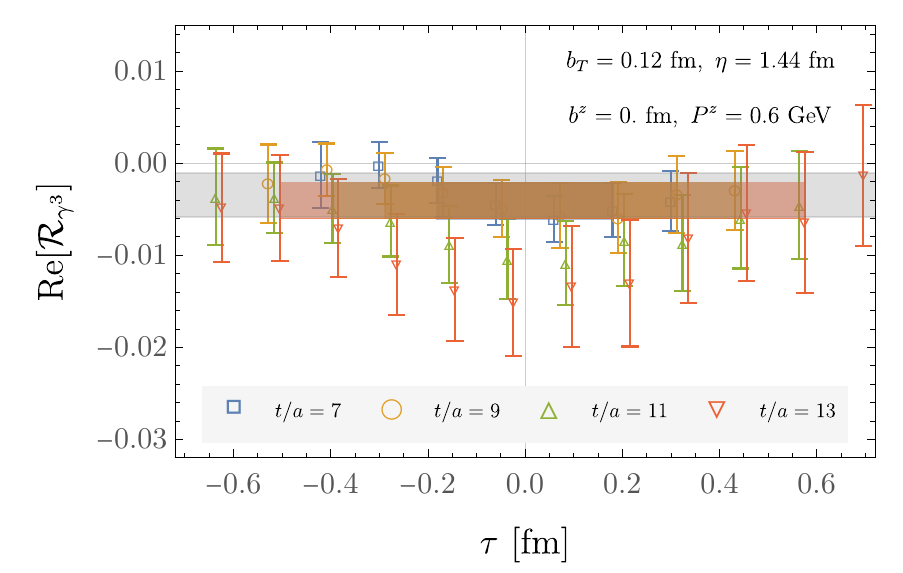}
	\includegraphics[width=0.46\linewidth]{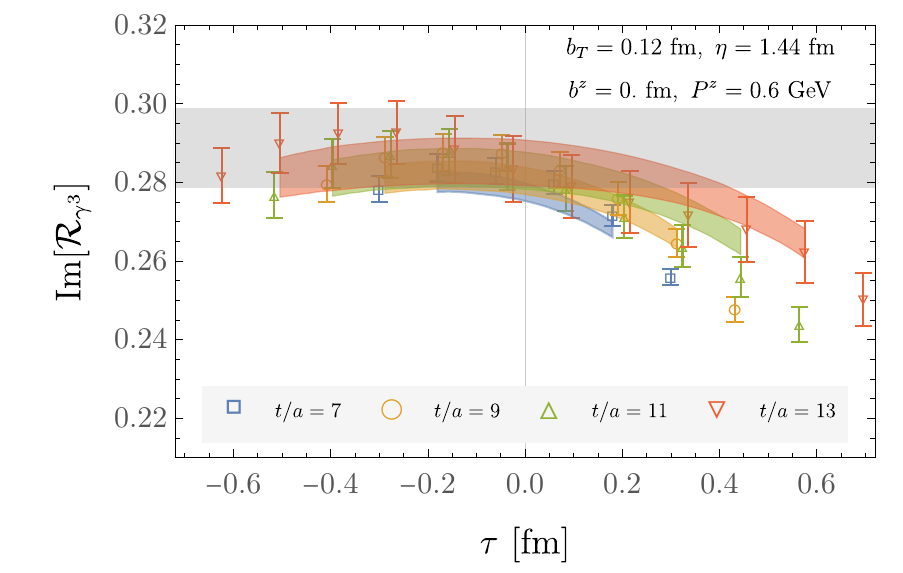} 
	\includegraphics[width=0.46\linewidth]{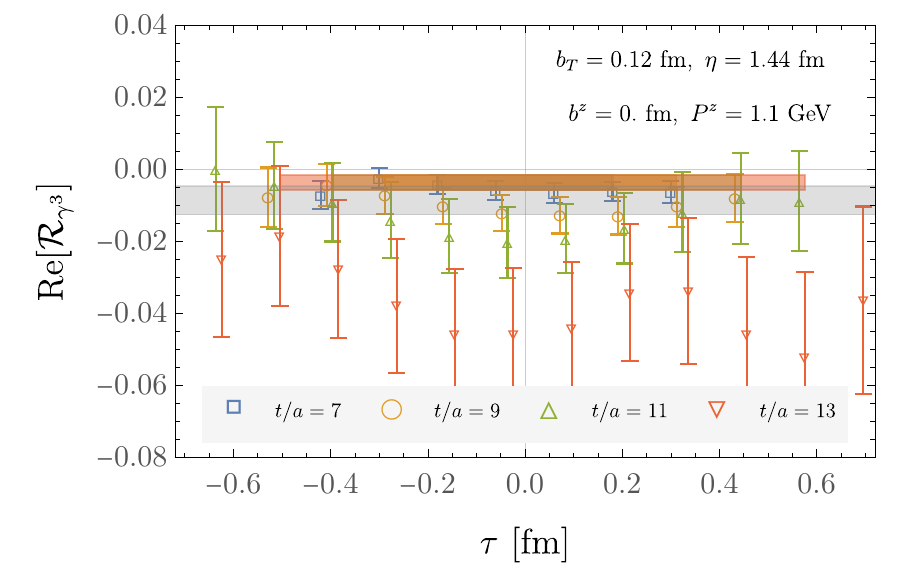}
	\includegraphics[width=0.46\linewidth]{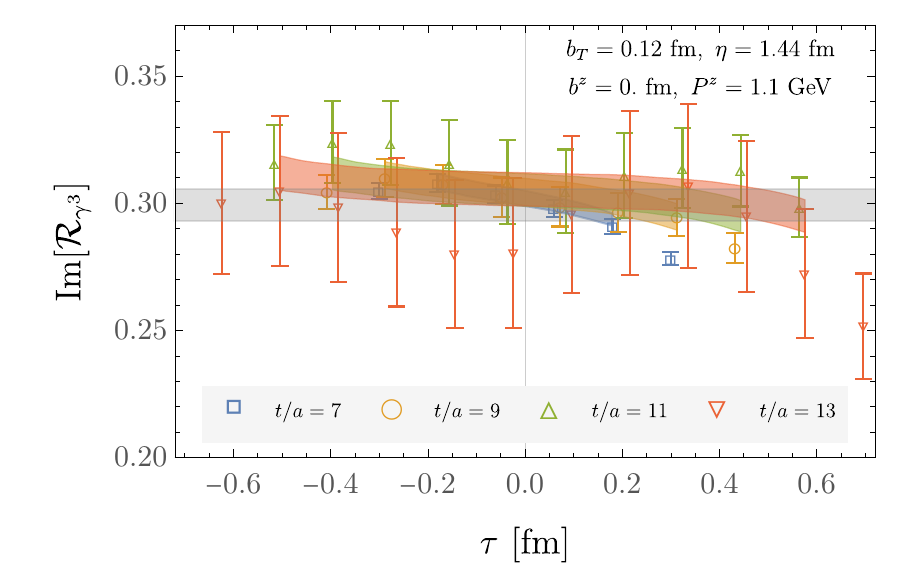} 
	\caption{Examples of fits to ratios of three and two-point functions $\mathcal{R}_\Gamma(t,\tau, b^\mu, a, \eta, P^z)$ defined in Eq.~\eqref{eq:Rcal}, performed as described in the text. The shaded bands matching the color of each set of points show 68\% bootstrap confidence intervals of the fits from the fit range that has the highest weight in the weighted average that determines the final result, which is indicated by the gray horizontal bands. These bands show the total uncertainty on the bare quasi beam functions, including both the statistical uncertainty and the systematic uncertainty which arises from variation of the results between different fit range choices.
\label{fig:3pt_2pt_ratios}}
\end{figure*}

As detailed in Sec.~\ref{subsec:barebeam}, ratios $\mathcal{R}_\Gamma(t,\tau, b^\mu, a, \eta, P^z)$ (Eq.~\eqref{eq:Rcal}) of three-point and two-point correlation functions asymptote to the bare quasi beam functions in the limit $\{\tau, t-\tau\}\rightarrow \infty$, with contamination from matrix elements in excited states present at finite $t$ and $\tau$. The $t$ and $\tau$-dependence of the ratios is fit precisely as defined in Appendix A of Ref.~\cite{Shanahan:2020zxr} to extract the bare quasi beam functions: fits are performed for all different possible cuts on source/operator/sink separations, with the AIC~\cite{1100705} used to select the number of states included in the spectral representation for each fit. The results are combined via a weighted averaging procedure. Some examples of the results of this fitting procedure are given in Fig.~\ref{fig:3pt_2pt_ratios}.

\section{Additional beam function results}
\label{app:barebeams}

\begin{figure*}[t]
	\centering
   \includegraphics[width=0.46\linewidth]{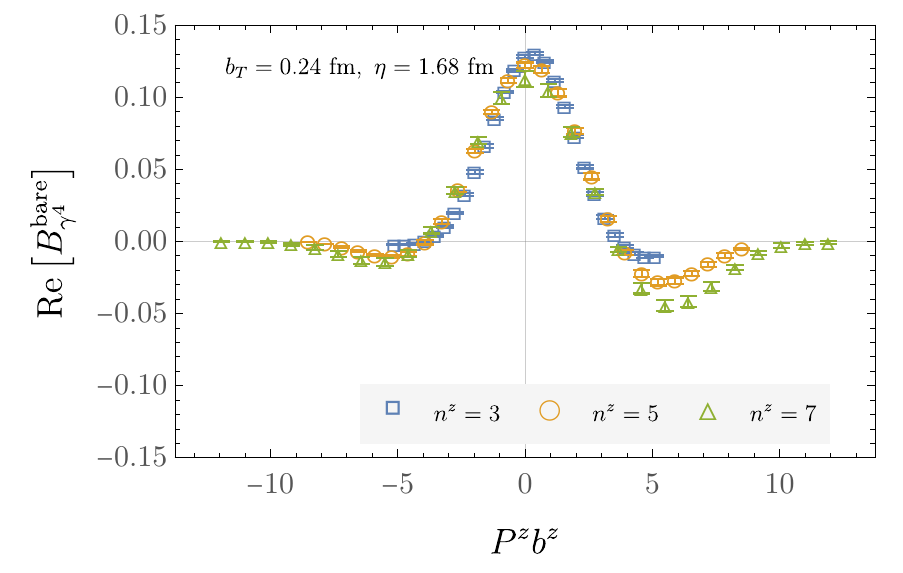}
	\includegraphics[width=0.46\linewidth]{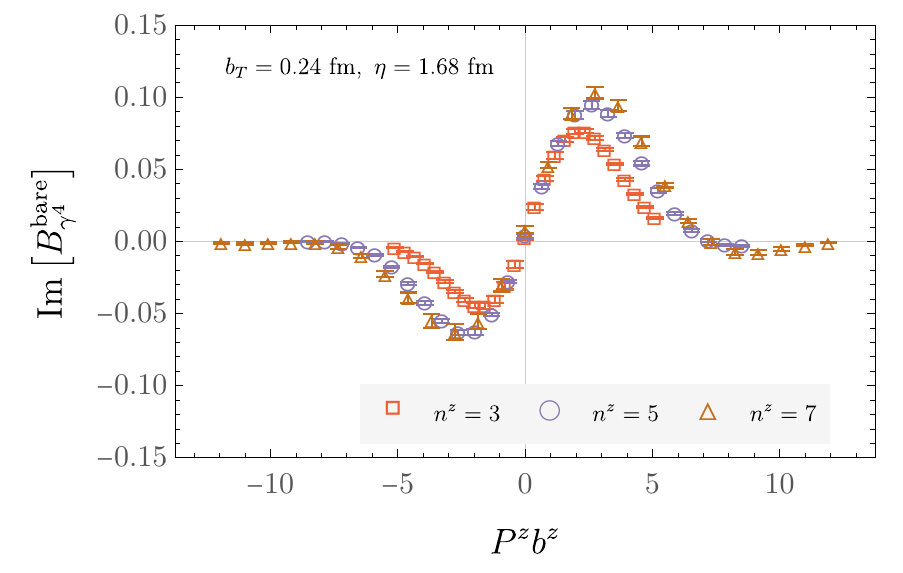} 
	\includegraphics[width=0.46\linewidth]{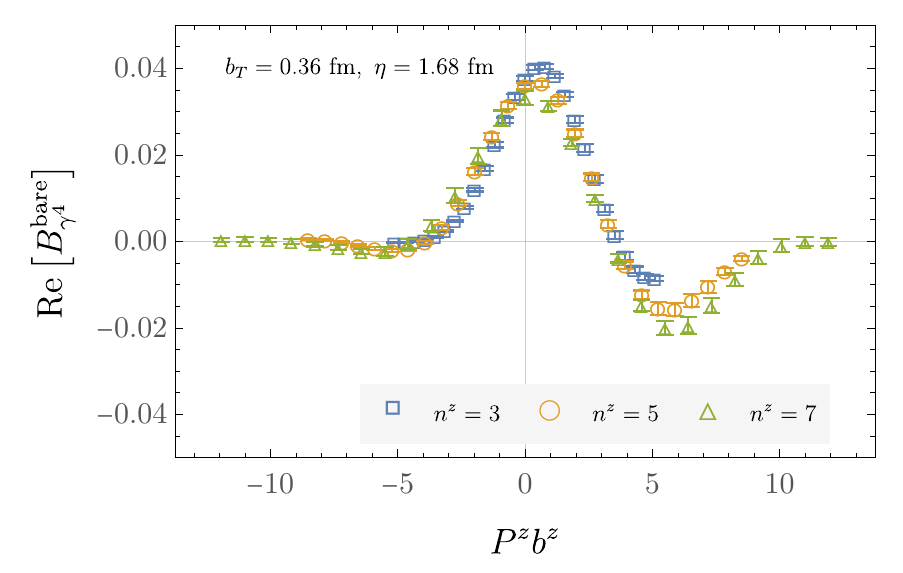}
	\includegraphics[width=0.46\linewidth]{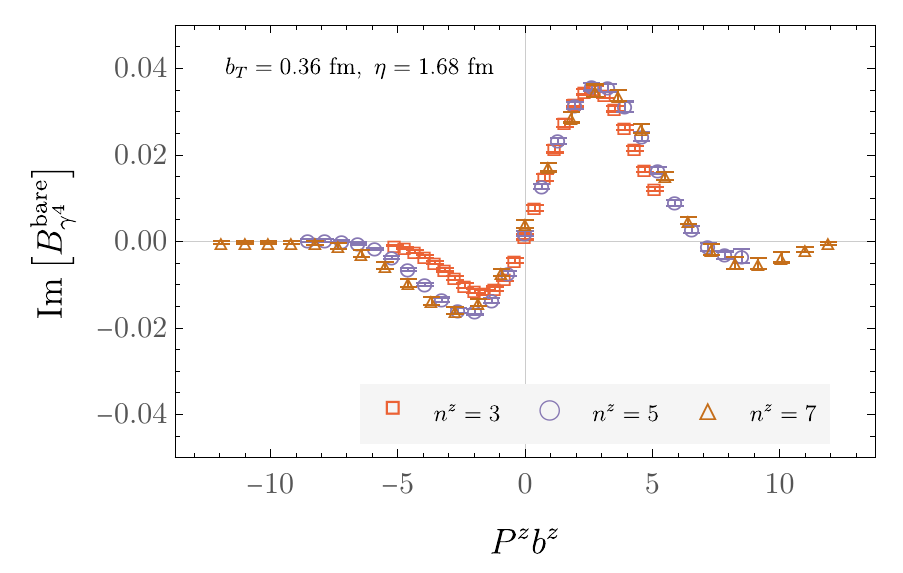} 
	\includegraphics[width=0.46\linewidth]{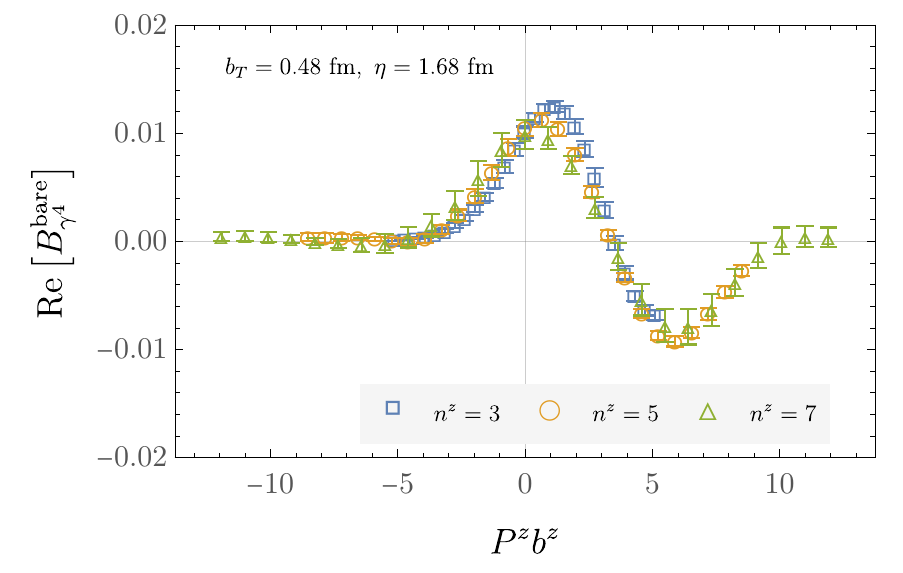}
	\includegraphics[width=0.46\linewidth]{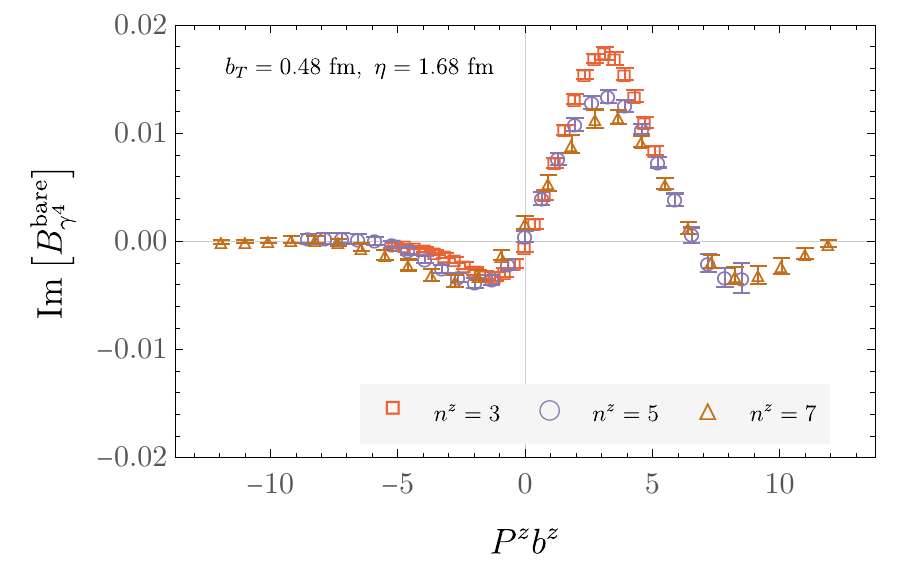} 
	\caption{Examples of the bare quasi beam functions $B^\text{bare}_{\gamma^4}$ determined as described in Sec.~\ref{subsec:barebeam} (note that $B^\text{bare}_{\gamma^4}$ for $b_T=0.12$~fm and $\eta=1.68$~fm is shown in Fig.~\ref{fig:bare_beam_eg} in the main text).
		\label{fig:bare_beam}}
\end{figure*}

\begin{figure*}[p]
	\centering
	\includegraphics[width=0.46\linewidth]{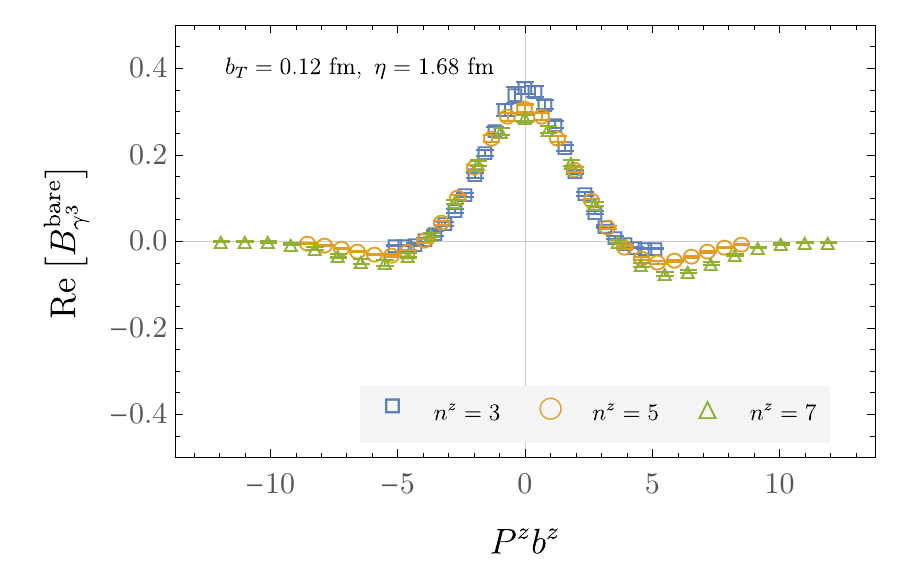}
	\includegraphics[width=0.46\linewidth]{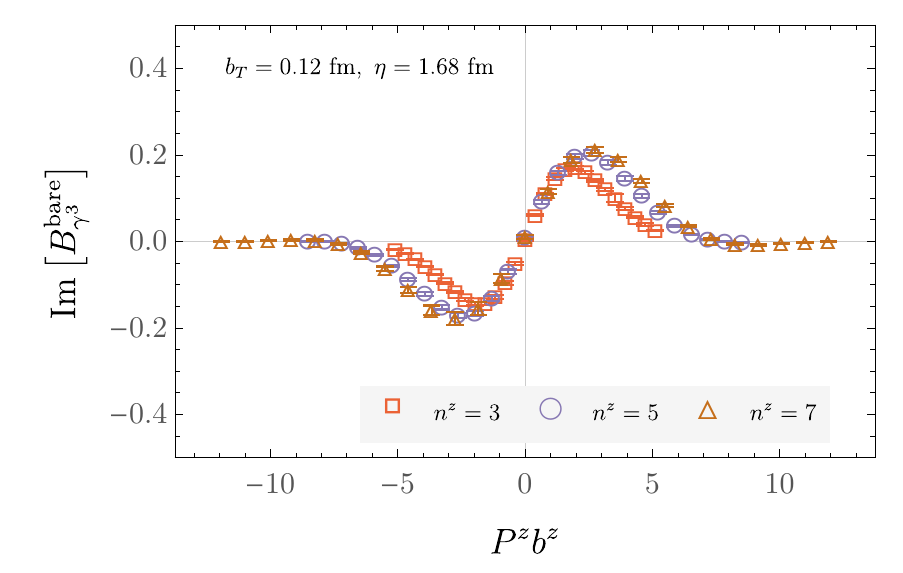} 
   \includegraphics[width=0.46\linewidth]{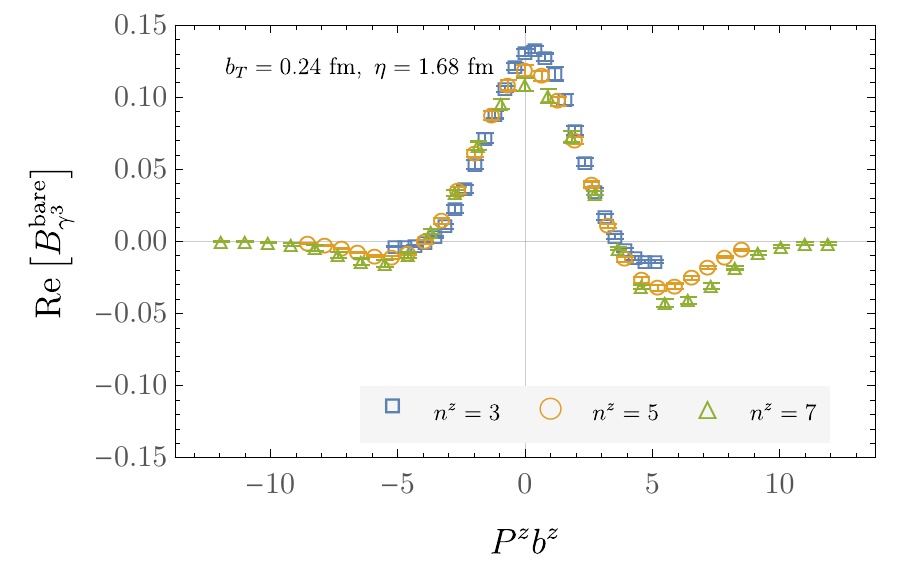}
	\includegraphics[width=0.46\linewidth]{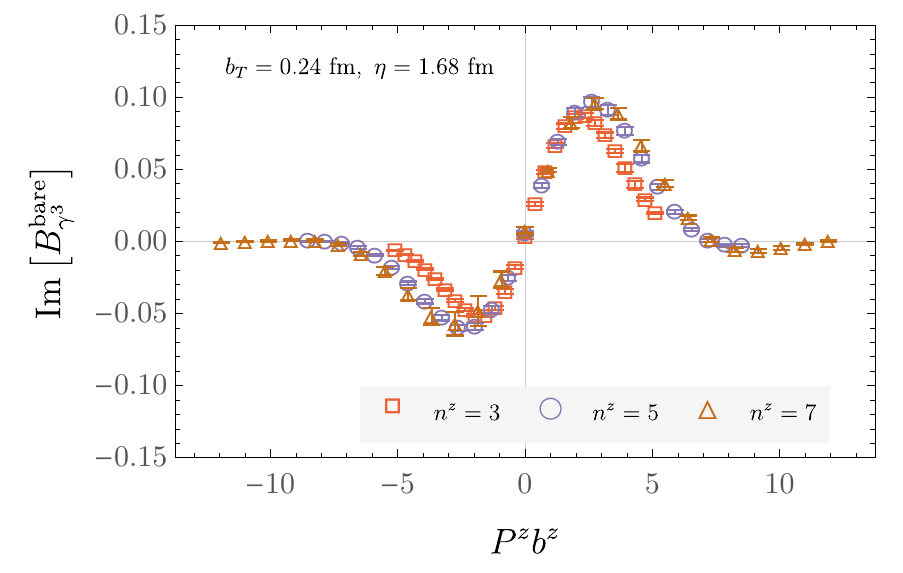} 
	\includegraphics[width=0.46\linewidth]{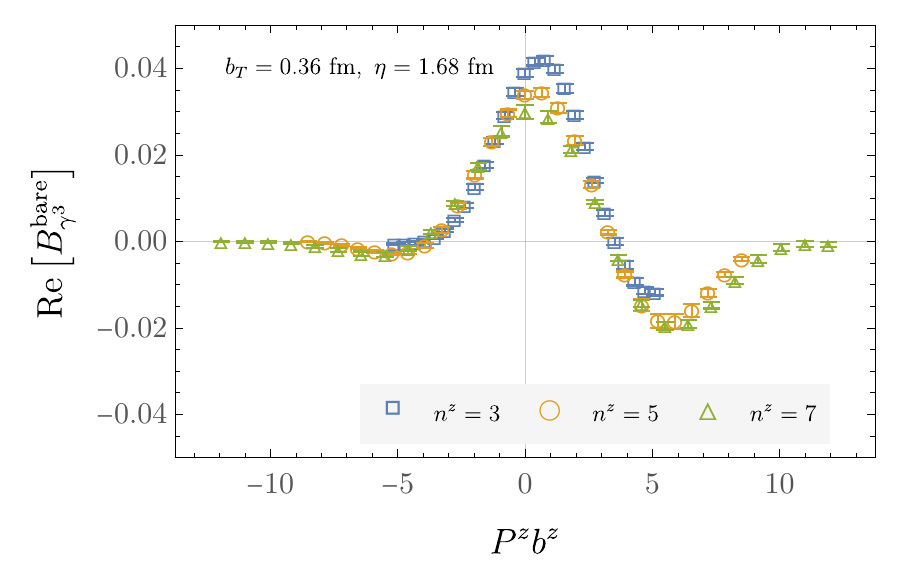}
	\includegraphics[width=0.46\linewidth]{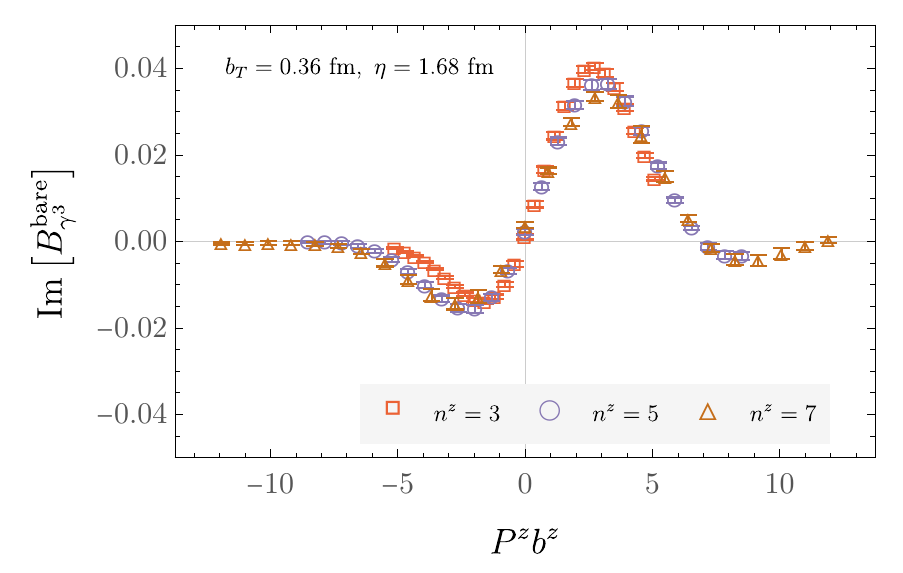} 
	\includegraphics[width=0.46\linewidth]{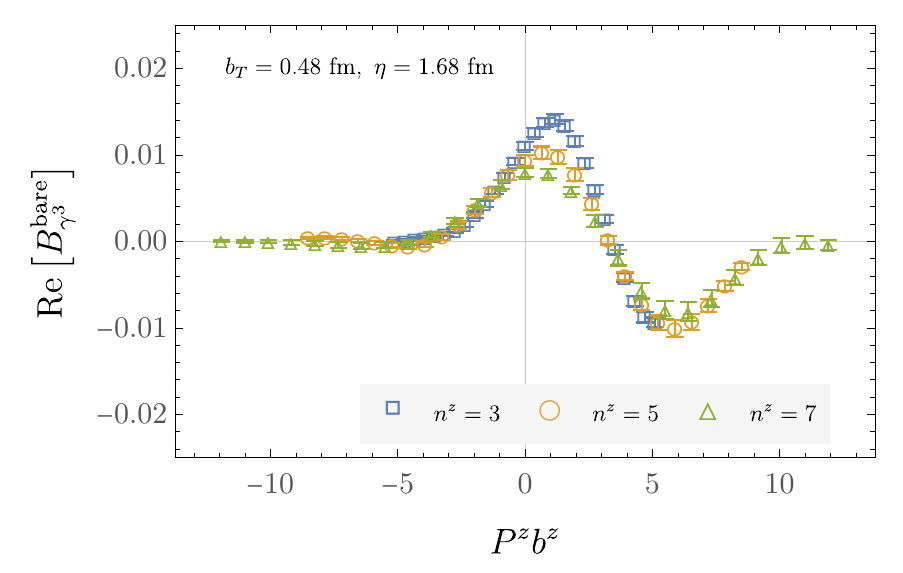}
	\includegraphics[width=0.46\linewidth]{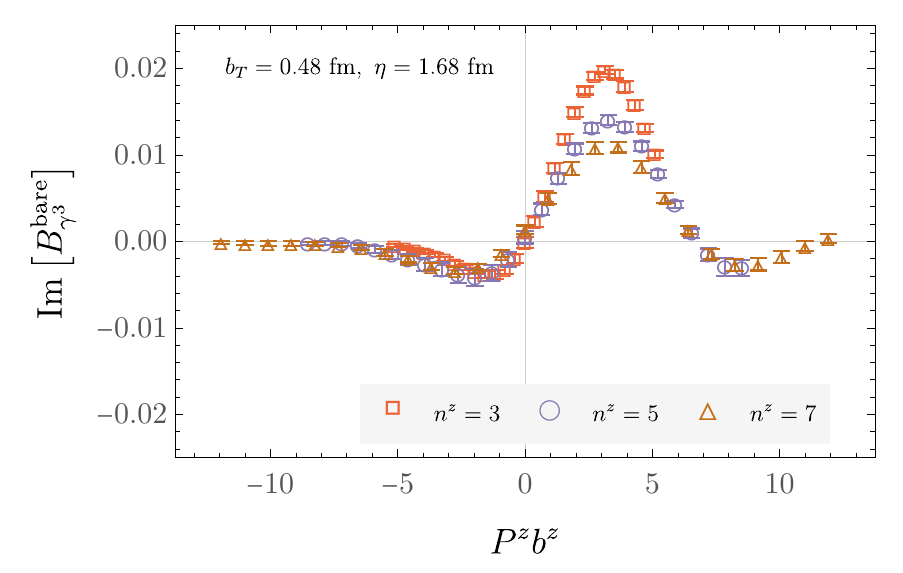} 
	\caption{ Examples of the bare quasi beam functions $B^\text{bare}_{\gamma^3}$ determined as described in Sec.~\ref{subsec:barebeam}.
		\label{fig:bare_beam_gz}}
\end{figure*}

\begin{figure*}[!p]
	\centering
	\includegraphics[width=0.46\linewidth]{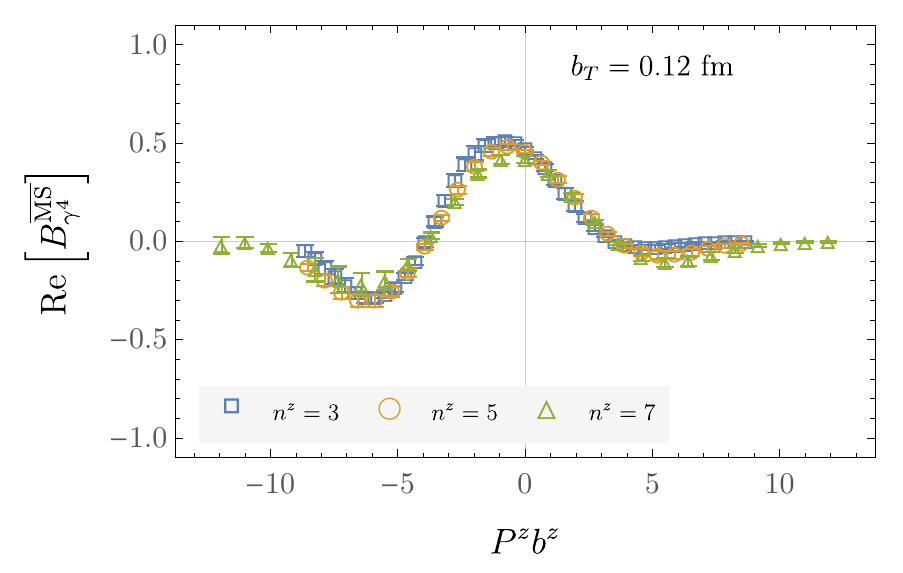}
	\includegraphics[width=0.46\linewidth]{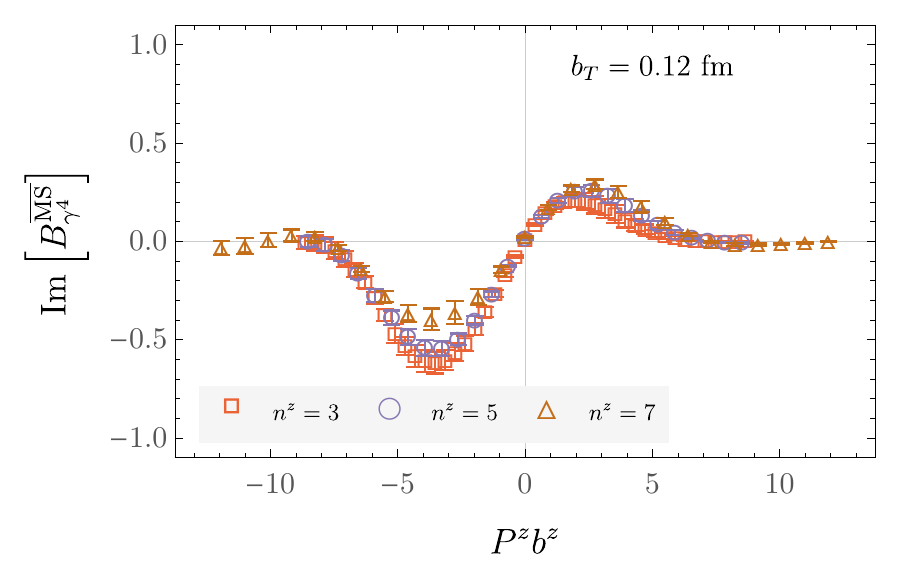}
	\includegraphics[width=0.46\linewidth]{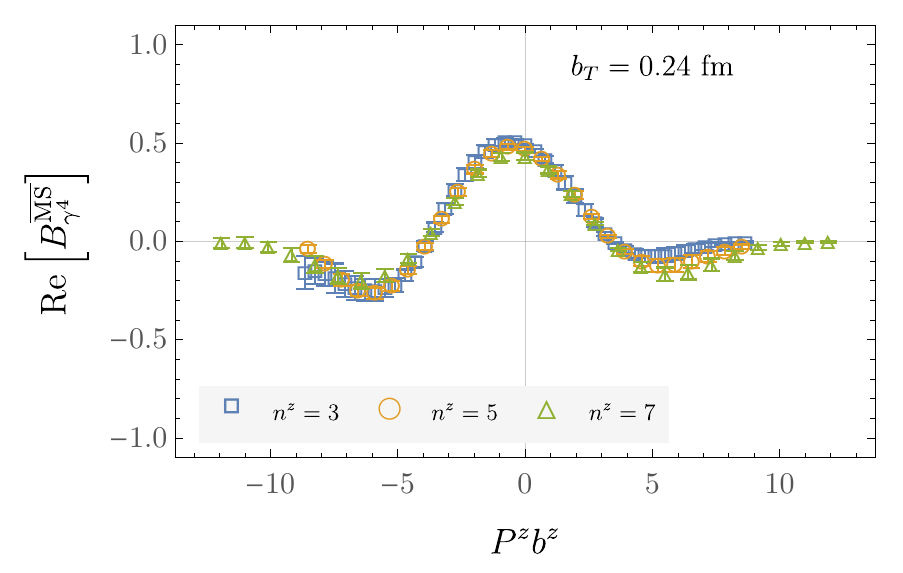}
	\includegraphics[width=0.46\linewidth]{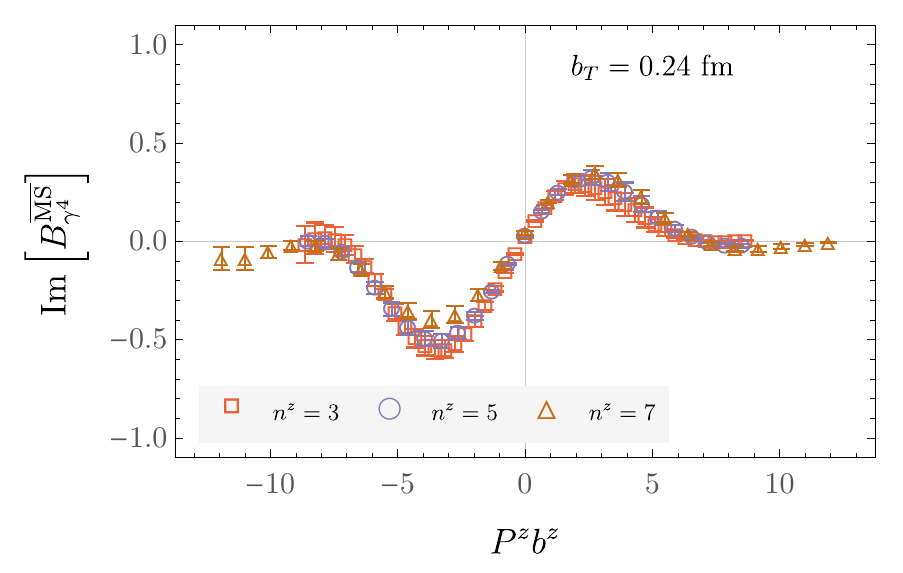}
	\includegraphics[width=0.46\linewidth]{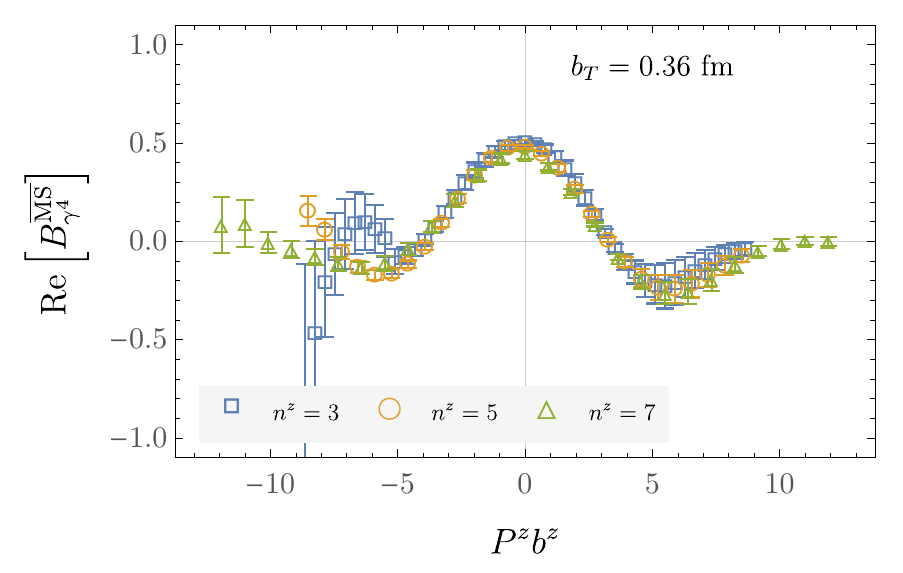}
	\includegraphics[width=0.46\linewidth]{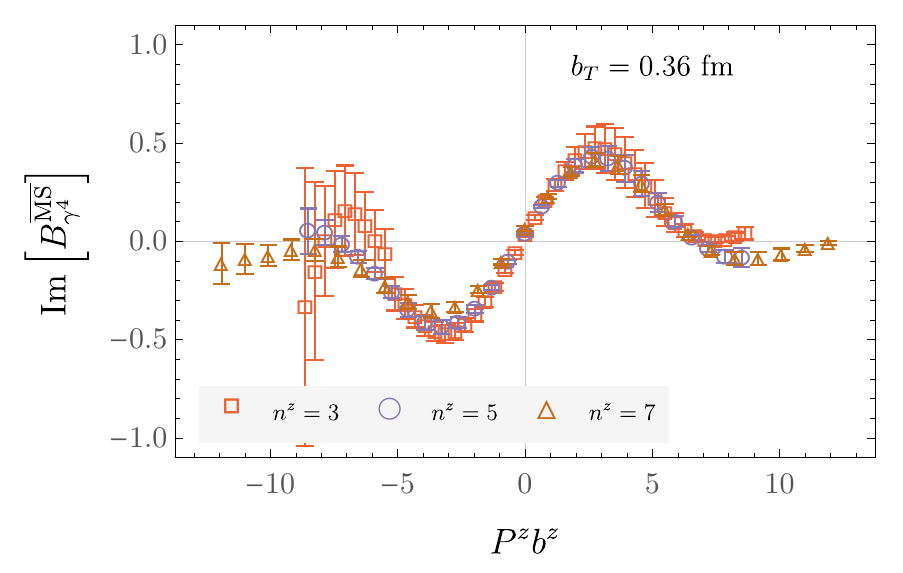}
	\includegraphics[width=0.46\linewidth]{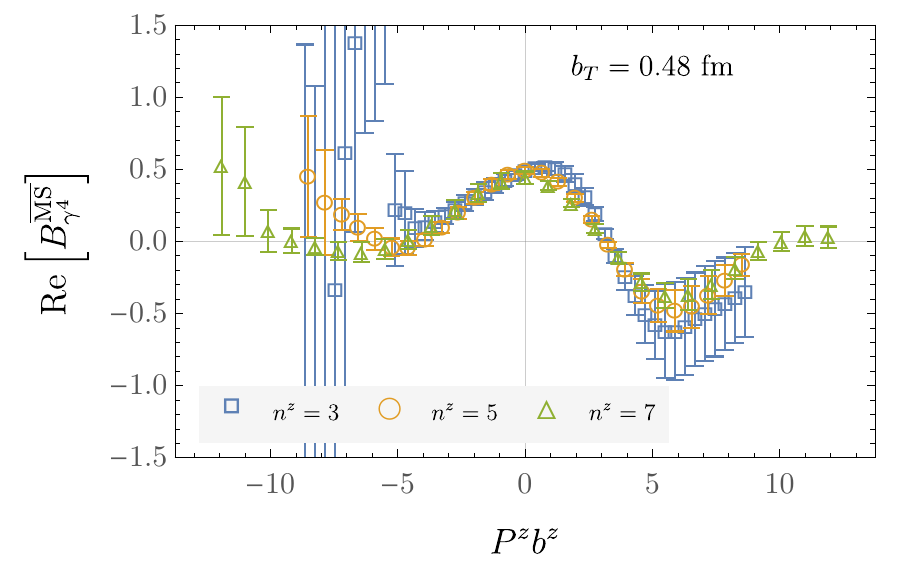}
	\includegraphics[width=0.46\linewidth]{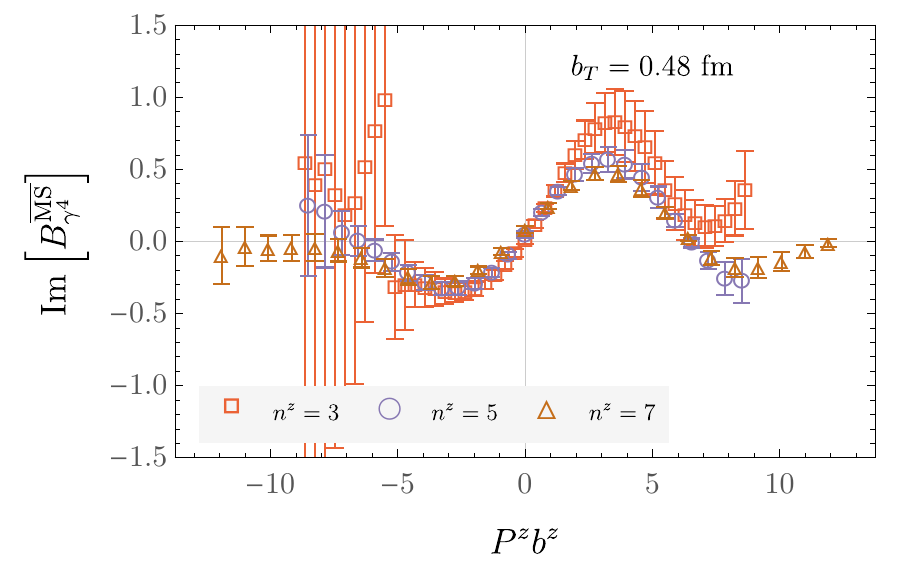}
   \caption{ Examples of the modified $\MS$-renormalized quasi beam functions $B^{\MS}_{\gamma^4}$ determined as described in Sec.~\ref{subsec:renbeam}.
	\label{fig:renorm_beam}}
\end{figure*}

\begin{figure*}[!p]
	\centering
	\includegraphics[width=0.46\linewidth]{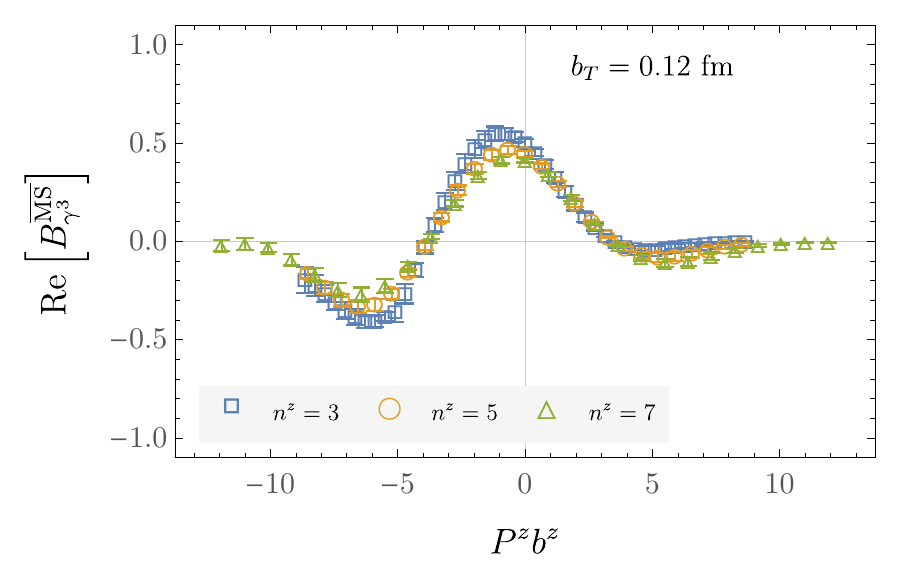}
	\includegraphics[width=0.46\linewidth]{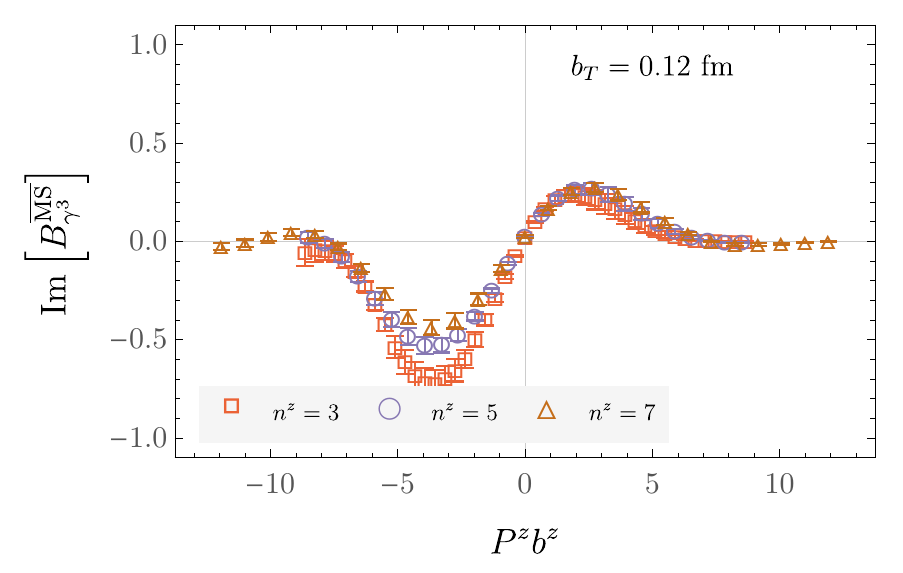}
	\includegraphics[width=0.46\linewidth]{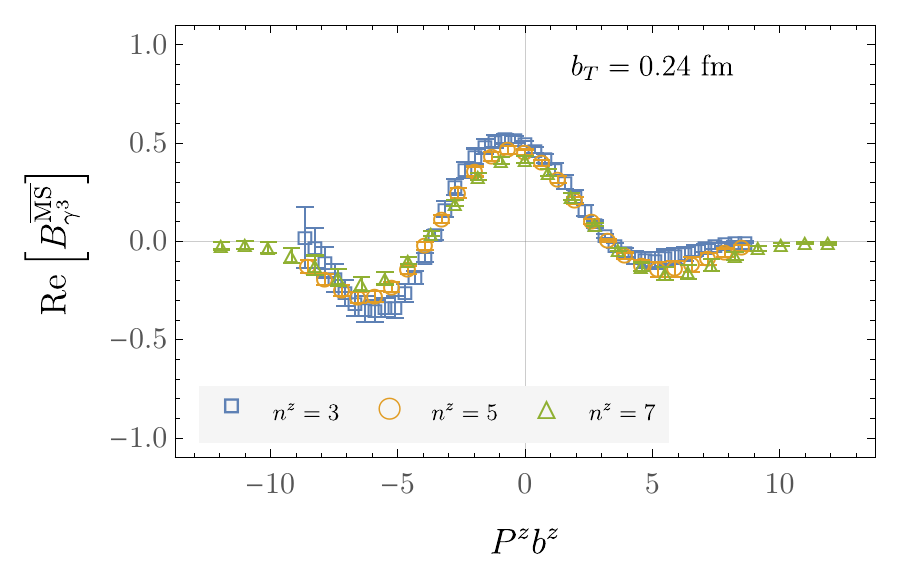}
	\includegraphics[width=0.46\linewidth]{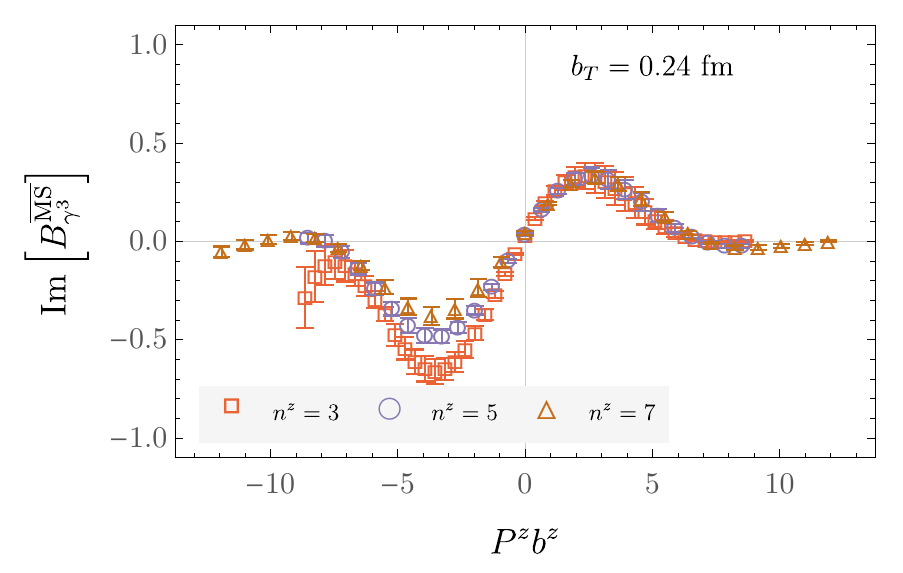}
	\includegraphics[width=0.46\linewidth]{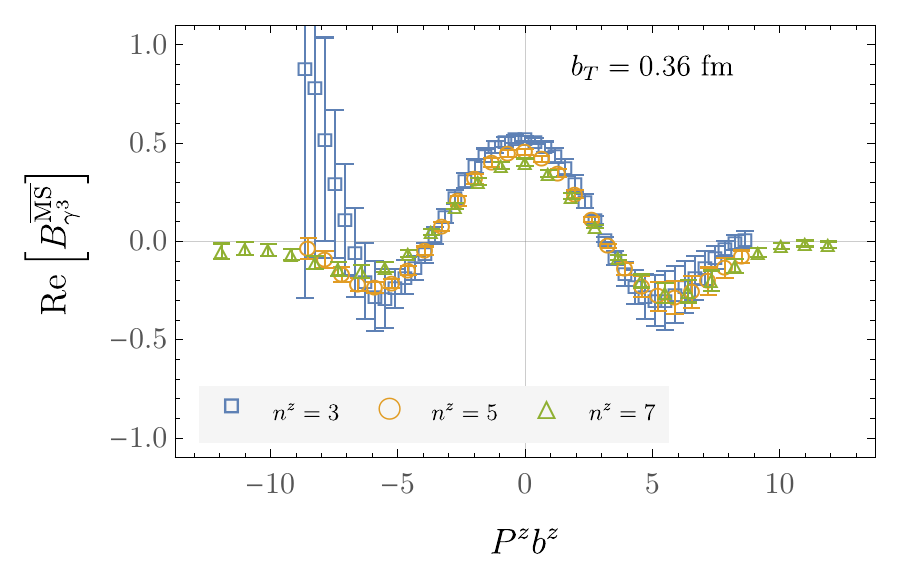}
	\includegraphics[width=0.46\linewidth]{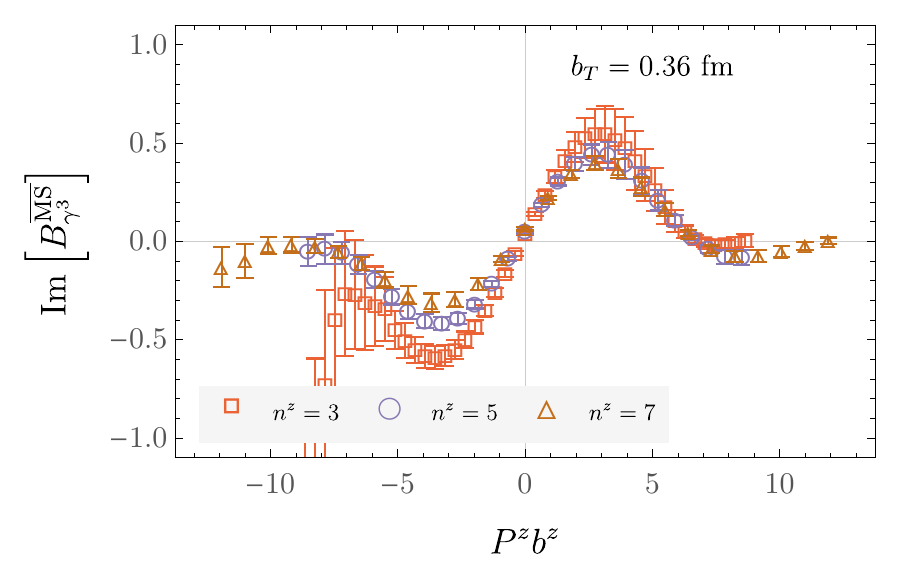}
	\includegraphics[width=0.46\linewidth]{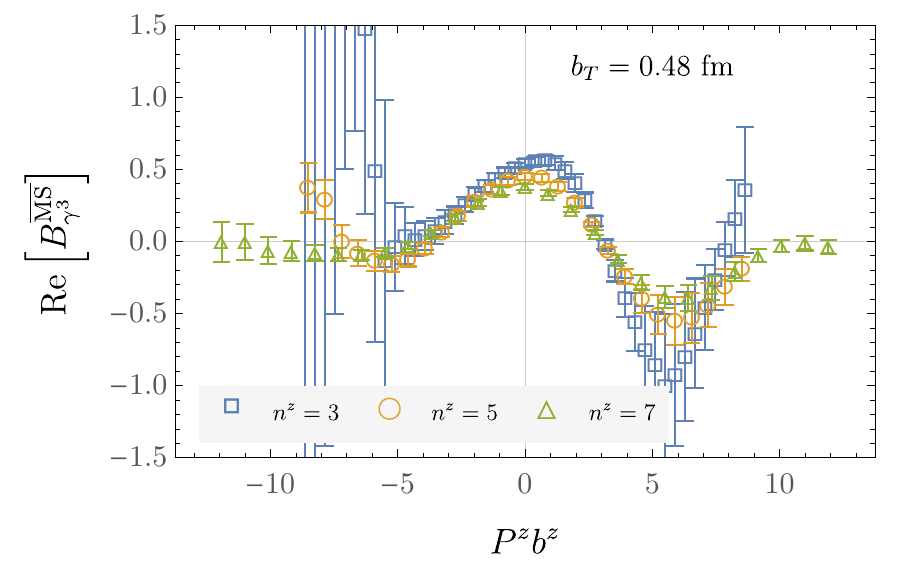}
	\includegraphics[width=0.46\linewidth]{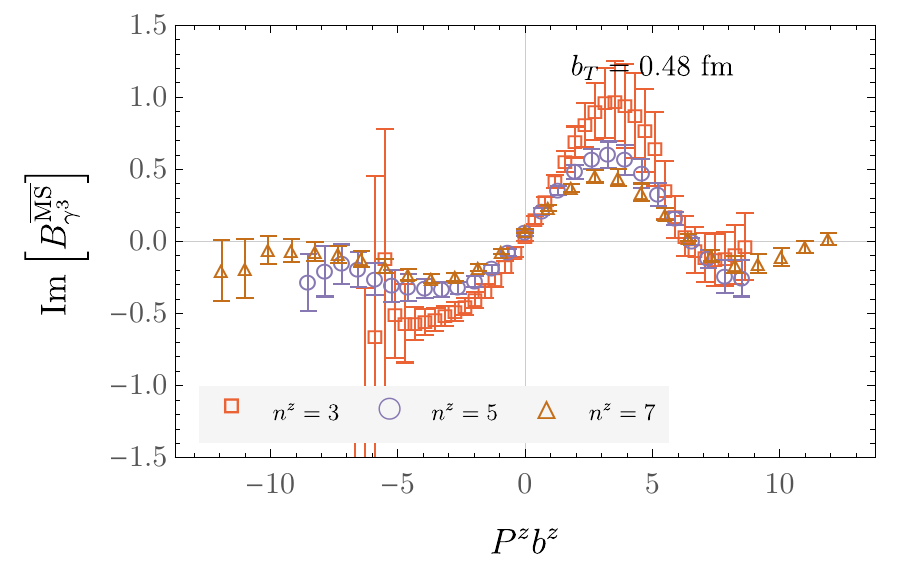}
   \caption{ Examples of the modified $\MS$-renormalized quasi beam functions $B^{\MS}_{\gamma^3}$ determined as described in Sec.~\ref{subsec:renbeam}.
		\label{fig:renorm_beam_gz}}
\end{figure*}

\begin{figure*}[!p]
	\centering
	\includegraphics[width=0.46\linewidth]{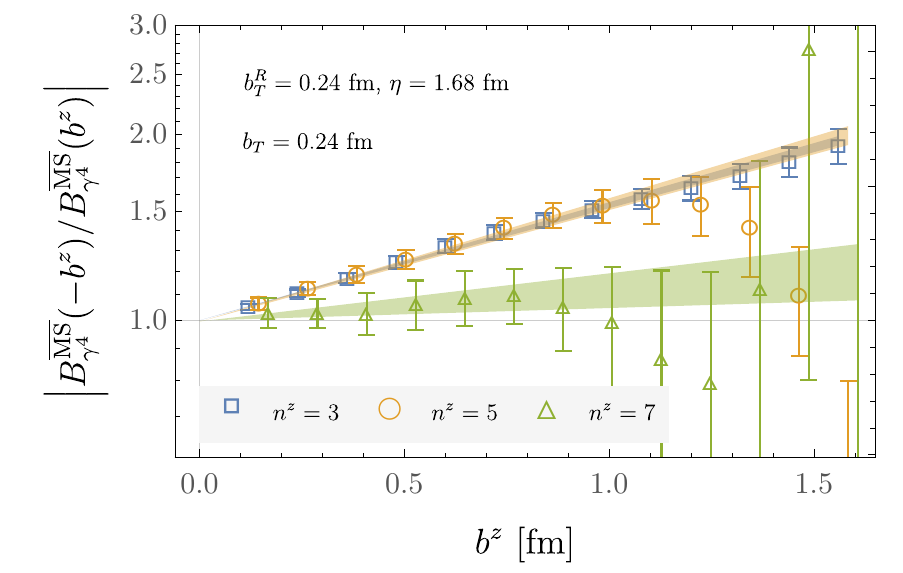}
	\includegraphics[width=0.46\linewidth]{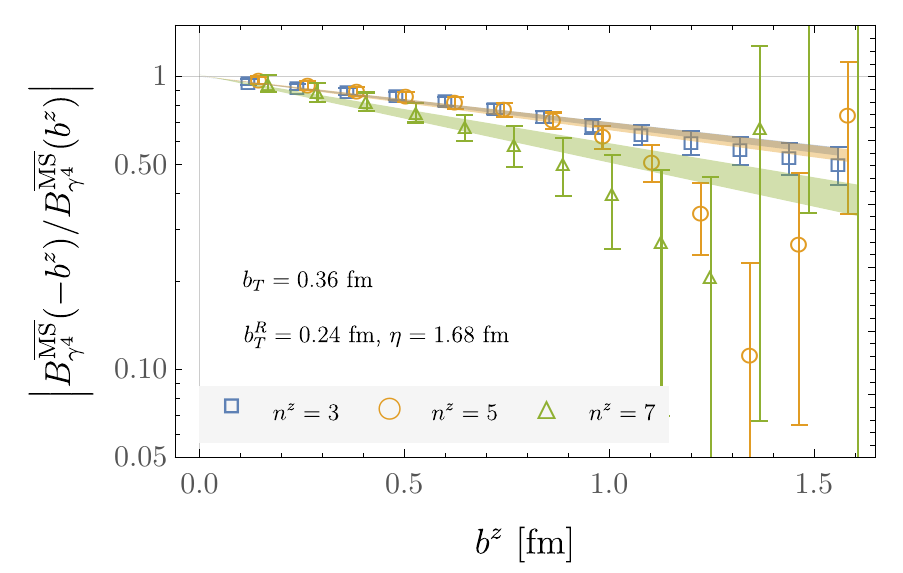}
	\includegraphics[width=0.46\linewidth]{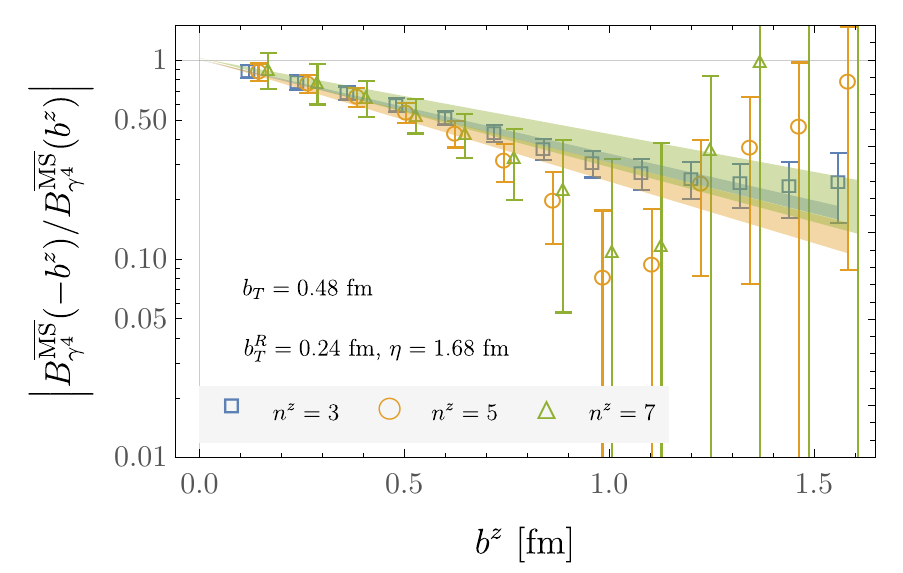}
	\includegraphics[width=0.46\linewidth]{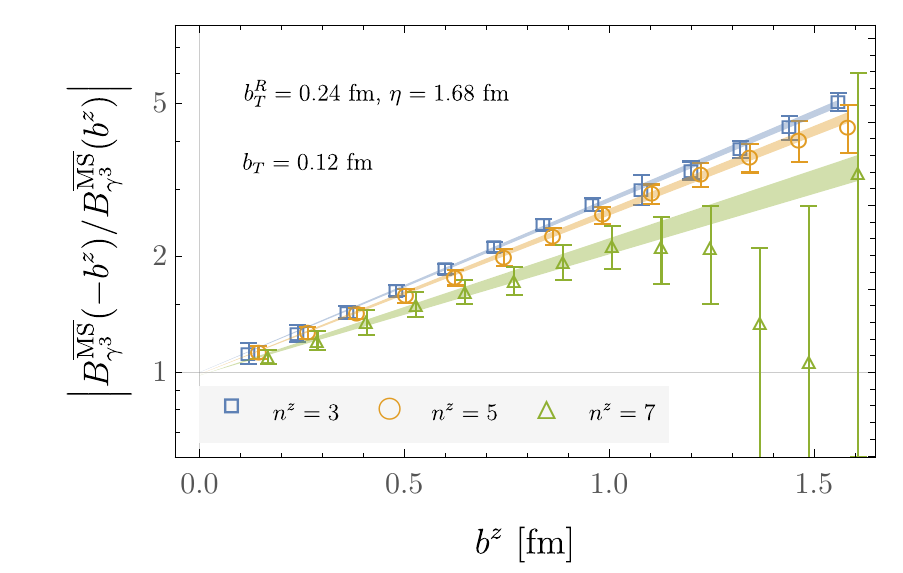}
	\includegraphics[width=0.46\linewidth]{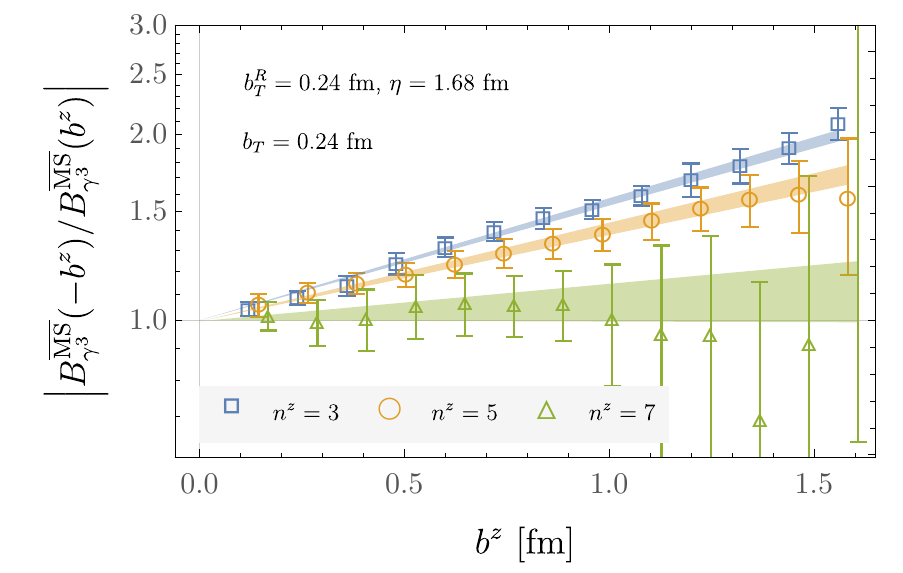}
	\includegraphics[width=0.46\linewidth]{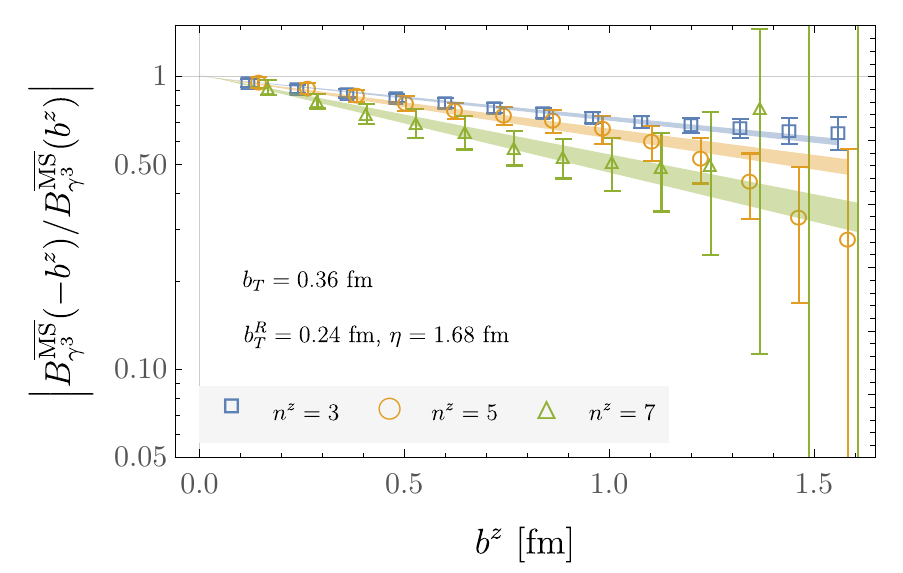}
	\includegraphics[width=0.46\linewidth]{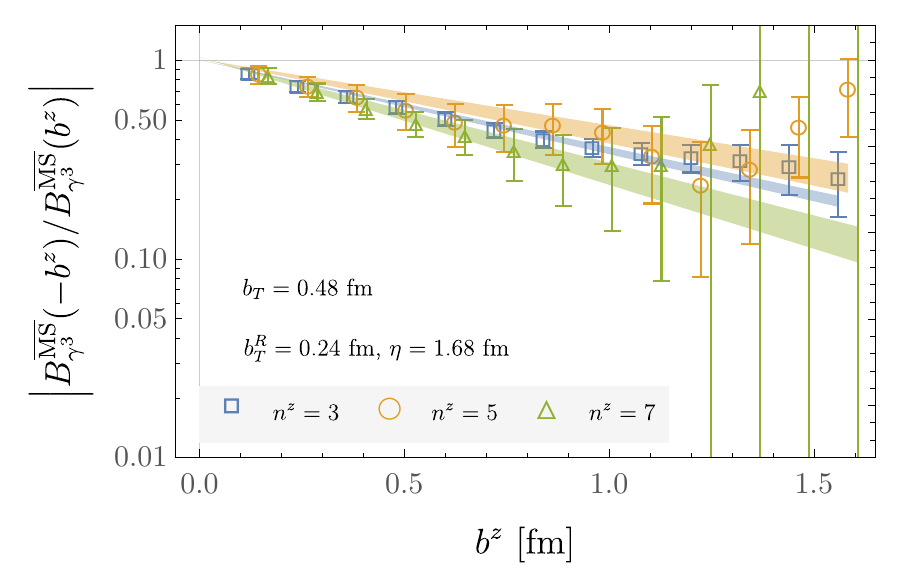}\\
   \caption{Fits to the $b^z$-dependent asymmetry in the modified $\MS$-renormalized quasi beam functions, as detailed in Sec.~\ref{subsec:renbeam} (the asymmetry in the ratio of beam functions $B^{\MS}_{\gamma^4}$ with $b_T$=0.12~fm, $\eta=1.68$~fm is provided in Fig.~\ref{fig:asymfig} of the main text). 
		\label{fig:asymm_beam_rat}}
\end{figure*}

\begin{figure*}[!t]
        \includegraphics[width=0.46\textwidth]{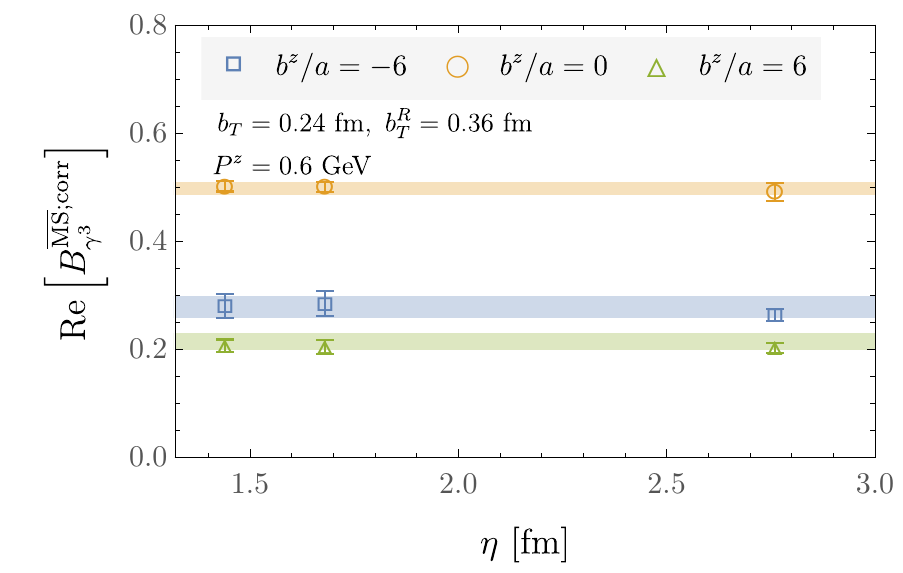}
        \includegraphics[width=0.46\textwidth]{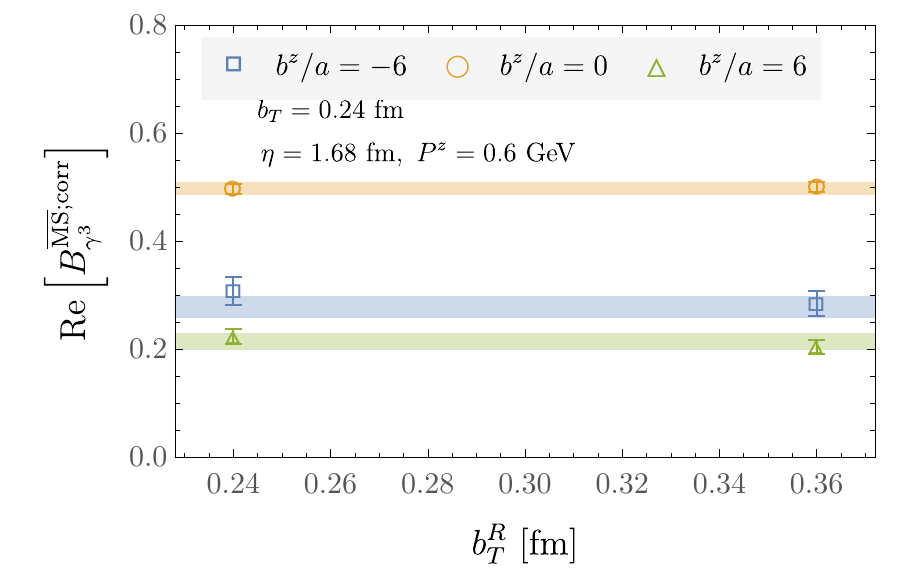}
        \caption{\label{fig:renorm_vs_bTR_gz}  Example of the asymmetry-corrected modified $\MS$-renormalized quasi beam function $B^{\MS;\text{corr}}_{\gamma^3}$, including the results of weighted averages of this quantity over choices of $b_T^R$ and $\eta$ (as a function of $b^z$, $b_T$, and $P^z$), as described in the main text. Fig.~\ref{fig:renorm_vs_bTR} displays the analogous figure for $B^{\MS;\text{corr}}_{\gamma^4}$. }
\end{figure*}

\begin{figure*}[!tp]
	\centering
	\includegraphics[width=0.46\linewidth]{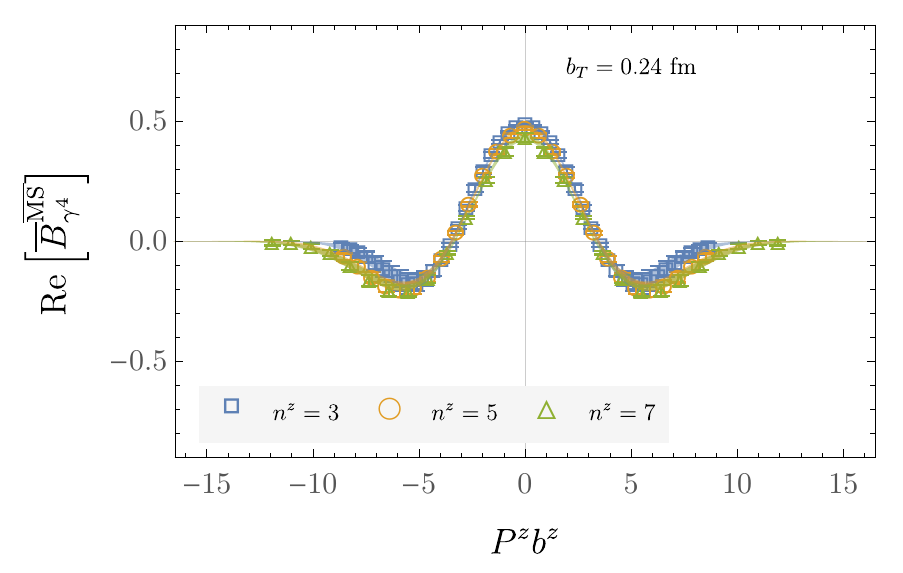}
	\includegraphics[width=0.46\linewidth]{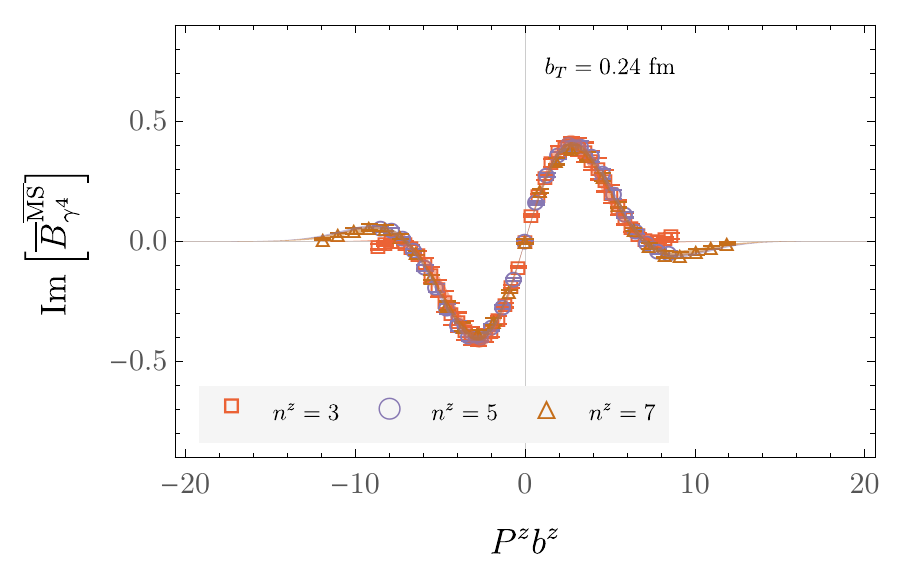}
	\includegraphics[width=0.46\linewidth]{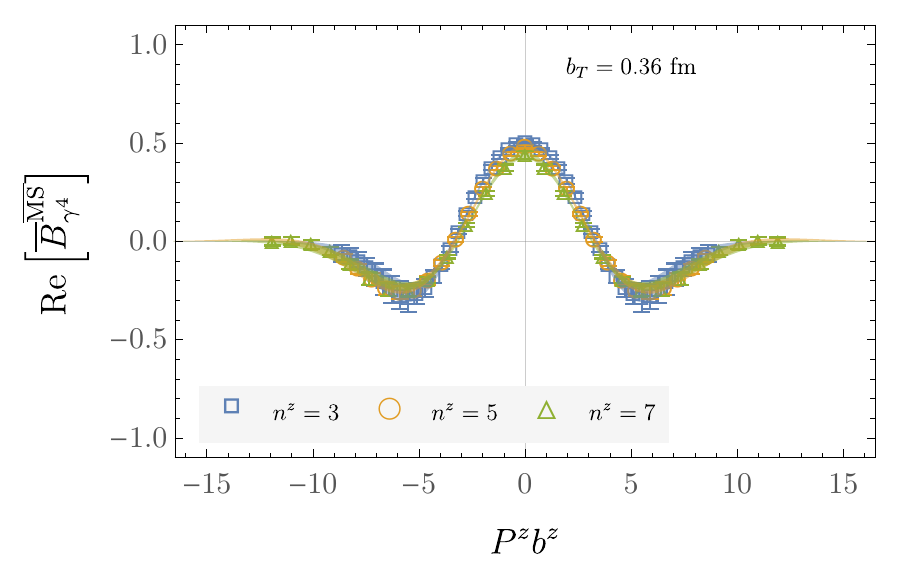}
	\includegraphics[width=0.46\linewidth]{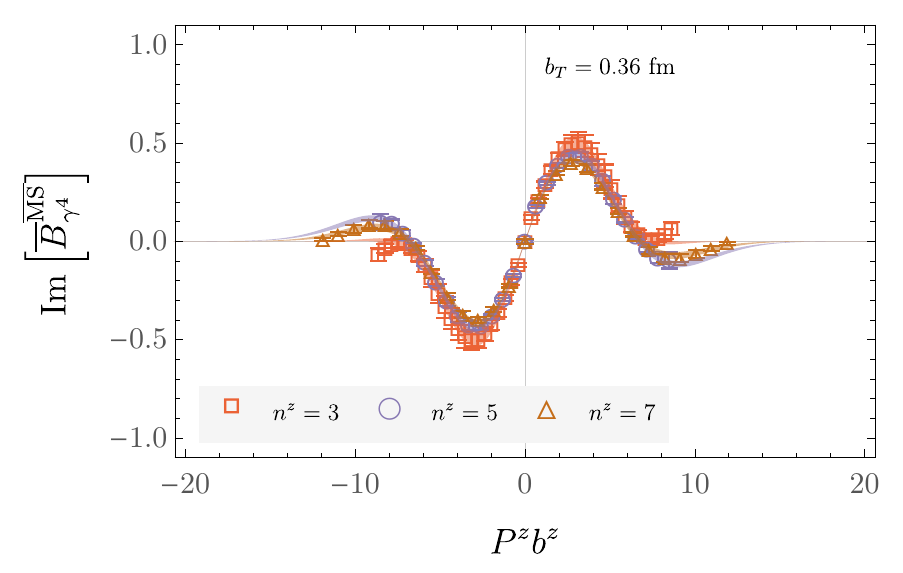}
	\includegraphics[width=0.46\linewidth]{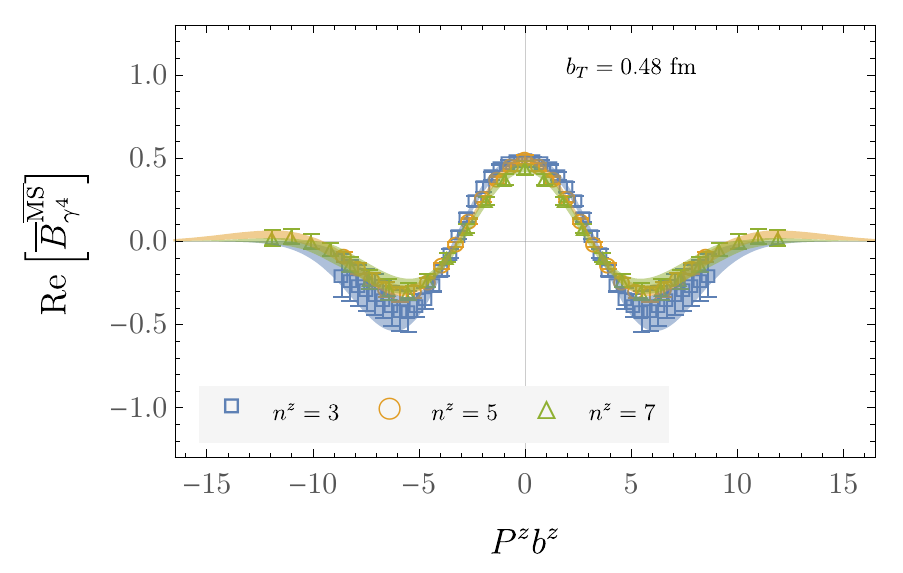}
	\includegraphics[width=0.46\linewidth]{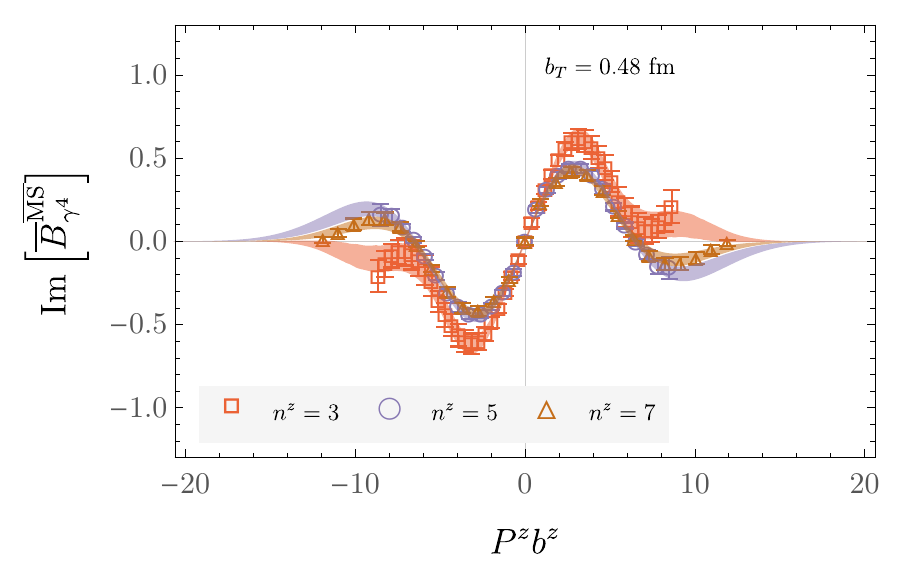}
   \caption{Examples of the averaged asymmetry-corrected modified $\MS$-renormalized quasi beam functions $\overline{B}^{\MS}_{\gamma^4}$, including fits by Eqs.~\eqref{eq:refit} and \eqref{eq:imfit} to the real and imaginary parts, shown as shaded bands. Fig.~\ref{fig:beam-fits} of the main text shows the example for $b_T=0.12$~fm. 
		\label{fig:symm_beam}}
\end{figure*}

\begin{figure*}[p]
	\centering
	\includegraphics[width=0.46\linewidth]{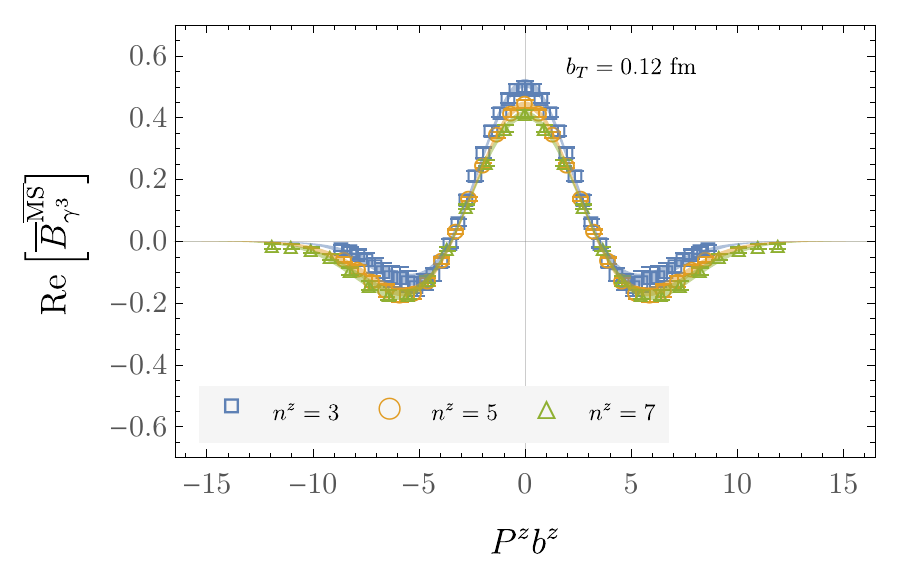}
	\includegraphics[width=0.46\linewidth]{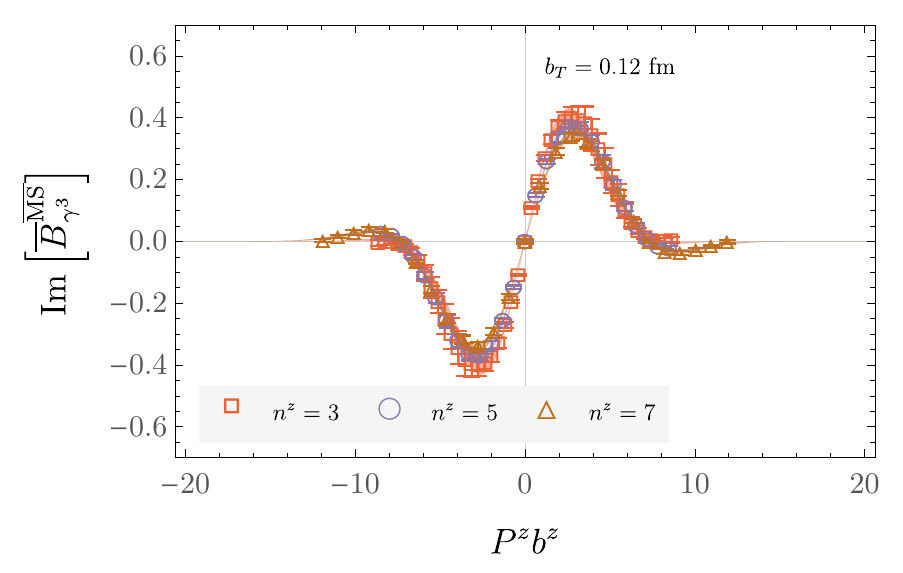}
	\includegraphics[width=0.46\linewidth]{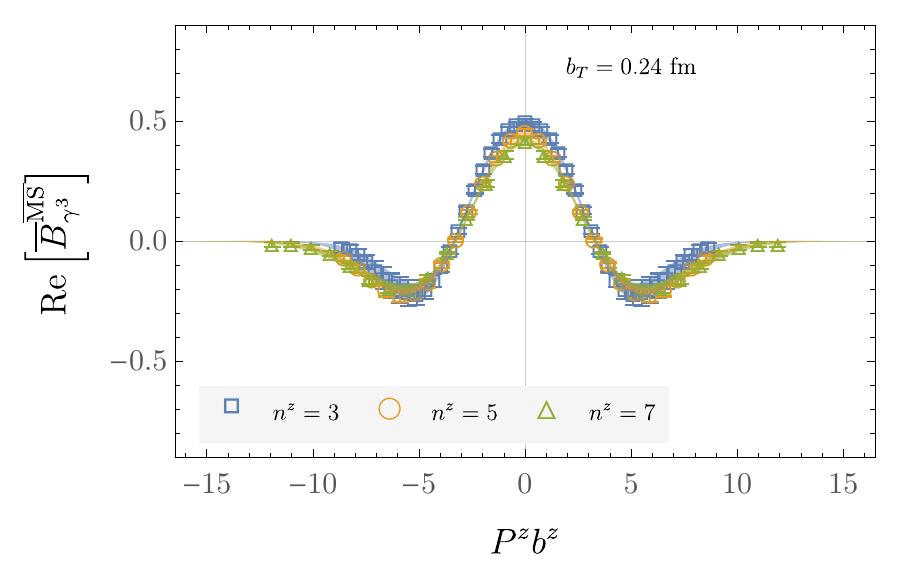}
	\includegraphics[width=0.46\linewidth]{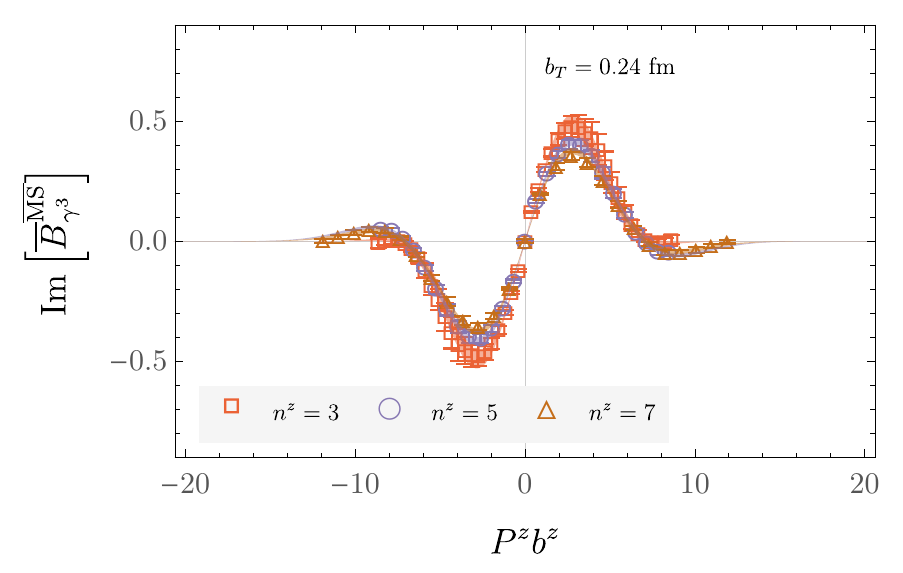}
	\includegraphics[width=0.46\linewidth]{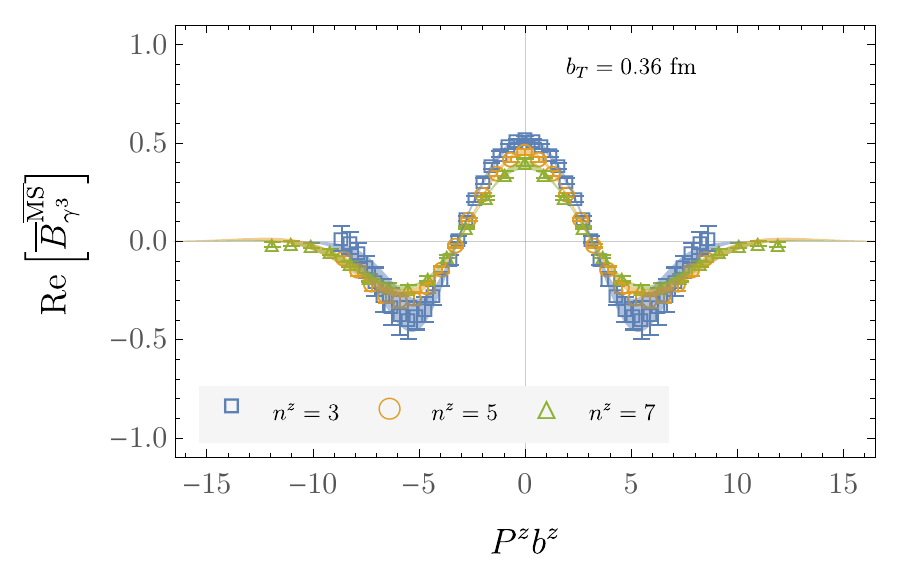}
	\includegraphics[width=0.46\linewidth]{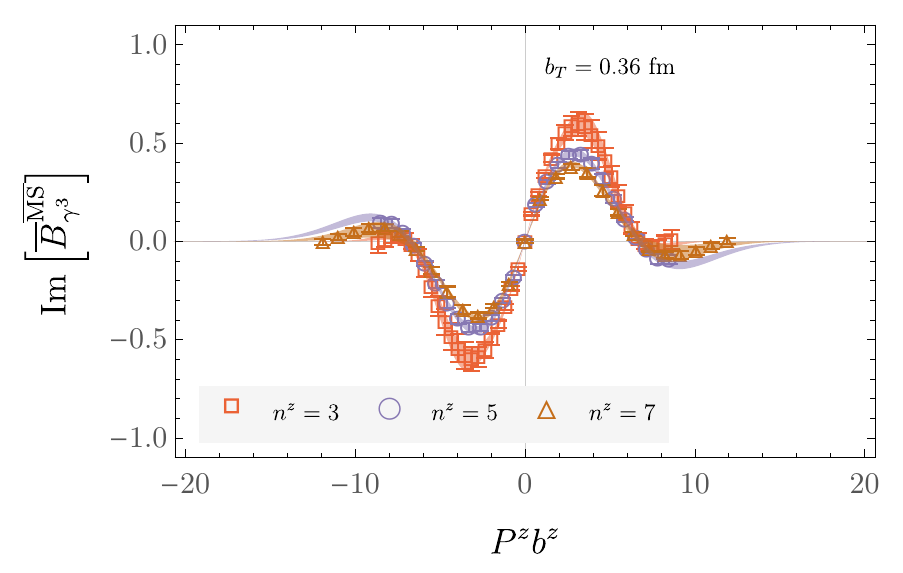}
	\includegraphics[width=0.46\linewidth]{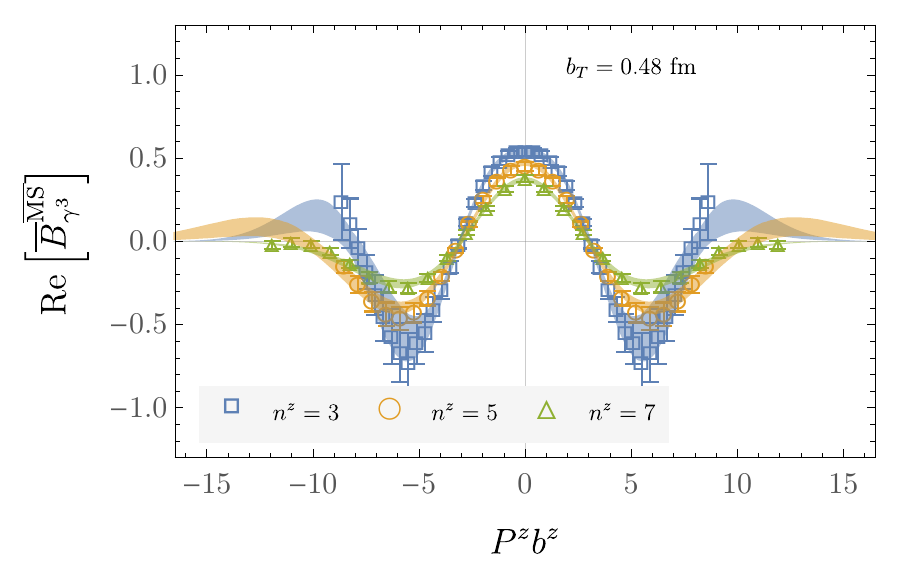}
	\includegraphics[width=0.46\linewidth]{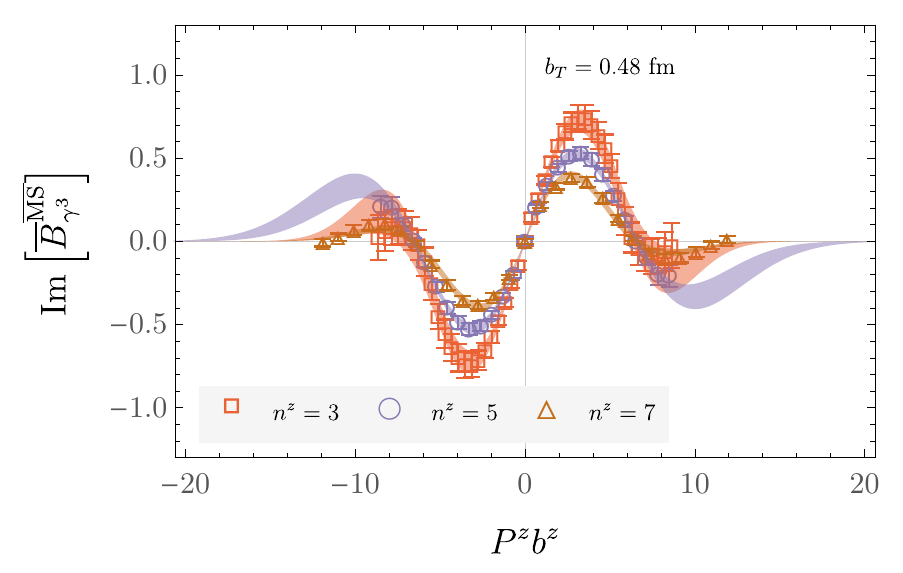}
   \caption{As in Fig.~\ref{fig:symm_beam}, for $\overline{B}^{\MS}_{\gamma^3}$.  
		\label{fig:symm_beam_gz}}
\end{figure*}

This section collates additional examples of results at intermediate states of the numerical calculation and analysis presented in Sec.~\ref{sec:extr}:
\begin{itemize}
    \item Bare quasi beam functions $B_\Gamma^\text{bare}(b^z,\vec{b}_T,a,\eta,P^z)$: supplementing Fig.~\ref{fig:bare_beam_eg} of the main text, additional examples of the bare quasi beam functions are provided in Fig.~\ref{fig:bare_beam} for $B_{\gamma^4}^\text{bare}$ and Fig.~\ref{fig:bare_beam_gz} for $B_{\gamma^3}^\text{bare}$.
    \item Renormalized quasi beam functions $B^{\MS}_{\Gamma}(\mu, b^z,\vec{b}_T,a,\eta,P^z)$: examples of the modified $\MS$-renormalized quasi beam functions $B^{\MS}_{\gamma^4}$ and ${B}^{\MS}_{\gamma^3}$, defined in Sec.~\ref{subsec:renbeam}, are given in Figs.~\ref{fig:renorm_beam} and ~\ref{fig:renorm_beam_gz} respectively.
    \item Renormalized quasi beam function asymmetry fits: supplementing Fig.~\ref{fig:asymfig} of the main text, fits to the $b^z$-dependence of the asymmetry in the renormalized quasi beam functions, performed as detailed in Sec.~\ref{subsec:renbeam}, are illustrated in Fig.~\ref{fig:asymm_beam_rat}.
    \item Asymmetry-corrected modified $\MS$-renormalized quasi beam functions $B^{\MS;\text{corr}}_{\Gamma}(\mu,b^z, \bt, a, \eta, P^z)$: an example of the dependence of $B^{\MS;\text{corr}}_{\gamma^3}$ on $b_T^R$ and $\eta$ is provided in Fig.~\ref{fig:renorm_vs_bTR_gz}, supplementing the analogous figure for $B^{\MS;\text{corr}}_{\gamma^4}$ which is presented in Fig.~\ref{fig:renorm_vs_bTR} in the main text.
    \item Averaged asymmetry-corrected modified $\MS$-renormalized quasi beam functions $\overline{B}^{\MS}_{\Gamma}(\mu,b^z,b_T,P^z)$: in addition to the example provided in Fig.~\ref{fig:beam-fits} of the main text, Figs.~\ref{fig:symm_beam} and \ref{fig:symm_beam_gz} give examples of $\overline{B}^{\MS}_{\gamma^4}$ and $\overline{B}^{\MS}_{\gamma^3}$, including fits to the $P^zb^z$-dependence of these functions and extrapolations to larger $P^zb^z$, performed as described in Sec.~\ref{sec:extr}.
    \item Estimator $\hat{\gamma}^q_\zeta(\mu, b_T; P_1^z,P_2^z,x)$ for the Collins-Soper kernel (Eq.~\eqref{eq:gammahat}): supplementing Fig.~\ref{fig:CSxfits} of the main text, the remaining numerical results for $\hat{\gamma}^q_\zeta$ as a function of $x$ at different values of $b_T$ are displayed in Fig.~\ref{fig:CS_x_dependence_all}.
\end{itemize}

\section{Discrete Fourier transform analysis}
\label{app:DFT}

\begin{figure*}[!p]
	\centering
	\includegraphics[width=.46\linewidth]{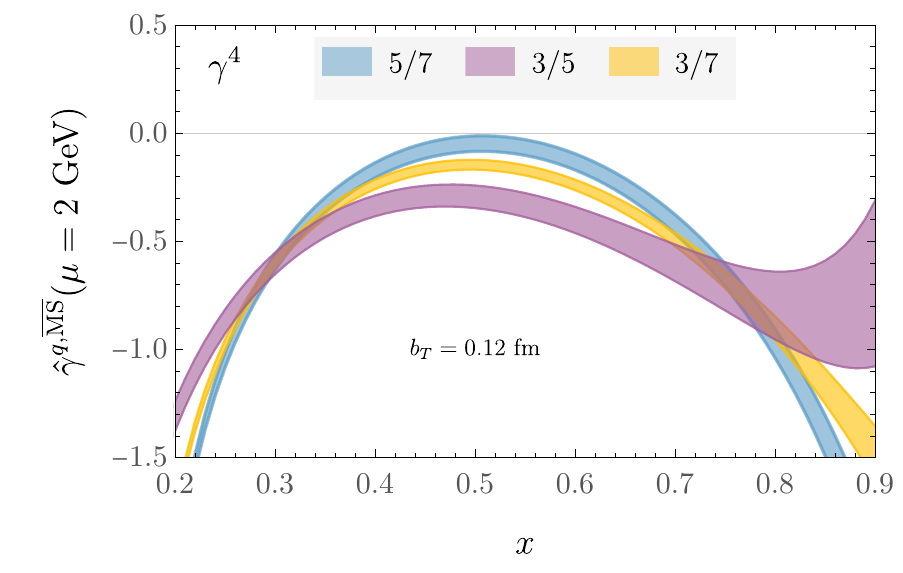}
	\includegraphics[width=0.46\linewidth]{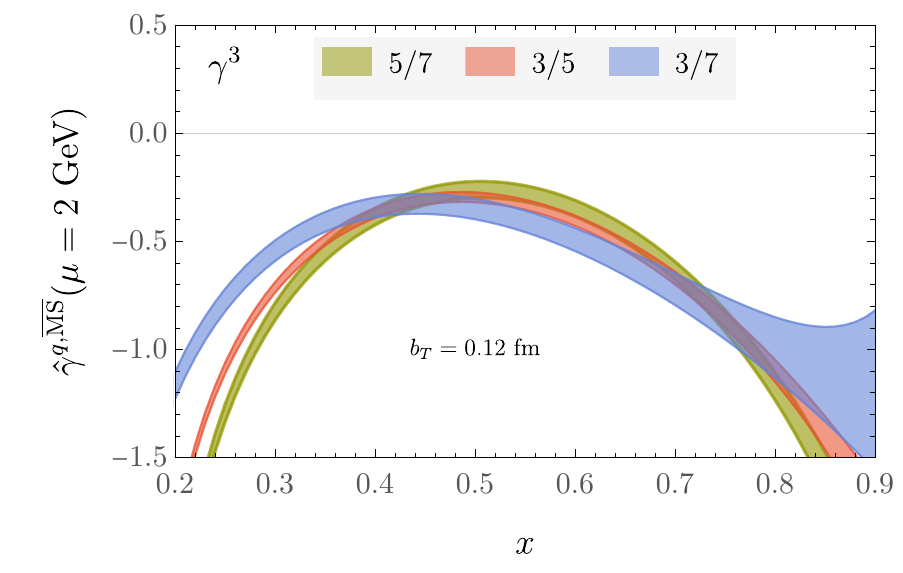}
	\includegraphics[width=.46\linewidth]{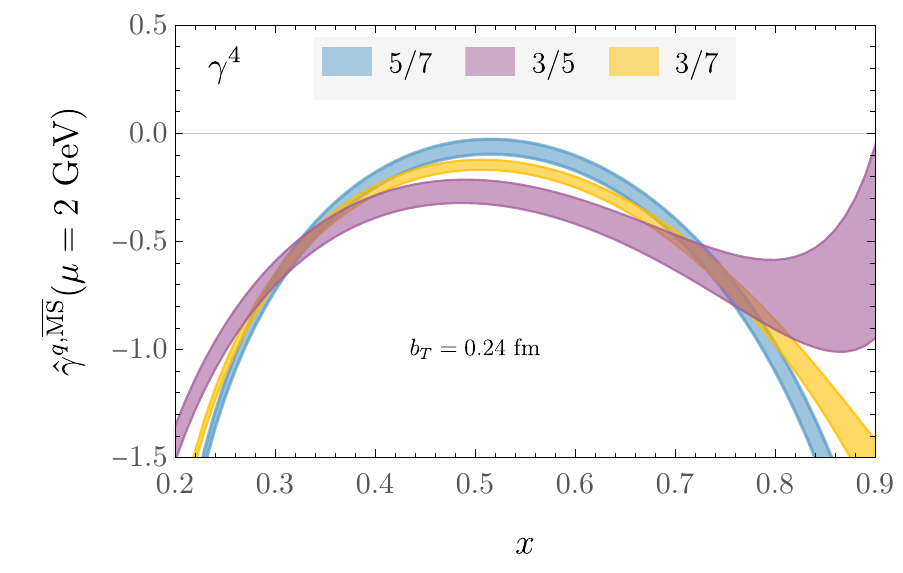}
	\includegraphics[width=0.46\linewidth]{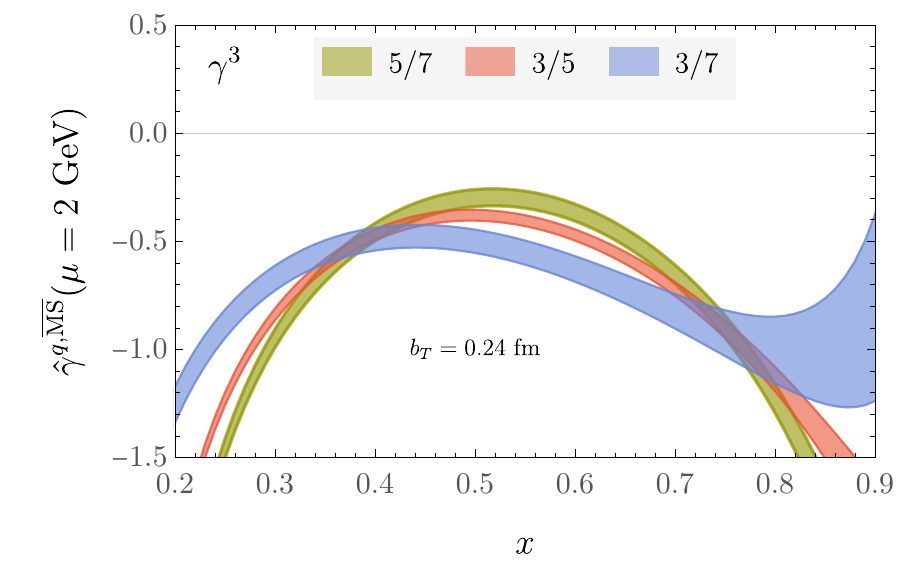}
	\includegraphics[width=.46\linewidth]{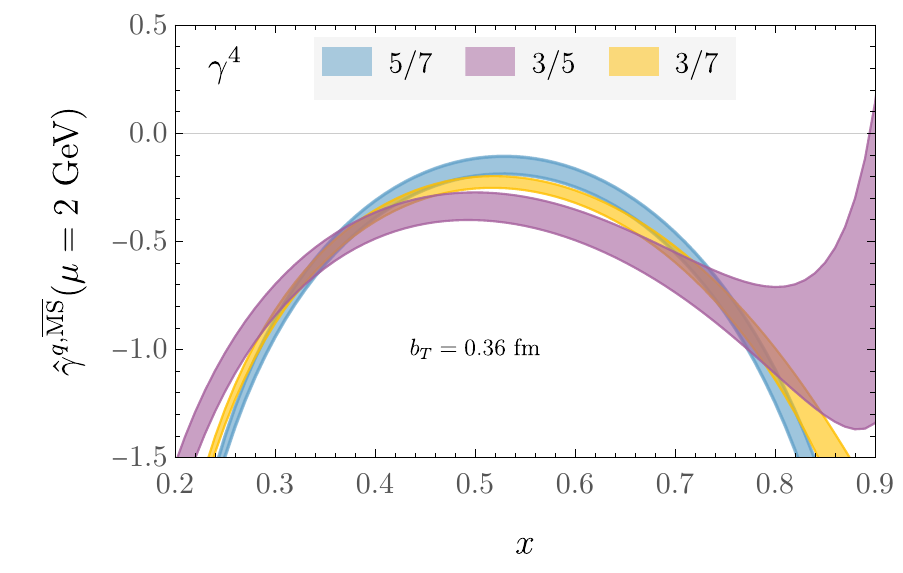}
	\includegraphics[width=0.46\linewidth]{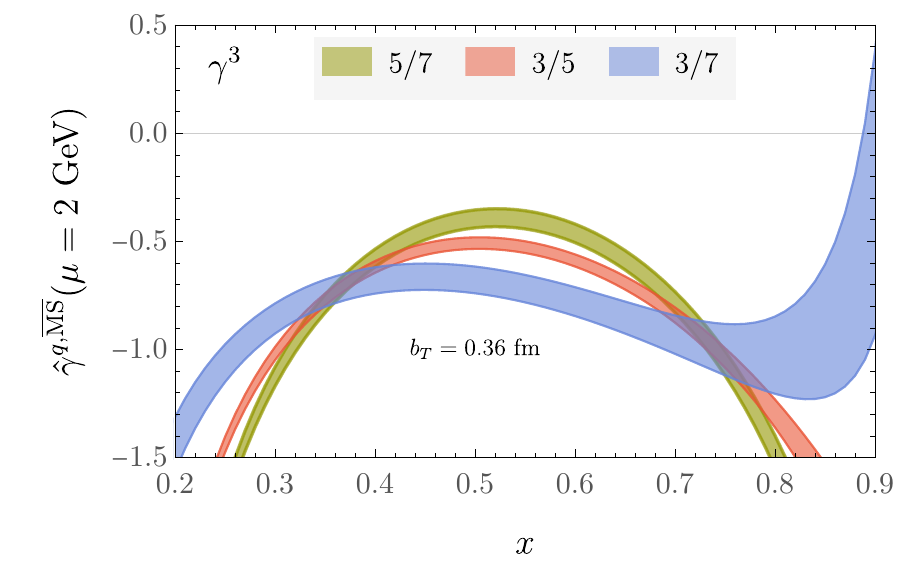}
	\includegraphics[width=.46\linewidth]{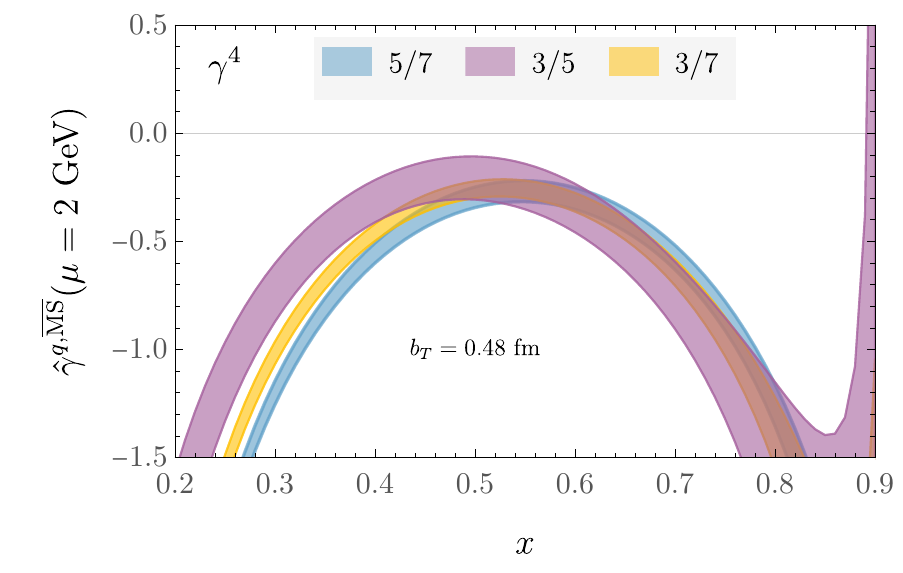}
	\includegraphics[width=0.46\linewidth]{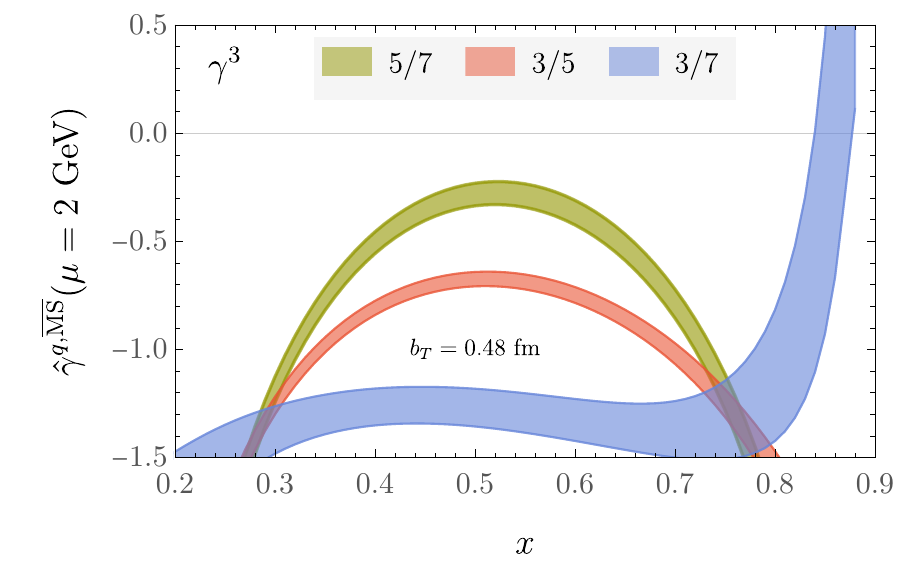}
	\caption{$\hat{\gamma}^q_\zeta$ computed as defined in Eq.~\eqref{eq:gammahat} with a DFT of $\overline{B}^{\MS}_{\Gamma}$ replacing the Fourier transform of analytic fits $\hat{B}^{\MS}_{\Gamma}$ to the $P^zb^z$-dependence of the quasi beam functions, for momentum pairs $\{P_1^z,P_2^z\}$ denoted by $P_1^z/P_2^z$ in the legend. 
		\label{fig:CS_x_dependence_DFT}}
\end{figure*}

As discussed in Sec.~\ref{sec:extr}, a model-independent lattice QCD extraction of the Collins-Soper kernel by the approach followed here would require that Eq.~\eqref{eq:gammahat} is evaluated with a DFT of $\overline{B}^{\MS}_{\Gamma}$ replacing the Fourier transform of analytic fits $\hat{B}^{\MS}_{\Gamma}$ to the $P^zb^z$-dependence of the quasi beam functions, and that the results are stable under truncation of the data in $P^zb^z$. The $P^zb^z$-range of the data presented here is, however, not sufficient for the tails of the quasi beam functions at large $|P^zb^z|$ to decay to plateaus consistent with zero, particularly at the largest $b_T$ and smallest $P^z$ values used in this study. It is thus to be expected that a DFT-based analysis has qualitative and quantitative differences from the analytic model approach. These differences can be seen by comparison of Fig.~\ref{fig:CS_x_dependence_DFT}, which displays the results of a DFT-based analysis, with Figs.~\ref{fig:CSxfits} and \ref{fig:CS_x_dependence_all}, which show the results of the analysis of Sec.~\ref{sec:extr}. As anticipated, the DFT approach yields numerical values which are significantly different from those achieved by modeling rather than truncating the tails of the quasi beam functions in $P^zb^z$, particularly for evaluations including quasi beam functions computed with the smallest boost corresponding to $n^z=3$. As a result, the values obtained with different choices of $P^z$ are not consistent within uncertainties at intermediate values of $x$. The differences are, naturally, less significant for results computed with the largest choices of $P^z$, supporting the expectation that future studies constraining larger values of $P^zb^z$ will achieve model-independent results via the DFT approach.

\section{Power corrections and higher-twist effects}
\label{app:powerandtwist}

\begin{figure*}[!p]
	\centering
	\includegraphics[width=.46\linewidth]{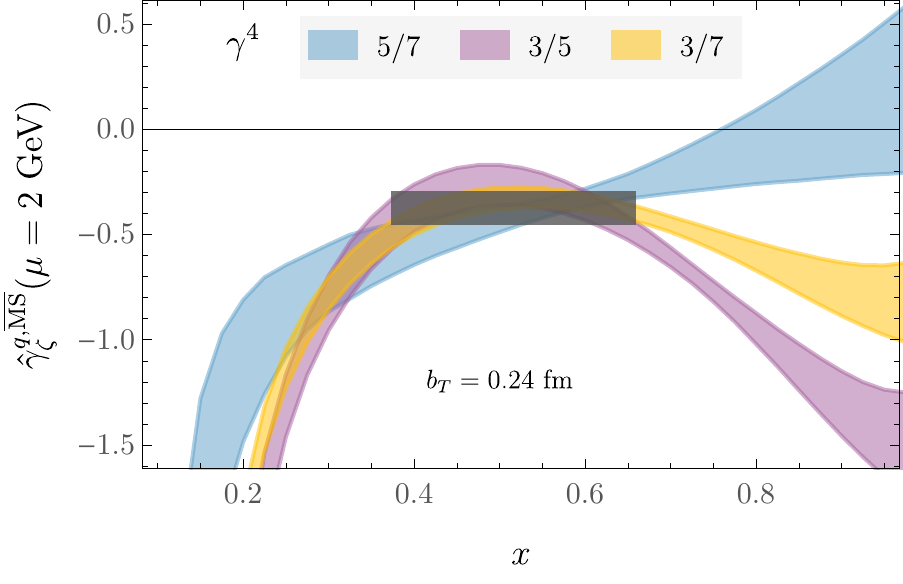}
	\includegraphics[width=.46\linewidth]{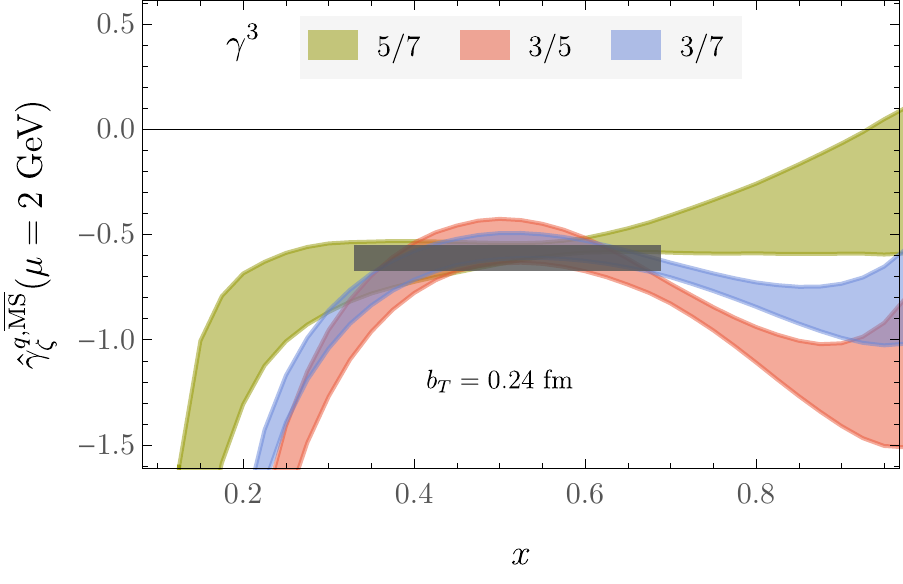}
	\includegraphics[width=.46\linewidth]{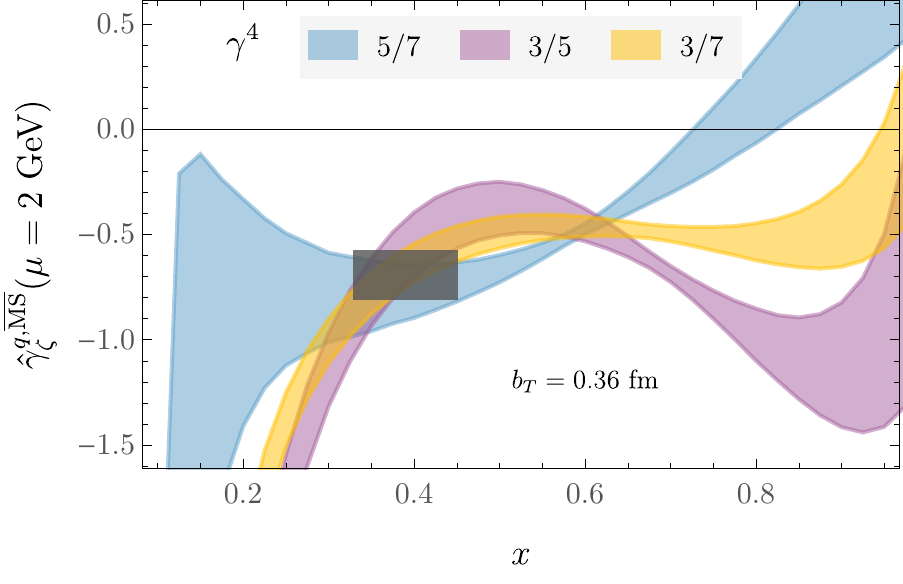}
	\includegraphics[width=.46\linewidth]{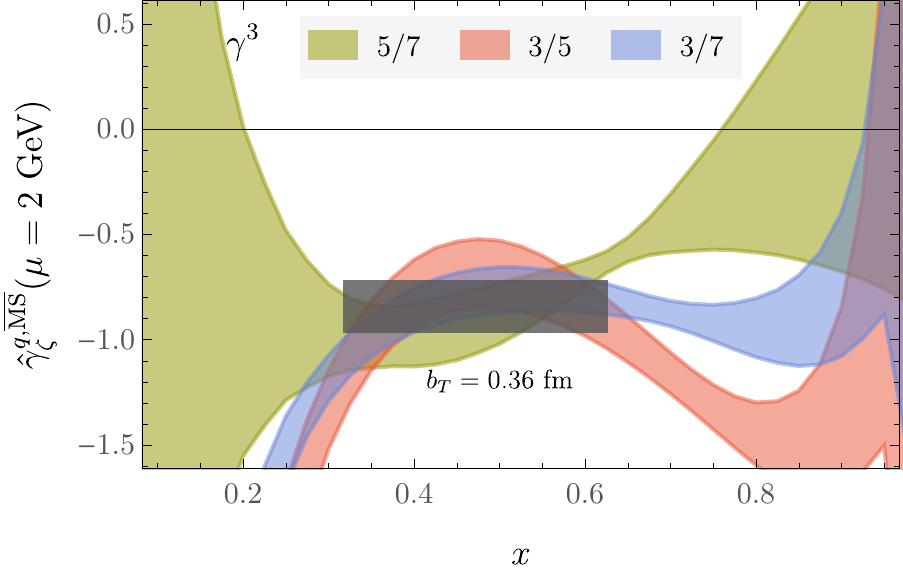}
	\includegraphics[width=.46\linewidth]{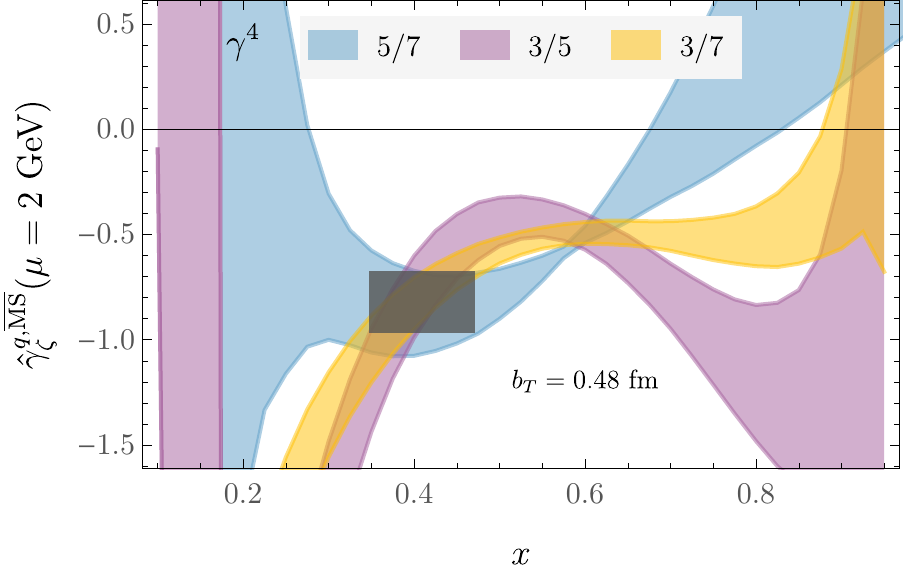}
	\includegraphics[width=.46\linewidth]{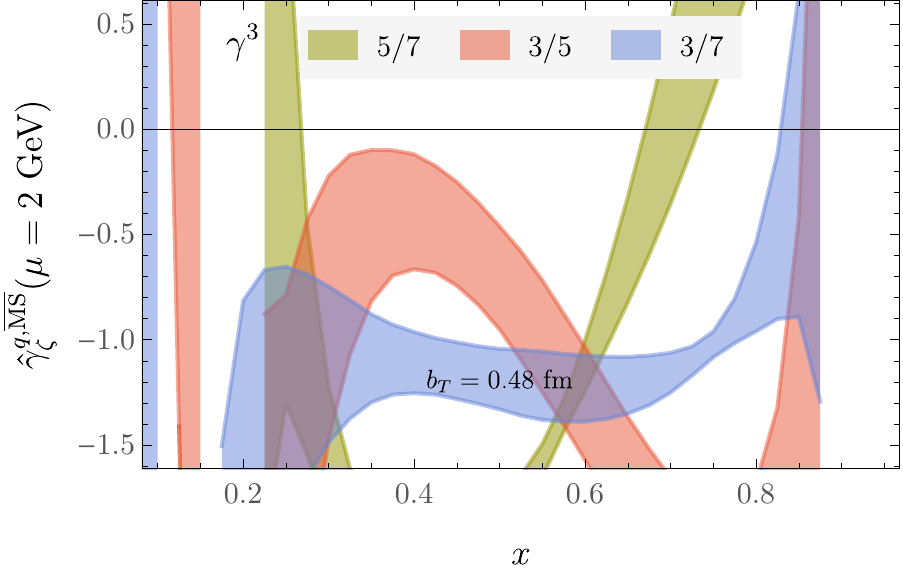}
   \caption{ $\hat{\gamma}^q_\zeta$ computed as defined in Eq.~\eqref{eq:gammahat} for momentum pairs $\{P_1^z,P_2^z\}$, denoted by $P_1^z/P_2^z$ in the legend. The horizontal shaded band shows the total uncertainty of the best result, and the corresponding $x$-window, determined as described in the text. Fig.~\ref{fig:CSxfits} of the main text shows the analogous results for $b_T=0.12$~fm.
		\label{fig:CS_x_dependence_all}}
\end{figure*}

The estimator $\hat{\gamma}^q_\zeta(\mu, b_T; P_1^z,P_2^z,x)$ (Eq.~\eqref{eq:gammahat}) coincides with the Collins-Soper kernel up to power corrections, such as higher-twist corrections in the factorization formula for the quasi TMDPDF, and discretization artifacts; in the absence of contamination from these effects, $\hat{\gamma}^q_\zeta$ should be independent of $x$, $P_1^z$ and $P_2^z$. Clearly, the results shown in Figs.~\ref{fig:CSxfits} and \ref{fig:CS_x_dependence_all} indicate that this contamination is not negligible relative to the uncertainties of this calculation. As discussed in Sec.~\ref{sec:extr}, it is natural to attempt to model, fit, and subtract this contamination in order to determine a best value for the Collins-Soper kernel. 

One possible approach is to consider a simple model of corrections to the factorization formula~\cite{Ebert:2019okf,Ji:2019ewn} for the quasi TMDPDF, for example through the inclusion of free parameters $\lambda_1$ and $\lambda_2$ parameterizing power corrections as: 
\begin{align}\label{eq:power}
    \tilde f_{\ns}^\TMD(x, \bt, &\mu, P^z) = \left[ C^\TMD_\ns (\mu,x P^z) + {\lambda_1 \Lambda_{\text{QCD}}^2 \over (xP^z)^2}  \right] \nn\\
   &\times \exp\left[{1\over 2}\gamma^q_\zeta(\mu,b_T) \ln{\frac{(2xP^z)^2}{ \zeta_0}}\right]g_S^q(b_T,\mu) \nn\\
    &\times f_{\ns}^\TMD(x, \bt, \mu, \zeta_0)\left[ 1  + {\lambda_2 \over (xP^zb_T)^2} \right]\,
\end{align}
(here $g_S^q(b_T,\mu)$ is defined as in Ref.~\cite{Ebert:2018gzl} as the mismatch between the lightlike and quasi soft factors). This form is chosen since in the $\MS$ scheme the higher-twist corrections must appear in even powers, with a suppression through $k_T^2/k_z^2$, which in Fourier space becomes $1/(xP^zb_T)^2$. At tree-level, the factor $\lambda_2$ is a constant. The $1/(xP^z)^2$ power correction comes from the renormalon ambiguity in the perturbative series for the matching coefficient.

The relationship between the quasi TMDPDF and the quasi beam functions (Eq.~\eqref{eq:quasiTMD}) then suggests a model of the Fourier transform of the quasi beam functions, defined as
\begin{equation}
    \widetilde{B}^{\MS}_{\gamma^4}(\mu,x,b_T,P^z) \equiv \int db^z\, e^{i b^z x P^z} P^z\hat{B}^{\MS}_{\gamma^4}(\mu,b^z,b_T,P^z),
\end{equation}
of the form

\begin{align}
&\widetilde{B}^\text{fit}(\mu, x, b_T, P^z) \nn\\
   &= \left[ C^\TMD_\ns (\mu,x P^z) + {\lambda_1 \Lambda_{\rm QCD}^2 \over (xP^z)^2} \right]\left({(2xP^z)^2\over \zeta_0}\right)^{{1\over 2}\gamma_\zeta(\mu,b_T)}\nn\\
    &\quad\times F(x,b_T,\mu,\zeta_0)\left[ 1  + {\lambda_2 \over (xP^zb_T)^2} \right]\,,
\end{align}
where $\gamma_\zeta$, $F$, $\lambda_1$ and $\lambda_2$ are free parameters and $\zeta_0$ can be chosen freely.
For each $x$, the model can be fit to $\widetilde{B}^{\MS}_{\gamma^4}$ (and separately to $\widetilde{B}^{\MS}_{\gamma^3}$) for all choices of $P^z$ and $b_T$ simultaneously, and the Collins-Soper kernel extracted as the fit parameter $\gamma_\zeta$. 

The results of this procedure yield results for the Collins-Soper kernel which are not more consistent with a constant in $x$ than the results without the correction applied. A global fit at discretized $x$ values is of poor quality, with $\chi^2/\text{d.o.f.}\gtrsim 2$. There are a range of reasons that the model form above may not a good description of the numerical data; for example, the assumption that the $1/(xP^zb_T)^2$ corrections are proportional to the leading power contribution in \eq{power}, may not be a good approximation. This approach is thus not taken in the main analysis presented here, but may be worthwhile to consider for future studies with larger values of $P^z$ where the power corrections will be suppressed relative to those in the current work.

\section{NLO matching effects}
\label{app:NLOmatching}

\begin{figure}[t]
	\centering
   \includegraphics[width=\linewidth]{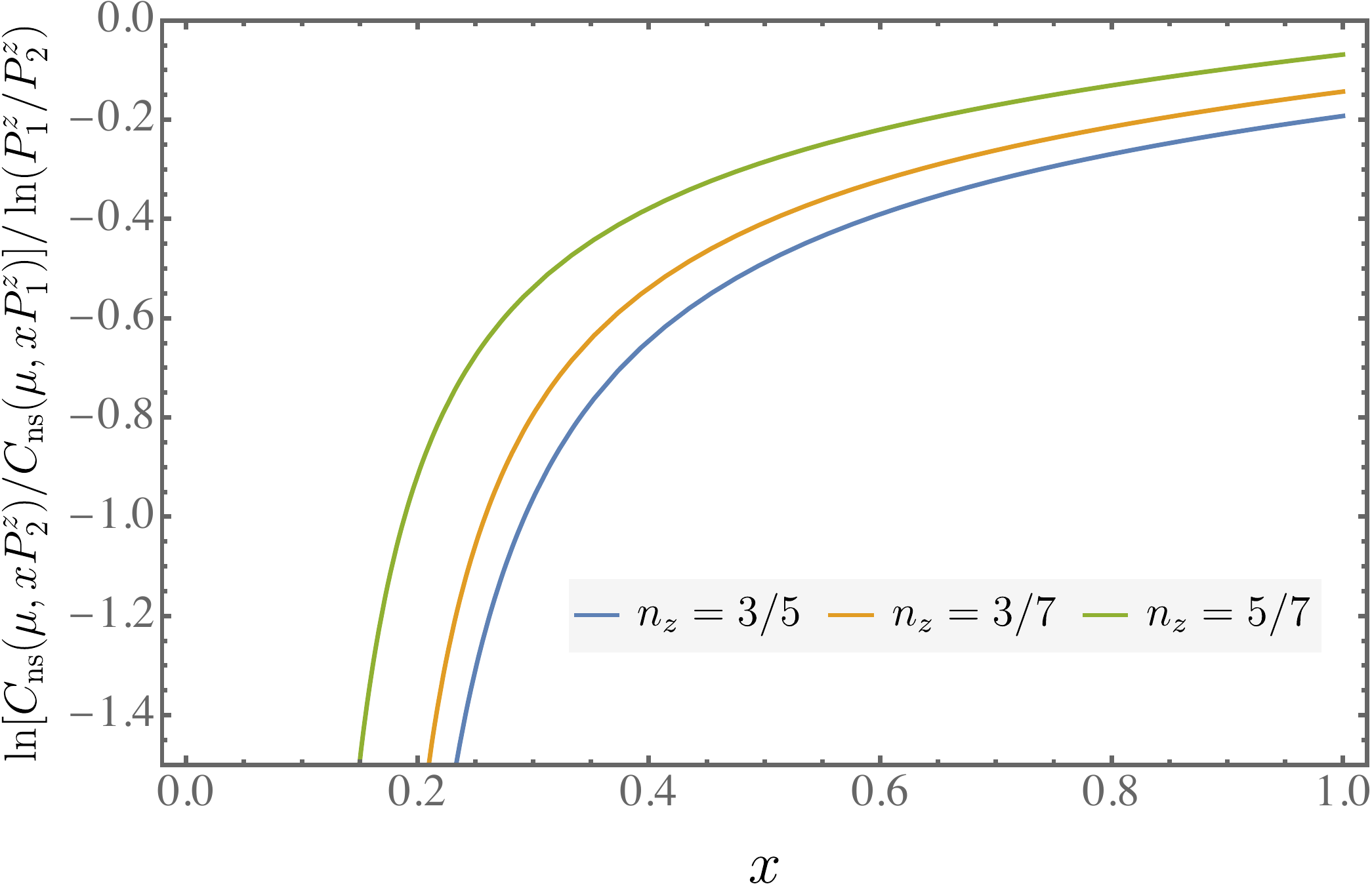}
	\caption{Contribution of the NLO matching coefficient to the estimator $\hat{\gamma}^q_\zeta$ for momentum pairs $\{P_1^z,P_2^z\}$ denoted by $P_1^z/P_2^z$ in the legend. The contribution of this term with LO matching, i.e., $C_\text{ns}(\mu,xP^z)=1$, would be zero.
	\label{fig:NLOmatching}}
\end{figure}

A key difference between the approach followed to obtain the primary results of this work and a number of the alternative approaches explored in Sec.~\ref{sec:extr} is the inclusion of the perturbative matching coefficient $C^\TMD_\ns$ computed to NLO, instead of to LO, in the calculation of the estimator $\hat{\gamma}^q_\zeta(\mu, b_T; P_1^z,P_2^z,x)$ via Eq.~\eqref{eq:gammahat}. To illustrate the importance of this effect, Fig.~\ref{fig:NLOmatching} displays the relevant contribution from the NLO matching coefficient to $\hat{\gamma}^q_\zeta$, computed in Refs.~\cite{Ebert:2018gzl,Ebert:2019okf}, for each of the momentum combinations used in the numerical study of this work. Clearly this contribution, which is of the order of the Collins-Soper kernel itself for $x<0.5$, is significant, and its inclusion affects not only the value but also the $x$-dependence of the estimator $\hat{\gamma}^q_\zeta$.

\bibliography{cs500}
\end{document}